\documentclass[
reprint,
superscriptaddress,
nobibnotes,
amsmath,amssymb,
aps,
prm,
]{revtex4-2}

\usepackage{graphicx}
\usepackage{dcolumn}
\usepackage{bm}
\usepackage{subfig}
\usepackage{booktabs}
\usepackage{multirow}
\usepackage{colortbl}
\usepackage{xcolor}
\usepackage{ulem}
\newcommand{\CT}{Cr$_2$Te$_3$}

\definecolor{armygreen}{rgb}{0.29, 0.33, 0.13}

\begin{document}

\title{Epitaxial van~der~Waals heterostructures of Cr$_2$Te$_3$ on 2D materials}

\author{Quentin Guillet}
 \altaffiliation{QG and LV contributed equally to this work}
\author{Libor Vojáček}
 \altaffiliation{QG and LV contributed equally to this work}
\affiliation{Université Grenoble Alpes, CEA, CNRS, IRIG-SPINTEC, 38000 Grenoble, France
}
\author{Djordje Dosenovic}
\affiliation{Université Grenoble Alpes, CEA, IRIG-MEM, 38000 Grenoble, France
}
\author{Fatima Ibrahim}
\affiliation{Université Grenoble Alpes, CEA, CNRS, IRIG-SPINTEC, 38000 Grenoble, France
}
\author{Hervé Boukari}
\affiliation{Université Grenoble Alpes, CNRS, Institut Neel, 38000 Grenoble, France
}
\author{Jing Li}
\affiliation{Université Grenoble Alpes, CEA, Leti, F-38000 Grenoble, France
}
\author{Fadi Choueikani}
\affiliation{Synchrotron SOLEIL, L'Orme des Merisiers, 91190 Saint-Aubin, France
}
\author{Philippe Ohresser}
\affiliation{Synchrotron SOLEIL, L'Orme des Merisiers, 91190 Saint-Aubin, France
}
\author{Abdelkarim Ouerghi}
\affiliation{Université Paris-Saclay, CNRS, Centre de Nanosciences et de Nanotechnologies, Palaiseau, France
}
\author{Florie Mesple}
\affiliation{Université Grenoble Alpes, CEA, CNRS, IRIG-PHELIQS, 38000 Grenoble, France
}
\author{Vincent Renard}
\affiliation{Université Grenoble Alpes, CEA, CNRS, IRIG-PHELIQS, 38000 Grenoble, France
}
\author{Jean-François Jacquot}
\affiliation{Université Grenoble Alpes, CEA, CNRS, IRIG-SYMMES, 38000 Grenoble, France
}
\author{Denis Jalabert}
\affiliation{Université Grenoble Alpes, CEA, IRIG-MEM, 38000 Grenoble, France
}
\author{Hanako Okuno}
\affiliation{Université Grenoble Alpes, CEA, IRIG-MEM, 38000 Grenoble, France
}
\author{Mairbek Chshiev}
\affiliation{Université Grenoble Alpes, CEA, CNRS, IRIG-SPINTEC, 38000 Grenoble, France
}
\affiliation{Institut Universitaire de France, 75231 Paris, France}
\author{Céline Vergnaud}
\affiliation{Université Grenoble Alpes, CEA, CNRS, IRIG-SPINTEC, 38000 Grenoble, France
}
\author{Frédéric Bonell}
\affiliation{Université Grenoble Alpes, CEA, CNRS, IRIG-SPINTEC, 38000 Grenoble, France
}
\author{Alain Marty}
\affiliation{Université Grenoble Alpes, CEA, CNRS, IRIG-SPINTEC, 38000 Grenoble, France
}
\author{Matthieu Jamet}
\affiliation{Université Grenoble Alpes, CEA, CNRS, IRIG-SPINTEC, 38000 Grenoble, France
}
\date{\today}

\begin{abstract}
Achieving large-scale growth of two-dimensional (2D) ferromagnetic materials with high Curie temperature ($T_\mathrm{C}$) and perpendicular magnetic anisotropy (PMA) is highly desirable for the development of ultra-compact magnetic sensors and magnetic memories. In this context, van der Waals (vdW) \CT\ appears as a promising candidate. Bulk \CT\ exhibits strong PMA and a $T_\mathrm{C}$ of 180~K. Moreover, both PMA and $T_\mathrm{C}$ might be adjusted in ultrathin films by engineering composition, strain, or applying an electric field. In this work, we demonstrate the molecular beam epitaxy (MBE) growth of vdW heterostructures of five-monolayer quasi-freestanding \CT\ on three classes of 2D materials: graphene (semimetal), WSe$_2$ (semiconductor) and Bi$_2$Te$_3$ (topological insulator). By combining structural and chemical analysis down to the atomic level with \textit{ab initio} calculations, we confirm the single crystalline character of \CT\ films on the 2D materials with sharp vdW interfaces. They all exhibit PMA and $T_\mathrm{C}$ close to the bulk \CT\ value of 180~K. \textit{Ab initio} calculations confirm this PMA and show how its strength depends on strain. Finally, Hall measurements reveal a strong anomalous Hall effect, which changes sign at a given temperature. We theoretically explain this effect by a sign change of the Berry phase close to the Fermi level. This transition temperature depends on the 2D material in proximity, notably as a consequence of charge transfer. MBE-grown \CT/2D material bilayers constitute model systems for the further development of spintronic devices combining PMA, large spin-orbit coupling and sharp vdW interface. 
\end{abstract}

\maketitle


\section{Introduction}

The discovery of ferromagnetic order in two-dimensional (2D) materials like Cr$_2$Ge$_2$Te$_6$ \cite{gong_discovery_2017} and CrI$_3$ \cite{huang_layer-dependent_2017} has paved the way for the development of new van~der~Waals (vdW) heterostructures \cite{wang_magnetic_2022}. Combined with the large spin-orbit coupling and low crystal symmetries of 2D materials like transition metal dichalcogenides (TMD) \cite{cferrari_science_2015}, 2D ferromagnets represent a key ingredient to construct ultra-compact devices for spintronic applications \cite{yang_two-dimensional_2022} such as spin transfer torque (STT) or spin-orbit torque (SOT) magnetic random access memories (MRAMs). These technologies based on 2D materials would allow for the miniaturization of today’s devices as well as a sizeable reduction of energy consumption \cite{liu_two-dimensional_2020}.

For this purpose, 2D ferromagnets with Curie temperatures ($T_\mathrm{C}$) higher than room temperature and with perpendicular magnetic anisotropy (PMA) are required \cite{dieny_perpendicular_2017}. Fe$_x$GeTe$_2$ (x=3, 4 or 5) \cite{ribeiro_large-scale_2022} and Cr$_{1+\delta}$Te$_2$ ($0 \leq \delta \leq 1$) [6-19] have emerged recently as the two most promising families of materials to achieve such conditions. Cr$_{1+\delta}$Te$_2$ materials are composed of 1T-CrTe$_2$ monolayers (ML) separated by a variable amount of intercalated chromium atoms (from empty to fully occupied). CrTe$_2$ is a vdW ferromagnet with room temperature ferromagnetic order ($T_\mathrm{C}$ = 315~K) \cite{freitas_ferromagnetism_2015,purbawati_-plane_2020,zhang_room-temperature_2021}, whereas Cr$_{1+\delta}$Te$_2$ ($\delta > 0$) are quasi vdW ferromagnets with $T_\mathrm{C}$ ranging from 160~K to 350~K and varying magnetic anisotropy from out-of-plane to in-plane easy axis of magnetization [9-14]. Magnetic properties of Cr$_{1+\delta}$Te$_2$ have been shown to depend on its stoichiometry \cite{fujisawa_tailoring_2020,dijkstrat_band-structurecalculations_1989}, its thickness in case of thin films \cite{wen_tunable_2020}, the Cr-Te flux ratio during crystal growth \cite{li_molecular_2019} and strain in the layer \cite{zhou_structure_2022,li_magnetic_2021}. The stoichiometry of the stack could be adjusted by post-growth annealing \cite{fujisawa_tailoring_2020} or by changing elemental fluxes \cite{coughlin_van_2021}. Highly efficient control of magnetic properties is required for spintronic applications \cite{yang_two-dimensional_2022} and it is, therefore, necessary to understand the growth mechanisms of these materials, especially for the development of functional vdW heterostructures. Exotic topological phenomena such as the topological Hall effect have also been reported in Cr$_2$Te$_3$/Bi$_2$Te$_3$ bilayers \cite{chen_evidence_2019,chen_conformal_2022} or Cr$_2$Te$_3$/Cr$_2$Se$_3$ \cite{jeon_emergent_2022}. Moreover, non-collinear spin textures were shown in \CT\ as a consequence of antiferromagnetic coupling between neighboring chromium atoms \cite{pramanik_angular_2017} making it an interesting host for exotic, trivial or topological spin textures.

In this work, we report the vdW epitaxy \cite{ueno_epitaxial_1990, ohuchi_van_1990} of 5 ML of Cr$_2$Te$_3$ on three different 2D materials, namely graphene (a semimetal with exceptional electronic properties), WSe$_2$ (a transition metal dichalcogenide semiconductor exhibiting strong photo-luminescence and spin-valley locking in its monolayer form), and Bi$_2$Te$_3$ (a topological insulator with strong spin-orbit interaction). Particular care was given to their full structural and magnetic characterizations including the determination of the film stoichiometry. Those bilayers represent model systems to study proximity effects in vdW heterostructures, interface spin textures as well as spin-orbit torques. The \CT\ films were grown by molecular beam epitaxy (MBE) in ultrahigh vacuum (UHV) by depositing simultaneously Cr and Te atoms. They exhibit in-plane compression and out-of-plane expansion with respect to the bulk phase. This strain is shown to vary with the post-growth annealing, but it is almost independent of the 2D layer underneath. Indeed, Cr$_2$Te$_3$ films on graphene and WSe$_{2}$ annealed at 400°C show the same lattice parameters, which are equal to the ones of 5 ML free-standing Cr$_2$Te$_3$ calculated by \textit{ab initio} methods. This demonstrates that the vdW epitaxy of Cr$_2$Te$_3$ on 2D materials leads, after annealing, to the formation of quasi-freestanding films with negligible interaction with the substrate. We then correlate the PMA of Cr$_2$Te$_3$ with strain and confirm our experimental findings using \textit{ab initio} calculations. Finally, magnetotransport measurements reveal a change of sign of the anomalous Hall effect in \CT\ with temperature and point out a charge transfer from the substrate to the film changing the p-type doping level of Cr$_2$Te$_3$. This effect has already been observed in several vdW heterostructures \cite{dau_beyond_2018,Dappe_2020}. The charge transfer is shown to govern the temperature at which the anomalous Hall effect changes sign. We reproduce theoretically this effect by showing a sign change of the Berry phase close to the Fermi level. Finally, our work demonstrates the ability of MBE to synthesize model vdW heterostructures incorporating 2D materials and quasi vdW ferromagnets which are highly promising for future 2D-based spintronic devices.

\section{Methods}
\subsection{Experimental methods}

All the films were grown by MBE using a home-designed UHV system. Metallic elements (Cr, W, Bi and Al) were evaporated using an electron gun and the growth rate was controlled using a quartz microbalance, whereas chalcogens (Te, Se) were evaporated from Knudsen cells. Their elemental fluxes were measured by a pressure gauge. The substrates were attached to a molybloc by wetting In underneath. The temperature of the samples during growth was controlled by a thermocouple touching the backside of the molybloc. \\
Scanning transmission electron microscopy (STEM) measurements were performed using a Cs-corrected FEI Themis at 200 kV. HAADF-STEM (high-angle annular dark field) images were acquired using a convergence angle of 20~mrad and collecting electrons scattered at angles higher than 60~mrad. STEM specimens were prepared by the focused ion beam (FIB) lift-out technique using Zeiss Crossbeam 550. The sample was coated with protective carbon and platinum layers prior to the FIB cut. \\
The out-of-plane x-ray diffraction (XRD) measurements were performed using a Panalytical Empyrean diffractometer operated at 35 kV and 50 mA, with a cobalt source, (K$\alpha$ = 1.79 \AA). A PIXcel-3D detector allowed a resolution of 0.02° per pixel, in combination with a divergence slit of 0.125° on the source side. Grazing in-plane XRD measurements were performed with a SmartLab Rigaku diffractometer equipped with a copper rotating anode (K$\alpha$ = 1.54 \AA) operating at 45 kV and 200 mA. Collimators with a resolution of 0.5° were used both on the source and the detector sides. The grazing incidence close to the critical angle of the substrate was optimized to maximize the intensity of the \CT\ Bragg peaks. Both diffractometers were equipped with multi-layer mirrors on the source side and K$\beta$ filter on the detector side.\\
Raman measurements were performed with a Horiba Raman setup with a 632 nm laser
excitation source and a spot size of 0.5 $\mu$m. The signal was collected by
using a 1800 grooves/mm grating.\\
Rutherford backscattering (RBS) measurements were performed with a $^4$He$^{+}$ beam delivered by the SAFIR Platform at Sorbonne University in Paris at beam energies ranging from 1.5 to 2.0 MeV. For all samples, the scattering angle was set to 160° and the resolution of the detector was 13.5 keV. To avoid channeling effects, the samples were tilted with respect to the normal of the sample in two perpendicular directions.\\
The magnetic properties were measured by superconducting quantum interference device (SQUID) magnetometry with the magnetic field applied parallel or perpendicular to the film plane. The measurements were performed using a Quantum Design magnetic property measurement system. The diamagnetic contribution was subtracted using the data at high field ($\ge 3$T) and some parasitic contributions were corrected by subtracting signals measured well above the $T_\mathrm{C}$ of the systems (at 350K). This method has already been used successfully by Ribeiro~\textit{et al.}~\cite{ribeiro_large-scale_2022} and confirmed by comparing it with magnetic moments extracted from x-ray magnetic circular dichroism (XMCD) measurements. \\
The XMCD measurements were performed on the DEIMOS beamline \cite{ohresser2014deimos} of synchrotron SOLEIL (Saint Aubin, France). The signals were recorded using the Total Electron Yield (TEY) method. Each XMCD spectrum was obtained from four
measurements, where both the circular helicity and the direction of the
applied magnetic field were flipped. The XAS data are then averaged (the signals of opposite helicity and field) and normalized to the absorption at the pre-edge of chromium (565 eV). The XMCD spectra are normalized to their maximum for comparison. \\
In order to carry out magnetotransport measurements, we processed Hall bars out of Cr$_2$Te$_3$ films by laser lithography and argon etching. Electrical contacts were made of e-beam evaporated Ti(10 nm)/Au(100 nm) bilayers. The length and width of Hall bars were approximately 100 $\mu$m and 10 $\mu$m respectively. All the electrical measurements were performed using an OXFORD Spectromag setup working in the 1.6-300~K temperature range with magnetic fields up to 7 Tesla. The anomalous Hall contribution was obtained by fitting the experimental data with a hyperbolic tangent function.

\subsection{Calculation methods}

The \textit{ab initio} calculations were performed using density functional theory (DFT) as implemented in the Vienna \textit{ab initio} simulation package (\texttt{VASP}) \cite{kresse_ab_1993, kresse_efficiency_1996} with the generalized gradient approximation (GGA) pseudopotentials in the Perdew-Burke-Ernzerhof (PBE) parametrization \cite{perdew_generalized_1996-1}. The DFT+$U$ approach using Dudarev's formulation \cite{dudarev_electron-energy-loss_1998} was applied with an effective Hubbard correction $U_\mathrm{eff} = 2.1$~eV to localize the Cr $d$-orbitals. A Cr pseudopotential with semicore $p$ electrons was chosen and an energy cutoff of 330 eV was used for the plane-wave basis. The van~der~Waals interaction was approximated by the DFT-D3 method \cite{grimme_consistent_2010} with the Becke-Johnson damping \cite{grimme_effect_2011}.

To compute the relaxed heterostructures of \CT\ with 2D materials, the relative orientation of the two materials in the calculation was not taken from an experiment, but chosen in a systematic way \cite{choudhary_efficient_2020} 
to minimize the lattice mismatch. This captures more realistically the weak epitaxy of the heterostructure.

The magnetic anisotropy calculation procedure is described in \cite{hallal_impurity-induced_2014}. A 9x9x5 k-point mesh was found to be sufficient.  The volume was fixed at its calculated equilibrium bulk value, while the in-plane and out-of-plane lattice parameters (\textit{a} and \textit{c}) were varied. A demagnetizing energy contribution $E_\mathrm{demag} = -\mu_0 M_\mathrm{s}^2/2$ was added to the calculated magnetocrystalline energy, using the experimental $M_\mathrm{s}$ value $\approx$ 300 kA/m.

The anomalous Hall effect was computed \cite{wang_ab_2006} by constructing a tight-binding Hamiltonian based on maximally localized Wannier functions using the \texttt{Wannier90} package \cite{mostofi_updated_2014}. We verified carefully that the model reproduces well the band structure of \CT\ from the original DFT calculation. Using the \texttt{WannierBerri} package \cite{tsirkin_high_2021-1, destraz_magnetism_2020}, the Berry curvature was then calculated for a dense k-point mesh and integrated over the Brillouin zone to obtain the Berry phase,
proportional to the anomalous Hall conductivity, at various Fermi level positions.

\section{Sample preparation}

In this study, we have grown samples of 5 layers of Cr$_2$Te$_3$ corresponding to a thickness of 6.1 nm on three different vdW surfaces: 1 monolayer of WSe$_2$ deposited on GaAs, 10 layers of Bi$_2$Te$_3$ on Al$_2$O$_3$, which were both grown in situ by MBE and monolayer graphene, which was obtained by the controlled graphitization of 4H-SiC(0001) \cite{kumar_growth_2016} in another reactor.

\begin{figure*}[h]
\includegraphics[width=10cm]{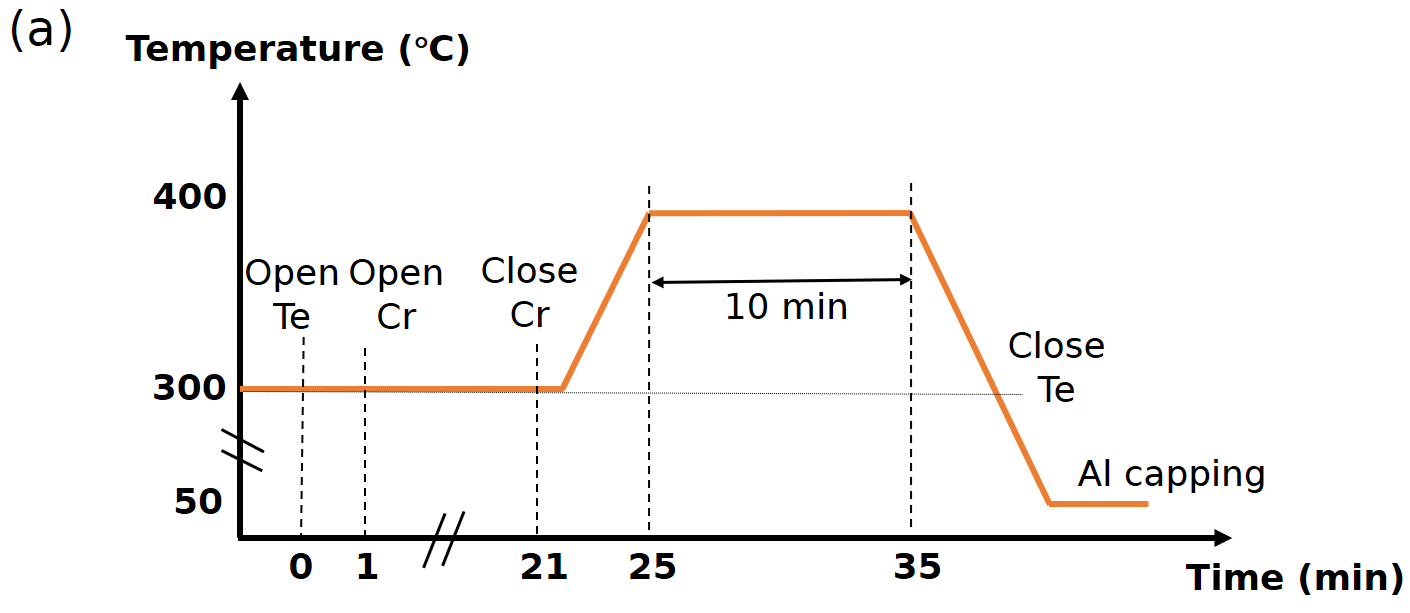}
\includegraphics[width=17.2cm]{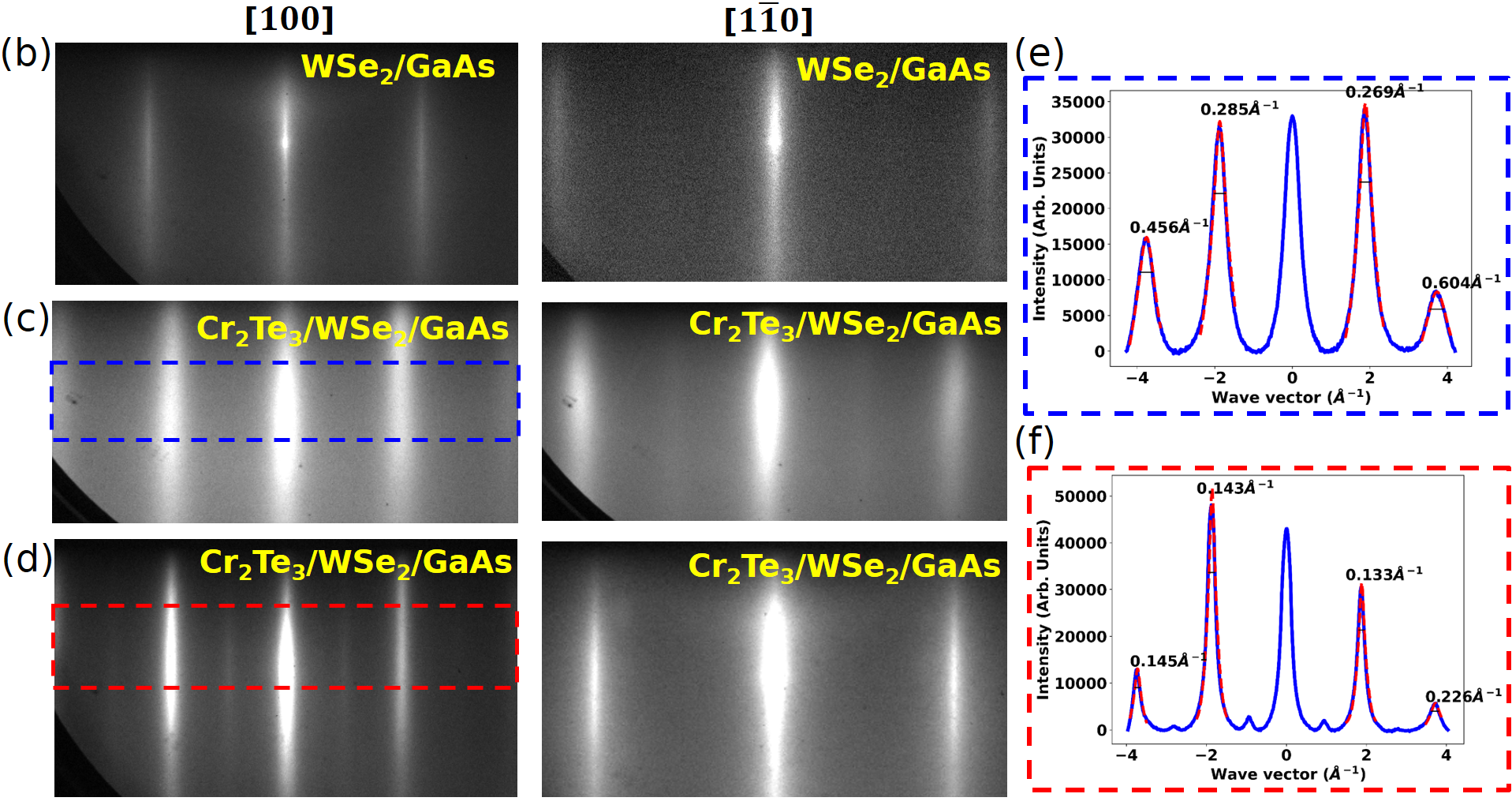}
\caption{MBE growth of Cr$_2$Te$_3$ on the 2D transition metal dichalcogenide WSe$_2$. (a)	Sketch of the deposition procedure. The growth temperature was 300°C and in situ annealing was performed at 400°C.
(b)	in situ RHEED images of 1 layer WSe$_2$ deposited on GaAs(111)B along two crystal directions (Time = 0 min). (c) RHEED pattern after the deposition of Cr$_2$Te$_3$ (Time = 21 min). (d) RHEED pattern after annealing (Time = 35 min). (e,f) Intensity profiles of the RHEED diffraction pattern for images c (blue dashed box) and d (red dashed box) respectively.
}
\label{growth}
\end{figure*}

WSe$_2$ was grown epitaxially on Se-passivated GaAs(111)B as detailed in \cite{pierucci_evidence_2022}. Bi$_2$Te$_3$ was grown epitaxially on sapphire. For this purpose, sapphire substrates were first annealed in air for one hour at 1000°C with a heating ramp of 40 minutes starting from room temperature. They were additionally annealed in situ for 30 minutes at 800°C. 10 quintuple layers of Bi$_2$Te$_3$ were then grown by co-evaporating Bi and Te from an electron gun and a cracker cell at deposition rates of 0.057 and 0.1 \AA/s, respectively. The substrate temperature was maintained at 250°C during the growth. Post-growth annealing at 300°C under Te flux was done for 10 minutes to improve the crystal quality. Finally, Gr/SiC layers were annealed in situ for 30 minutes at 650°C after their transfer.

The \CT\ films were grown using a two-step process as sketched in Fig.~\ref{growth}(a). The growth temperature, Te:Cr ratio and deposition rate were set at 300°C, 10, and 0.25 L/min respectively. The Te cell shutter was opened 1 minute before chromium deposition to ensure that the surface of the substrate was saturated in Te at the first stage of growth
[see Fig.~\ref{growth}(a)]. After the growth, the samples were annealed at 400°C for 10 minutes using the same Te flux as during the growth and a heating ramp of 40°C/min. The samples were then cooled down to 50°C and 3 nm of aluminum was deposited to prevent oxidation of the layers during transfers between experimental setups.

The film morphology was monitored in situ by reflection high-energy electron diffraction (RHEED) as can be seen in Fig.~\ref{growth}(b) in the case of WSe$_2$. A streaky diffraction pattern was observed indicating a flat and well-crystallized surface. The different diffraction patterns along the two high symmetry axes of the WSe$_2$ substrate [Fig.~\ref{growth}(c)-(e)] indicate a good alignment of Cr$_2$Te$_3$ grains with the underlying layer. After annealing, the width of the diffraction rods is approximately divided by 2 as a consequence of the larger grain size [Fig.~\ref{growth}(d)-(f)]. We made similar RHEED observations for the two other vdW substrates Bi$_2$Te$_3$ and graphene (see the Supplemental Material Fig.~S1), except an increased isotropic contribution on graphene attributed to a lower interaction with the substrate.

\section{Structural properties}

\begin{figure*}[h]
    \centering
    \includegraphics[width=17.2cm]{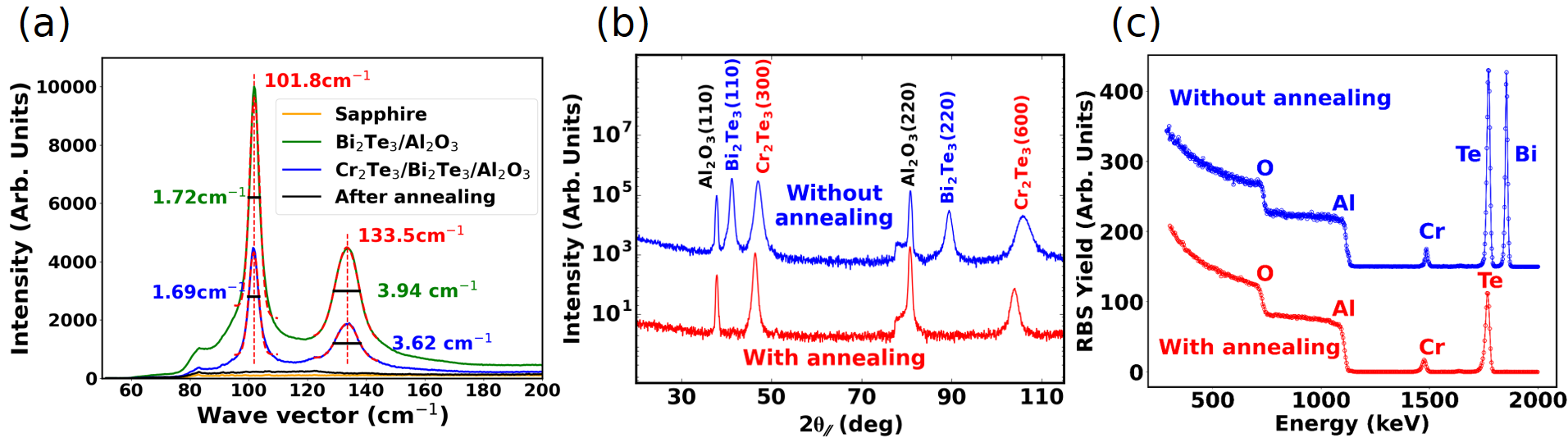}
    \caption{Structural properties of \CT\ on Bi$_2$Te$_3$/Al$_2$O$_3$. (a) Raman spectra of the sapphire substrate, Bi$_2$Te$_3$/sapphire, and Cr$_2$Te$_3$/Bi$_2$Te$_3$/sapphire with and without annealing. Positions and full widths at half maximum of Bi$_2$Te$_3$ peaks are indicated.  (b) Radial x-ray diffraction spectra for Cr$_2$Te$_3$/Bi$_2$Te$_3$/Al$_2$O$_3$ without (top in blue) and after (bottom in red) annealing. (c) RBS of Cr$_2$Te$_3$ grown on Bi$_2$Te$_3$/ Al$_2$O$_3$ without (top in blue) and with (bottom in red) annealing. No elemental Bi can be found after annealing.}    
    \label{Bi$_2$Te$_3$}
\end{figure*}

We found similar structural characteristics for \CT\ grown on WSe$_2$ and graphene. Therefore, we present the results of the growth on WSe$_2$ and the ones on graphene are given in the Supplemental Material. The results on Bi$_2$Te$_3$ are shown in Fig.~\ref{Bi$_2$Te$_3$}. 

\begin{figure*}[h]
    \center
    \includegraphics[width=17.2cm]{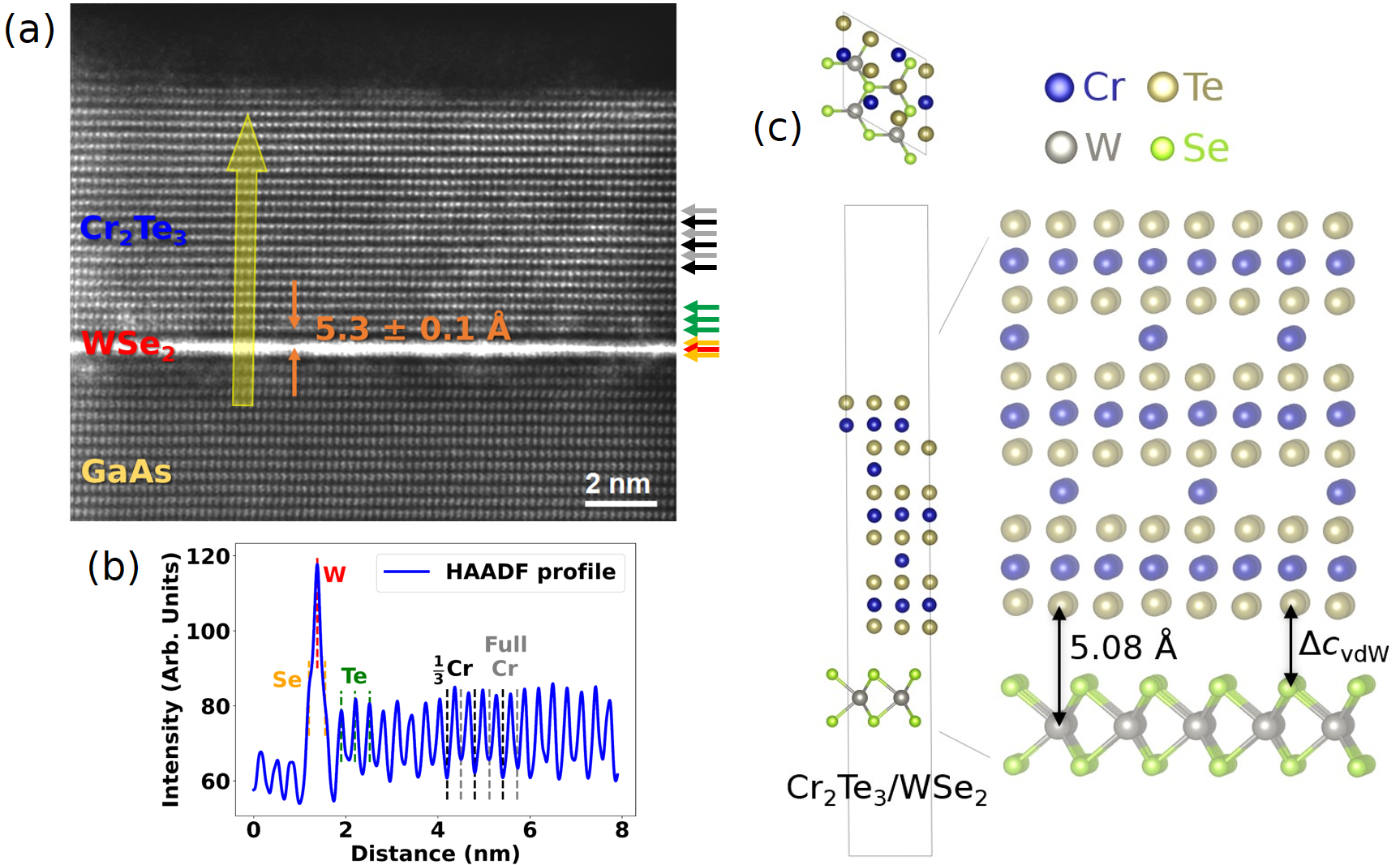}
    \caption{
    (a)	Low-pass filtered HAADF-STEM image of 5 layers Cr$_2$Te$_3$ grown on one monolayer WSe$_2$ deposited on a GaAs(111)B surface. The van~der~Waals gap between the layers is shown to highlight the high quality of the interface. Arrows on the right side indicate the position of the atomic planes noted in the line profile.
    (b) Line profile along the c-direction of \CT\ layers (yellow arrow) with intensity distinction between partially and fully occupied Cr planes [see crystal structure in (c)].
    (c) A unit cell of the calculated Cr$_2$Te$_3$/WSe$_2$ heterostructure - in the interstitial planes, only 1/3 of the lattice sites is occupied by the intercalated Cr atoms. The \textit{ab initio}-calculated W-Te distance is 5.08~\AA.
    }
\label{crystal_structure}
\end{figure*}

\begin{figure*}[h]
    \center
    \includegraphics[width=17.2cm]{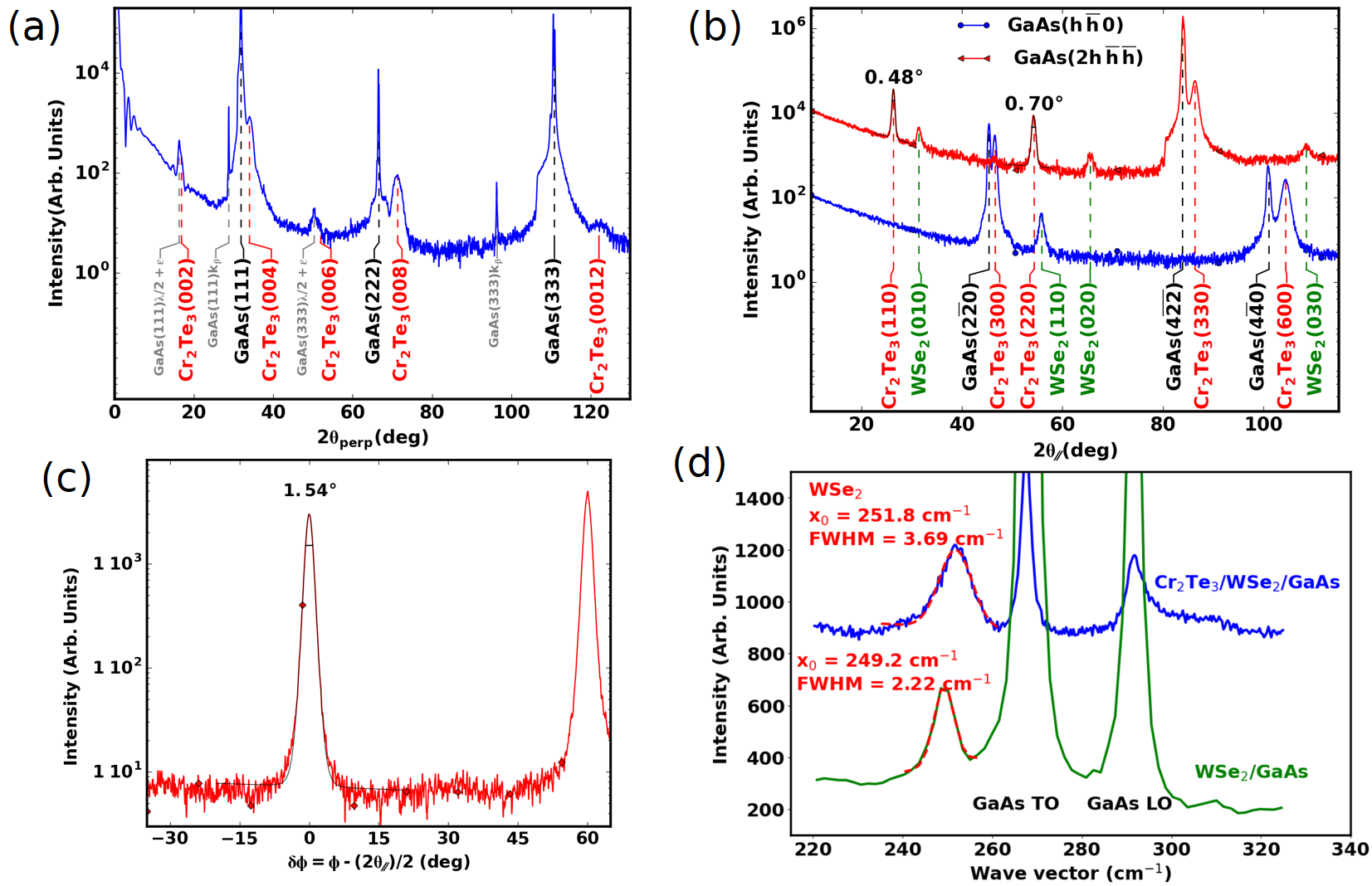}
    \caption{
    Post-growth characterization of the crystal structure of \CT/WSe$_2$/GaAs (sample 2).
    (a)	Out-of-plane $\Theta$/2$\Theta$ XRD scan shows, in addition to \CT~(00l) peaks (red), GaAs substrate peaks (black) with weak additional peaks due to spurious radiations not completely eliminated by the mirror and the K$\beta$ filter (grey).
    (b)	In-plane radial XRD scans performed along the GaAs substrate R=(\textit{h$\bar{h}$0}) direction, and R+30°=(\textit{2h$\bar{h}\bar{h}$}) direction. These scans show the substrate peaks (black), WSe$_2$ (green), and \CT\ peaks (red) labeled with their FWHM.
    (c)	In-plane azimuthal XRD scan of the (300) peak measured within a range of 100° shows thin peaks with a FWHM of 1.54° separated by 60° corresponding to the 6-fold symmetry of the crystal.
    (d) Raman spectra of WSe$_2$ before and after deposition of \CT.}
\label{WSe2_carac}
\end{figure*}

Figure~\ref{crystal_structure}(a) is a cross-section STEM image of the layers, revealing a sharp and well-defined interface between the vdW ferromagnet and the 2D layer as evidenced by the W-Te distance between W atoms of WSe$_2$ and the first Te atoms plane of \CT\ with a value of 5.3~\AA. This value is taken to obtain a better experimental determination of the gap (due to the large atomic number of W with respect to Se) as can be seen in Fig.~\ref{crystal_structure}(b) showing a line profile along the c-direction of the heterostructure. This corresponds to a vdW gap $\Delta c_\mathrm{vdW}$ of 3.5~\AA\ if we assume a relaxed WSe$_2$ layer, in agreement with our XRD data (see Fig.~\ref{WSe2_carac}). It is worth noting that we resolve an intensity difference between fully and partially filled Cr planes. Indeed, every two Cr planes, only one third of the lattice sites are occupied by intercalated Cr atoms in the case of \CT. We also observe that the monolayer of WSe$_2$ remains intact after the growth of Cr$_2$Te$_3$ on top. \\

The experimental W-Te distance was compared with the one obtained by \textit{ab initio} calculations performed on Cr$_2$Te$_3$/WSe$_2$ consisting of 1.5 unit cell thick Cr$_2$Te$_3$ on top of a single layer of WSe$_2$ [see Fig.~\ref{crystal_structure}(c)]. The calculated distance is 5.08~\AA, in good agreement with the experimental one of 5.3~\AA.\\

To accurately determine the lattice parameters of Cr$_2$Te$_3$ layers and crystal orientation with respect to each of the substrates, systematic XRD analysis were performed to extract the in-plane and out-of-plane lattice parameters. These measurements allowed us to measure accurately the strain in each layer when compared with the bulk lattice parameters. 

Figure~\ref{WSe2_carac}(a), \ref{WSe2_carac}(b) and \ref{WSe2_carac}(c) show XRD out-of-plane, in-plane radial and in-plane azimuthal scans of Cr$_2$Te$_3$/WSe$_2$/GaAs (sample 2, see Table~\ref{tab:data}) respectively. The diffraction patterns of Cr$_2$Te$_3$ deposited on WSe$_2$ reveal the single crystalline character of the film and the clear epitaxial relationship with WSe$_2$. The thin Bragg peaks in radial scans in Fig.~\ref{WSe2_carac}(b) (Full Widths at Half Maximum -FWHM- of 0.48° and 0.70°) indicate the large grain size and the uniformity of the lattice parameter. In Fig.~\ref{WSe2_carac}(c), the mosaic spread (FWHM of 1.54°) is negligibly small which confirms the perfect orientation of Cr$_2$Te$_3$ on WSe$_2$ .

All the XRD data are summarized in Table \ref{XDR_data}. Compared to bulk values of Cr$_2$Te$_3$ with $a=6.812$~\AA \ and $c=12.07$~\AA \ \cite{andresen1970magnetic}, we systematically found an in-plane compressive strain and a resulting out-of-plane expansion. We found similar lattice parameters regardless of the substrate underneath although the mismatch between inter-atomic distances is very large (WSe$_2$:+19.1\%/Gr:+56.3\%/Bi$_2$Te$_3$:-10.8\%). An in-plane compressive strain would be expected for \CT\ deposited on WSe$_2$ and graphene, whereas an in-plane tensile strain would be expected for the growth on Bi$_2$Te$_3$. Moreover, we could not find any commensurable relationship between the in-plane lattice parameter of Cr$_2$Te$_3$ and the one of the substrate. This was confirmed by \textit{ab initio} calculations: the lattice parameter of the vdW heterostructure corresponds to the one of bulk \CT\ above 7~MLs of \CT, as demonstrated for \CT/Gr (see the Supplemental Material Fig.~S2). We thus conclude about the pure vdW interaction between Cr$_2$Te$_3$ and the substrate. The slight difference between lattice parameters might be due to the surface topography (presence of steps, terrace, etc.) and the microscopic structure of Cr$_2$Te$_3$ (grain size, grain boundaries, etc.).

\begin{table*}[h]
\setlength{\tabcolsep}{0mm}
\renewcommand{\arraystretch}{1}
\begin{tabular}{*{11}{c}}
\hline
\textbf{N°}&
\textbf{\; 2D layer}&
\textbf{\; Temperature}&
\textbf{\; \textit{a} (\AA)}&
\textbf{\;$\frac{a-a_\mathrm{bulk}}{a_\mathrm{bulk}}$}&
\textbf{\; \textit{c} (\AA)}&
\textbf{\;$\frac{c-c_\mathrm{bulk}}{c_\mathrm{bulk}}$ \;}&
\textbf{\;$\frac{c}{a}$ \;}&
\textbf{\; $\Delta \theta_{//}$ \;}&
\textbf{\; $\Delta \phi$ \;}&
\textbf{\; Stoichiometry}\\
\hline \hline
1 & WSe$_2$ & 300°C & 6.731 & -1.2\% & 12.44 & +3.1\% & 1.848 & 1.00° & 2.36° & \\
2 & WSe$_2$ & 400°C & 6.760 & -0.76\% & 12.28 & +1.7\% & 1.817 & 0.62° & 1.54° & \\
3 & graphene & 300°C & 6.754 & -0.88\% & 12.18 & +0.91\% & 1.804 & 0.72° & 24.8° & Cr$_{1.88}$Te$_3$\\
4 & graphene & 400°C & 6.758 & -0.79\% & 12.30 & +1.9\% & 1.820 & 0.60° & 16.6° & \\
5 & Bi$_2$Te$_3$ & 300°C & 6.691 & -1.8\% & 12.50 & +3.6\% & 1.868 & 0.87° & 2.77° & Cr$_{1.97}$Te$_3$\\
\rowcolor[gray]{0.85}6 & Bi$_2$Te$_3$ & 400°C & 6.778 & -0.50\% & 12.18 & +0.91\% & 1.797 & 0.56° & 1.28° & Cr$_{2.07}$Te$_3$\\
\hline
\end{tabular}
\caption{\label{tab:data}Growth/annealing temperature and structural parameters measured by x-ray diffraction and chemical composition from RBS, with \textit{a} (\textit{c}) the in-plane (out-of-plane) lattice parameter, the radial width  ($\Delta \theta_{//}$) of the (300) diffraction peak and the mosaic spread ($\Delta \phi$) measured on the same Bragg peak.
}
\label{XDR_data}
\end{table*}

However, the energy given to the system by annealing seems to be the driving force to control the final crystal structure since all the lattice parameters converge to the same values after annealing at 400°C. Besides, layers grown on graphene exhibit a much larger mosaic spread even after annealing, indicating even lower interaction between Cr$_2$Te$_3$ and the substrate during growth. \\

\begin{figure}[h]
    \center
    \includegraphics[width=3.4in]{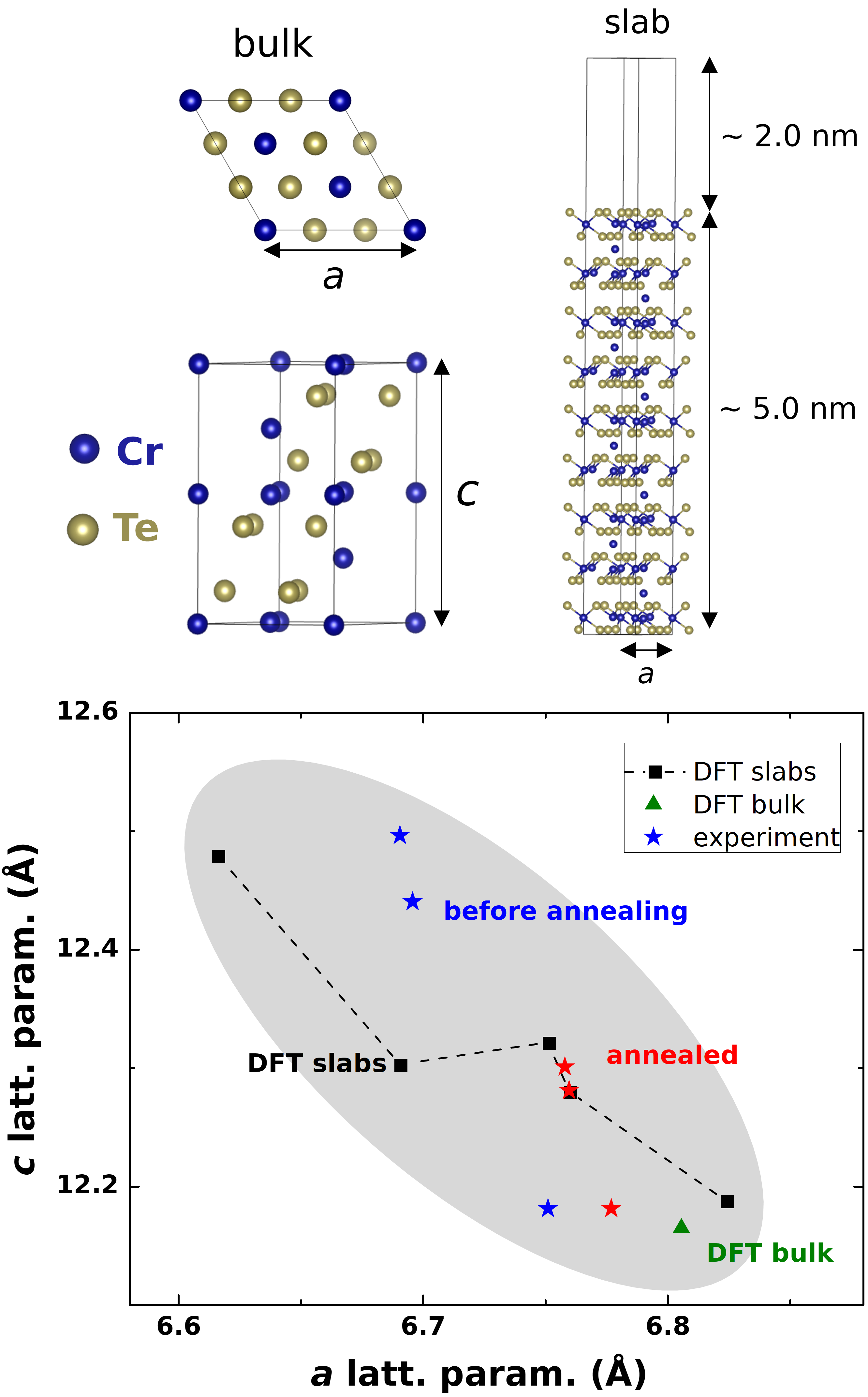}
    \caption{Top: \CT~bulk crystal structure and a thin free-standing film constructed from~it. The in-plane lattice parameter $a$ was fixed at a range of values while the atomic positions were relaxed to obtain the out-of-plane lattice parameter $c$. Bottom: The calculated and experimental lattice parameters for free-standing slabs and for bulk structures. The measured values follow well the trend calculated for free-standing \CT~films. 
}
    \label{fig:DFT_structure}
\end{figure}

The measured lattice parameters match well with the \textit{ab initio} calculations performed on $\sim 5$~nm-thick free-standing films which corresponds to the experimental thickness (see Fig.~\ref{fig:DFT_structure}). In particular, for the Cr$_2$Te$_3$ films annealed on graphene and WSe$_2$ (samples 2 and 4), the experimental lattice parameters fall exactly on the theoretical curve confirming a weak interaction between the film and the substrate. The case of \CT\ grown on Bi$_2$Te$_3$ is discussed later in Fig.~\ref{Bi$_2$Te$_3$}.

In Table \ref{tab:data}, we also show the measured composition of some selected samples by RBS (see the Supplemental Material Fig.~S4) and found compositions very close to Cr$_2$Te$_3$. No measurement could be performed on GaAs substrates as Ga and As are heavier than chromium, causing the Cr signal peak to lie in the substrate background, preventing any determination of the Cr:Te ratio in these samples.

Raman spectroscopy was also performed before and after the growth of Cr$_2$Te$_3$ to control the quality and integrity of the 2D layers. Figure~\ref{WSe2_carac}(d) depicts the Raman spectra of WSe$_2$/GaAs and Cr$_2$Te$_3$/WSe$_2$/GaAs. The width and position of WSe$_2$ peaks are preserved, indicating that the deposition of Cr$_2$Te$_3$ did not alter the WSe$_2$ layer. The reference signal of WSe$_2$/GaAs (green curve) was measured with a 532 nm laser instead of 633 nm as the other samples, explaining the intensity differences. Similar observations have been made on the Cr$_2$Te$_3$/graphene heterostructure as shown in the Supplemental Material Fig.~S5.

Figure~\ref{Bi$_2$Te$_3$}(a) shows the Raman spectra of Cr$_2$Te$_3$/Bi$_2$Te$_3$/Al$_2$O$_3$ at different stages of growth. We detected two characteristic peaks of Bi$_2$Te$_3$ at 101.8 cm$^{-1}$ and 133.5 cm$^{-1}$, which correspond to the E$_g^2$ and A$_{1g}^2$ vibrational modes and have also been reported in \cite{wang_situ_2013}. After the deposition of 5 layers of Cr$_2$Te$_3$ at 300°C, those peaks remained unchanged (the amplitude drop is explained by the partial absorption of the laser fluence in the metallic Cr$_2$Te$_3$ layer). However, when the sample was annealed at 400°C, the two characteristic peaks of Bi$_2$Te$_3$ disappeared. Indeed, x-ray diffraction measurements performed before and after annealing in Fig.~\ref{Bi$_2$Te$_3$}(b) clearly show the disappearance of Bi$_2$Te$_3$ after thermal annealing. Finally, in Fig.~\ref{Bi$_2$Te$_3$}(c), RBS measurements on the annealed sample show the absence of Bi in the heterostructure. This reveals that Bi$_2$Te$_3$ was evaporated during annealing leaving the Cr$_2$Te$_3$ film on the pristine sapphire substrate. We shaded the structural data of this sample (n°6) in Table \ref{tab:data} because Cr$_2$Te$_3$ is standing directly on the sapphire substrate after annealing. Moreover, in this case, the substrate is no more vdW and defects might have been created in Cr$_2$Te$_3$ by evaporation of the Bi$_2$Te$_3$ layer underneath.

\section{Magnetic properties}

\begin{figure*}[h]
    \centering
    \includegraphics[width=17.2cm]{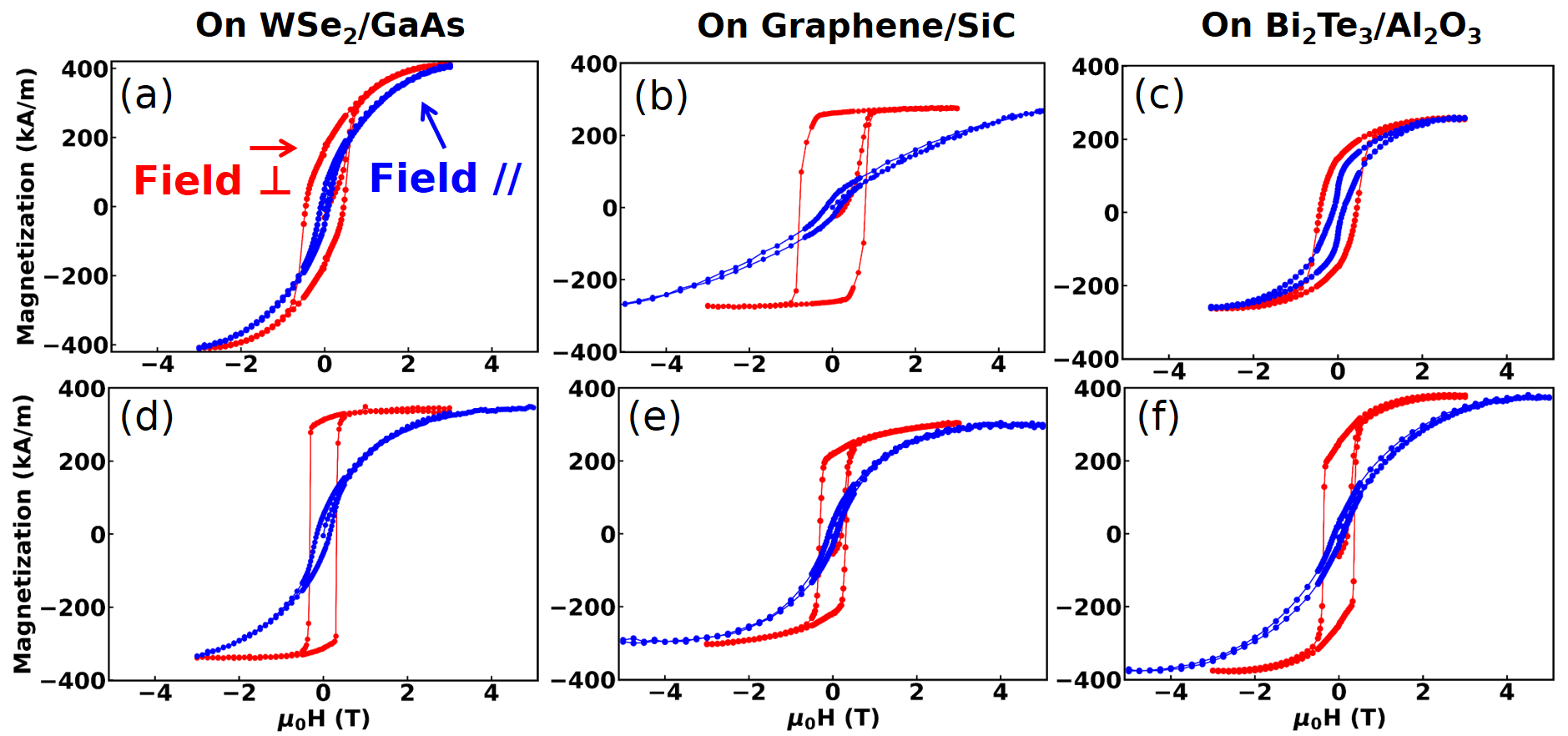}
    \caption{SQUID hysteresis loops with out-of-plane ($\perp$) and in-plane (//) applied magnetic field measured at 5~K are plotted after the removal of the substrate diamagnetic contribution. (a) (b) (c) Measurements on samples 1, 3 and 5 (without annealing on WSe$_2$/GaAs, graphene/SiC, and Bi$_2$Te$_3$/Al$_2$O$_3$). (d) (e) (f) Same measurements on samples 2, 4 and 6 (annealed at 400°C).
    }
    \label{hyst_loops}
\end{figure*}
Hysteresis loops were measured by SQUID magnetometry at 5~K and are displayed in Fig.~\ref{hyst_loops}. For all samples, the easy axis of magnetization was found along the c-axis and by integrating the difference of area between the out-of-plane and the in-plane magnetization curves, the magnetic anisotropy energy (MAE) could be experimentally derived for all the samples.

\begin{figure*}[h]
    \centering
    \includegraphics[width=17.2cm]{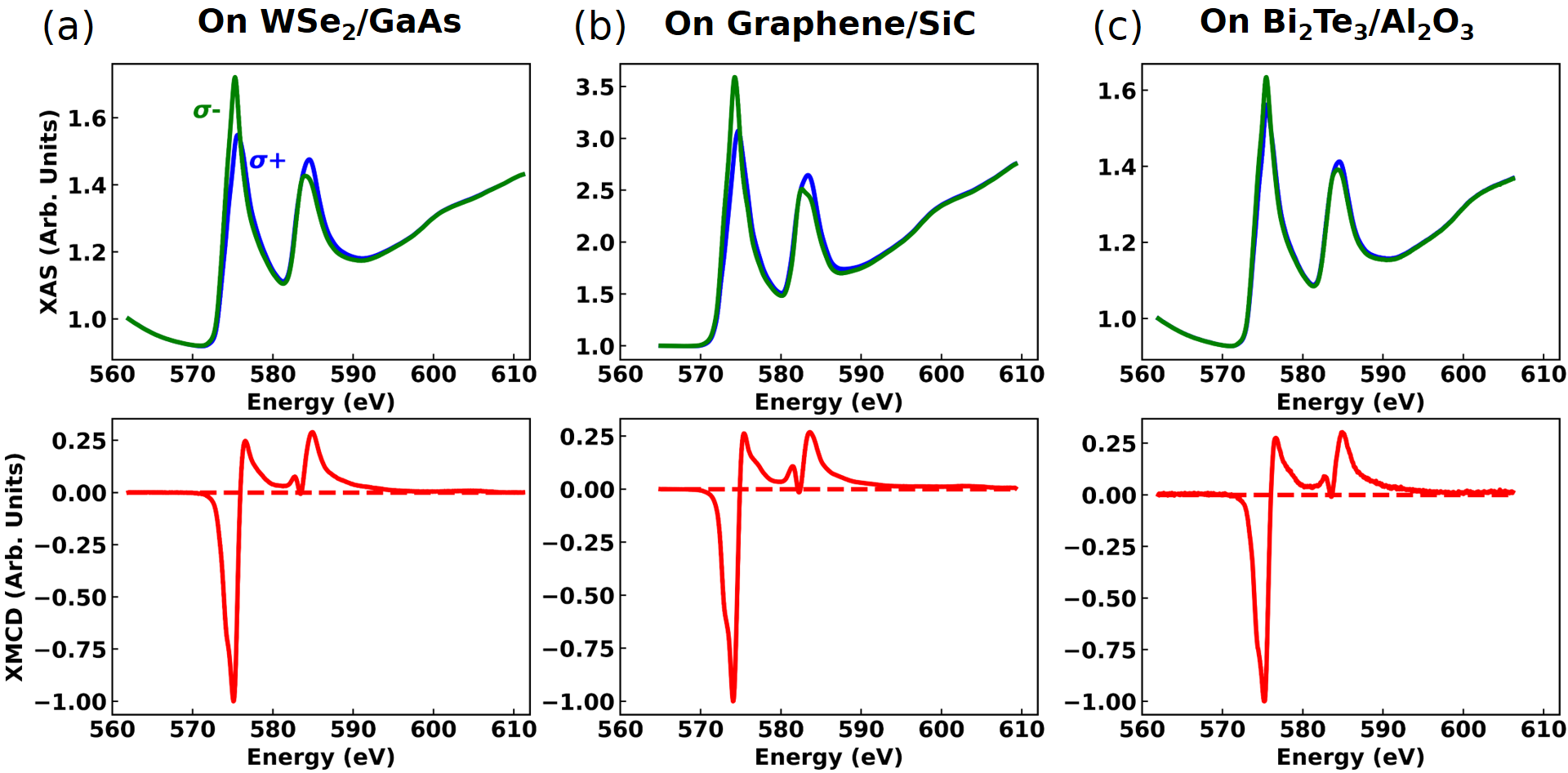}
    \caption{Top: x-ray absorption spectroscopy (XAS) and bottom: x-ray magnetic circular dichroism (XMCD) measurements performed on Cr$_2$Te$_3$ layers grown and annealed on (a) WSe$_2$ (sample 2), (b) graphene (sample 4) and (c) Bi$_2$Te$_3$ (sample 6).}
    \label{xmcd}
\end{figure*}

\begin{figure}[h]
    \center
    \includegraphics[width=3.4in]{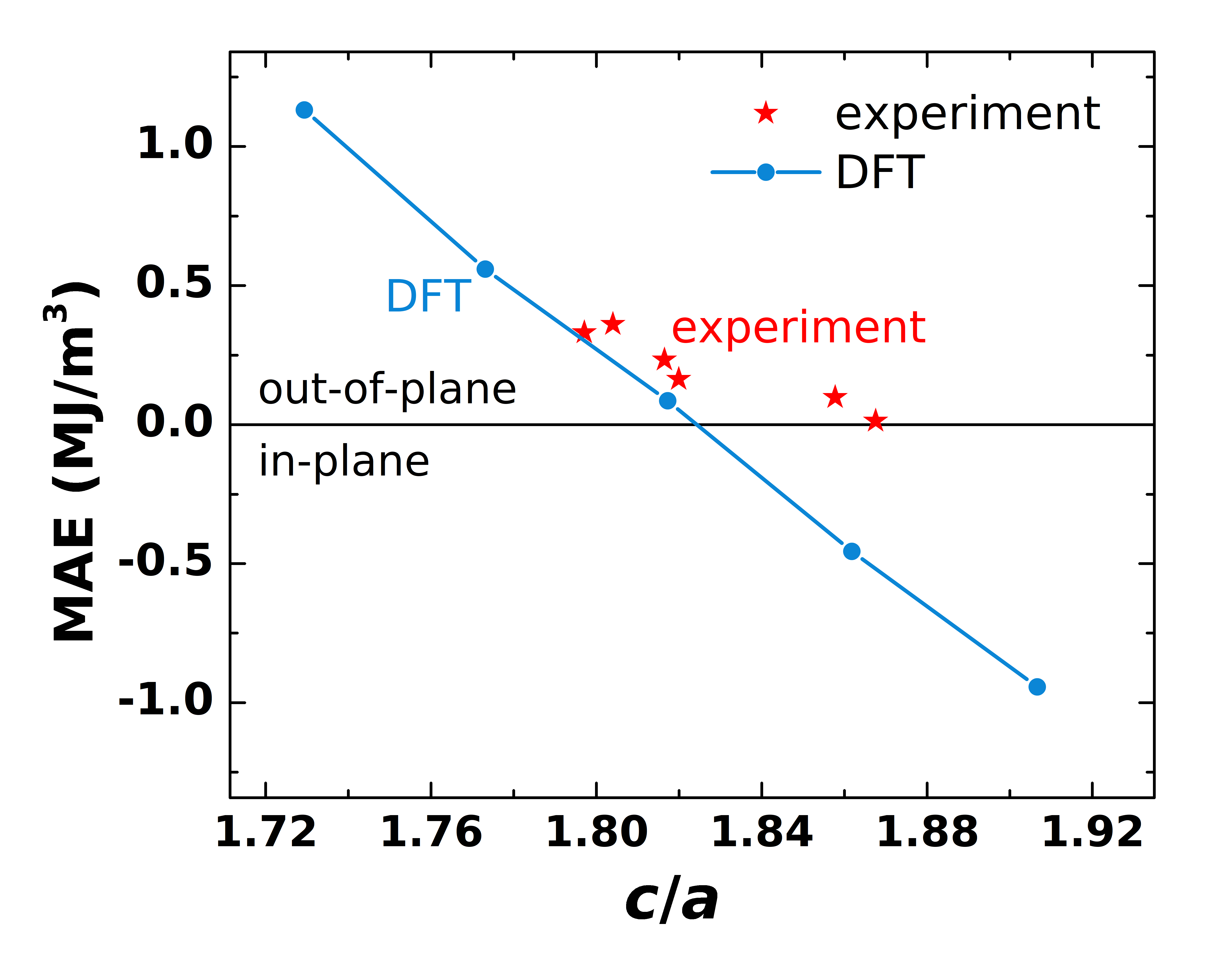}
    \caption{Magnetic anisotropy energy of bulk \CT\ as a function of strain compared for experiment and theory. It is a sum of the DFT-calculated magnetocrystalline energy with a demagnetizing energy contribution of -0.06~MJ/m$^3$ corresponding to the experimentally measured magnetization of $\approx$ 300 kA/m.} 
    \label{fig:DFT_anisotropy}
\end{figure}

The origin of ferromagnetism in our layers was confirmed by XMCD performed at the SOLEIL synchrotron radiation source. The energy spectra are shown in Fig.~\ref{xmcd} and a hysteresis loop is displayed in the Supplemental Material Fig.~S6. A clear magnetic dichroism signal with a similar spectral shape was obtained for all the three different substrates. This proves that the chemical environment of Cr atoms in Cr$_2$Te$_3$ films is essentially independent of the substrate. The lower magnetic moment for the sample on Bi$_2$Te$_3$/Al$_2$O$_3$ [Fig. \ref{xmcd}(c)] is explained by a lower sample thickness (three monolayers instead of five).\\
To better understand the magnetic properties, the magnetic anisotropy energy was calculated theoretically as a function of strain for bulk Cr$_2$Te$_3$ and was compared to experimental values in Fig.~\ref{fig:DFT_anisotropy}. The results reveal that the MAE is correlated to the strain of the layers. Overall, the trend and magnitude correspond well with experimental data. In particular, there is no sharp discontinuous change from positive to negative anisotropy values, as reported in~\cite{wen_tunable_2020}. However, the experimental data show larger PMA values compared to the theory. Since our calculations were performed in bulk Cr$_2$Te$_3$, we can attribute this shift to the presence of interfacial PMA at the Cr$_2$Te$_3$/substrate or Cr$_2$Te$_3$/AlO$_x$ capping layer interfaces.\\

\begin{figure}[h]
    \centering
    \includegraphics[width=8.6cm]{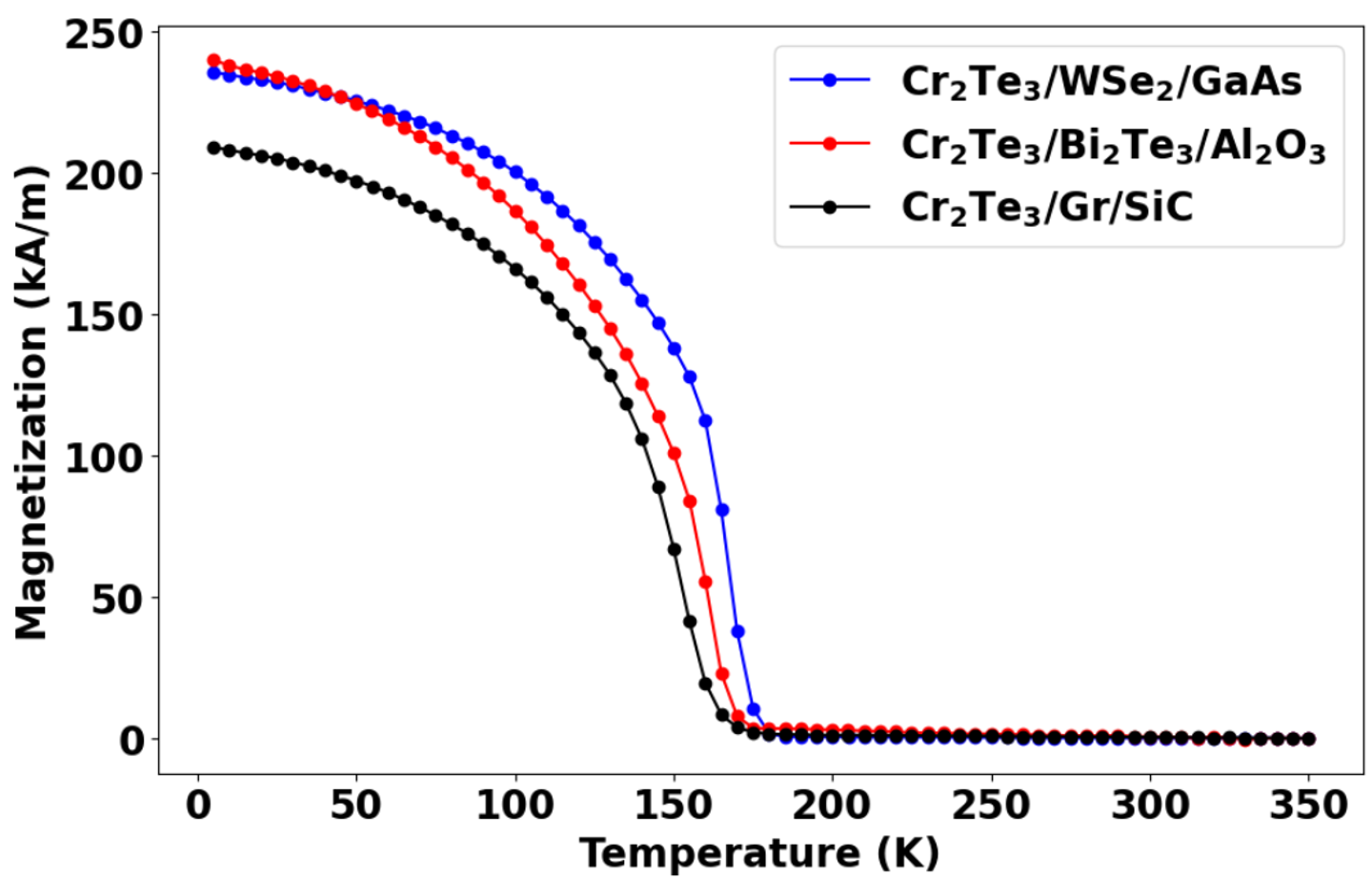}
    \caption{Remanent magnetization of Cr$_2$Te$_3$ layers grown and annealed on WSe$_2$ (sample 2), graphene (sample 4), and Bi$_2$Te$_3$ (sample 6) as a function of temperature with no external field.}
    \label{remanence}
\end{figure}

To determine the $T_\mathrm{C}$ of each annealed sample, we recorded the remanent magnetization after saturation at 5 T (with 5~K steps and no external field) as a function of temperature (Fig.~\ref{remanence}). A value close to 180~K was found for the three substrates demonstrating again the very weak interaction between Cr$_2$Te$_3$ and the vdW substrates. Here, we believe that the $T_\mathrm{C}$ is fully determined by the 2:3 stoichiometry of the films.

\section{Magnetotransport}
  
\begin{figure*}[h]
    \center
    \includegraphics[width=17.2cm]{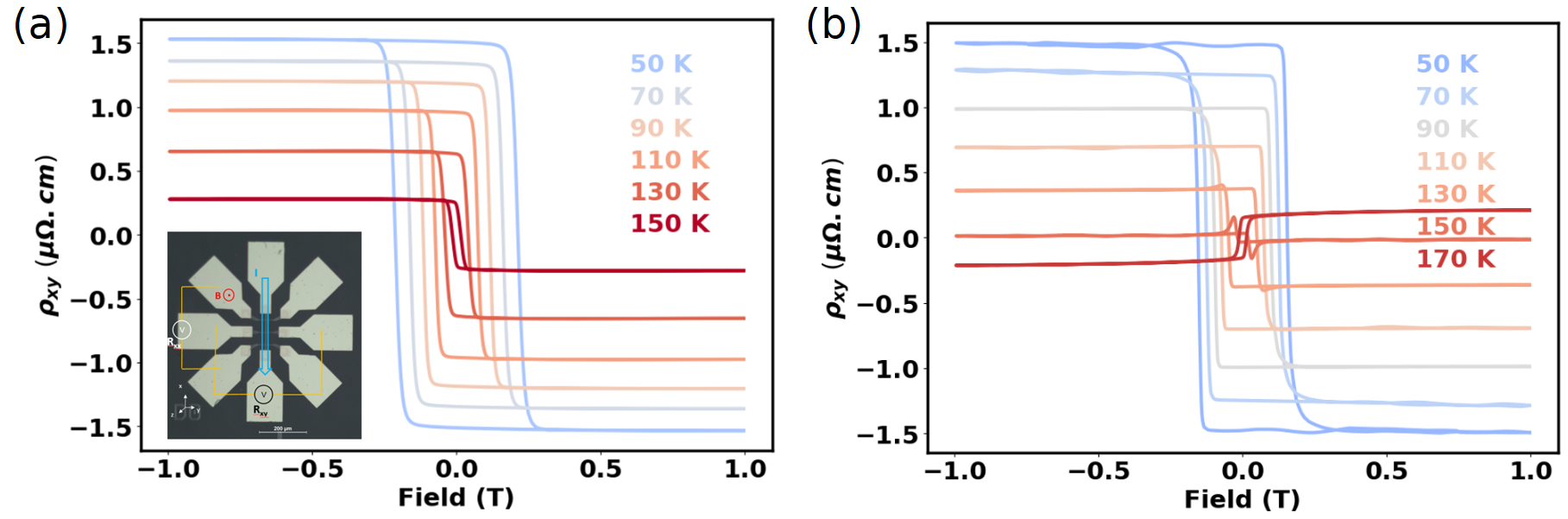}
    \caption{(a) Temperature-dependent Hall resistivity of Cr$_2$Te$_3$/WSe$_2$/GaAs (sample 2) after removal of the ordinary Hall slope with a magnetic field applied out-of-plane. Inset: an optical image of the Hall bar device processed by laser lithography with Ti(10nm)/Au(100nm) contacts. (b) Temperature-dependent Hall resistivity of Cr$_2$Te$_3$/Bi$_2$Te$_3$/Al$_2$O$_3$ (sample 6) after annealing (evaporation of Bi$_2$Te$_3$) and subtraction of the ordinary Hall slope.
}
\label{electric_wse2}
\end{figure*}

To study the magnetotransport properties, we performed four-probe resistance measurements and found an increasing longitudinal resistivity of Cr$_2$Te$_3$ layers with temperature indicating a metallic character (see the Supplemental Material Fig.~S8). The resistivity is of the order of 500 $\mu \Omega$.cm at 4~K. Figure~\ref{electric_wse2}(a) shows the Hall resistivity of 5 ML of Cr$_2$Te$_3$ deposited on WSe$_2$ (sample 2) as a function of the perpendicular magnetic field at different temperatures. For visibility, the ordinary Hall slope was subtracted and a carrier density of 1.6$\times$10$^{15}$ holes/cm$^2$ was extracted at 50~K, compared to 7.0$\times$10$^{15}$ holes/cm$^2$ for 5 ML of Cr$_2$Te$_3$ directly deposited on sapphire (see the Supplemental Material Fig.~S9), indicating a charge transfer from the WSe$_2$ layer. The clear anomalous Hall signal confirmed the strong PMA of the ferromagnet.

The same measurements were performed for a sample grown on Bi$_2$Te$_3$ (sample 6) and annealed at 400°C (resulting in Bi$_2$Te$_3$ evaporation), as shown in Fig.~\ref{electric_wse2}(b). The ordinary hall slope was removed and a carrier density of 4.5$\times$10$^{15}$ holes/cm$^2$ was extracted at 50~K. Since there is no charge transfer with sapphire, the difference in carrier density with Cr$_2$Te$_3$ directly grown on sapphire could be explained by the presence of defects at the interface introduced during the evaporation of the Bi$_2$Te$_3$ layer. 

\begin{figure*}[h]
    \center
    \includegraphics[width=17.2cm]{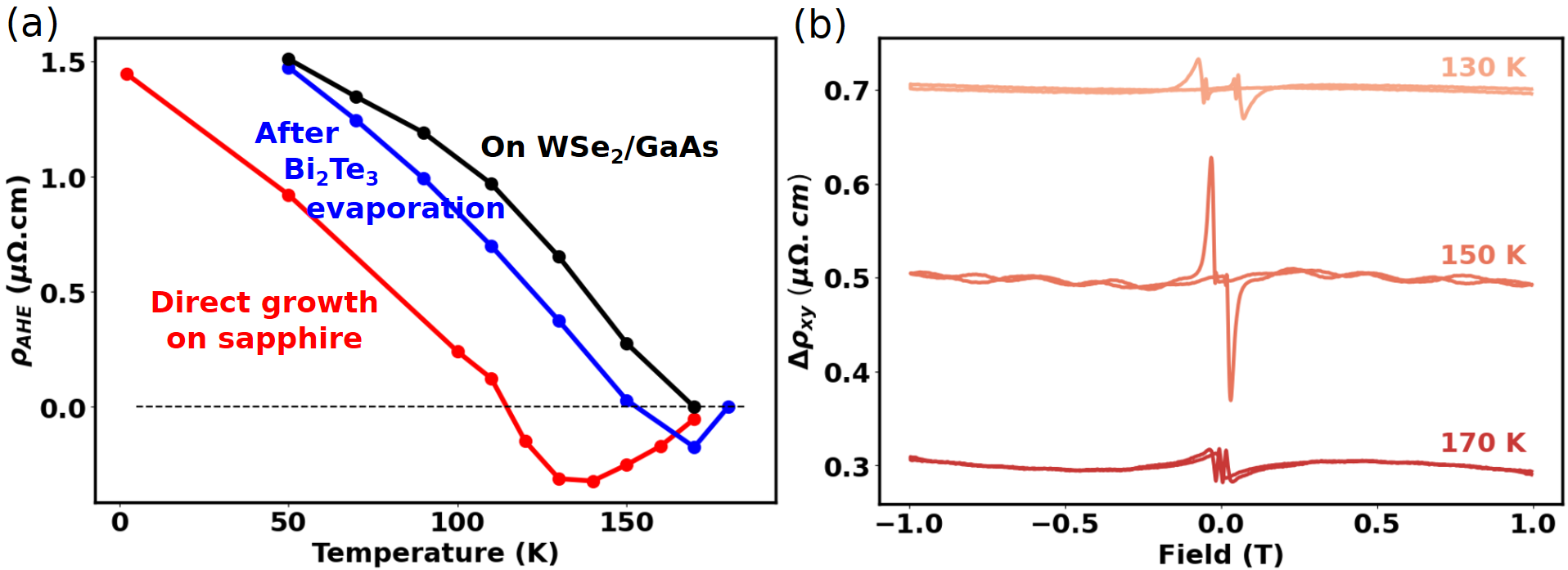}
    \caption{(a) Anomalous Hall resistivity of Cr$_2$Te$_3$/Al$_2$O$_3$ as a function of temperature for direct growth on sapphire, after thermal removal of the Bi$_2$Te$_3$ layer and growth on WSe$_2$. 
(b)	Hall resistivity of Cr$_2$Te$_3$ deposited on Bi$_2$Te$_3$ and annealed after subtracting the ordinary and anomalous Hall contributions. The curves are vertically shifted for clarity.
}
\label{electric-analysis}
\end{figure*}

In Fig.~\ref{electric-analysis}(a), the anomalous Hall resistivity is extracted for the two samples on sapphire as a function of temperature. We observe in both cases a sign change of the anomalous Hall resistivity below the Curie temperature of 180~K. 
Similar observations were reported in \cite{chi2022strain, fujisawa2022widely}. The possible origin of this effect is discussed in the following as a consequence of the energy-dependent Berry phase of Cr$_2$Te$_3$.\\

In the temperature range of the AH resistivity sign reversal, a resonance of the Hall signal manifested as peaks at the coercive fields can be observed. Figure~\ref{electric-analysis}(b) shows the Hall resistivity after subtraction of the ordinary and anomalous Hall effect at two temperatures below the sign change and one above. The bumps are enhanced when the temperature is closer (but still lower) than the temperature of the sign change and disappear above it. The width of the bumps also decreases with temperature, which could be related to the shrinking of the coercive field. The physical origin of such an effect is still under debate. In a similar heterostructure, Chen~\textit{et al.}~\cite{chen_evidence_2019} interpreted it as the topological Hall effect, which would originate from the presence of magnetic skyrmions. Skyrmions nucleate during the magnetization reversal and give rise to an extra transverse transport channel inducing a  peak in the Hall resistivity. Imaging such spin textures has been performed by Lorentz-TEM in Cr$_3$Te$_4$ layers \cite{zhang_room_2022}. Nevertheless, another explanation has been put forward by other groups as two anomalous Hall contributions with opposite signs \cite{tai_distinguishing_2022}. The origin could be thickness variations, inhomogeneities in the film or interface effects leading to the sign of the AHE being different \cite{wang_controllable_2020,kimbell_challenges_2022}. In the case of \CT\, these peaks appear close to the anomalous Hall resistivity sign change. If the thickness of the layer is not strictly constant over the Hall bar, some areas could have slightly different temperatures at which the anomalous Hall signal changes sign. In this case, for intermediate temperatures, two AHE components with opposite signs would indeed add up and could explain the observed behavior.

\begin{figure}
    \center
    \includegraphics[width=3.4in]{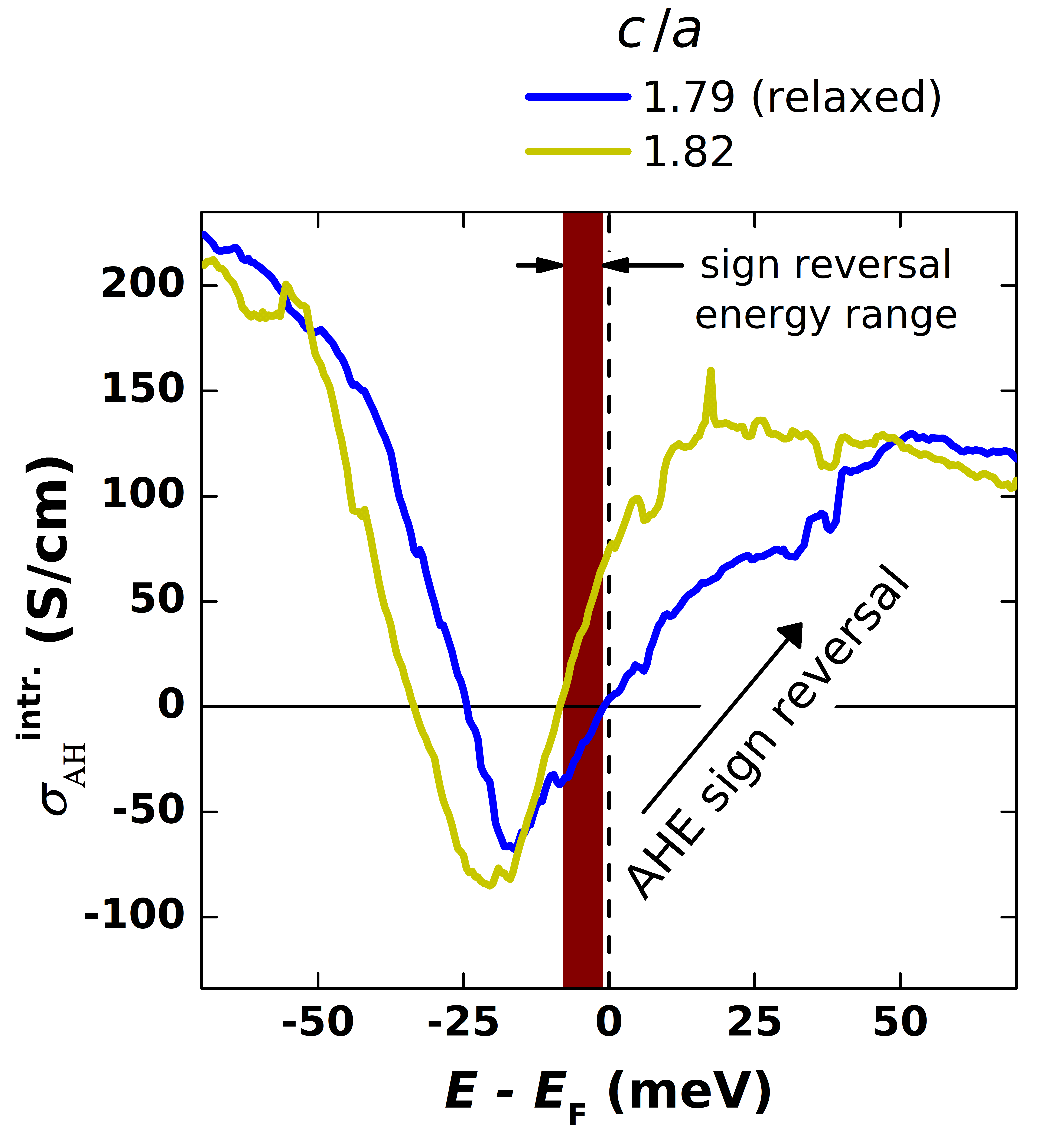}
    \caption{Intrinsic Anomalous Hall conductivity in bulk \CT~as a function of energy. A sign reversal occurs close to $E_\mathrm{F}$. Two different strains ($c/a$ ratios) are considered, in agreement with the experimental ones. Films with $E_\mathrm{F}$ within the red energy range will experience $\sigma^\mathrm{int.}_\mathrm{AH}$ sign reversal upon this possible strain change with temperature.}
    \label{fig:DFT_AHE}
\end{figure}

The sign reversal of the anomalous Hall effect observed experimentally can be elucidated by \textit{ab initio} calculations. The longitudinal resistivity is in the range where the contribution to AHE from intrinsic and impurity scattering components coexist, while the intrinsic part stays significant \cite{fujisawa2022widely}. We thus calculated the intrinsic contribution to AHE for bulk \CT\ (see Methods). As shown in Fig.~\ref{fig:DFT_AHE}, it exhibits a clear sign reversal very close to $E_\mathrm{F}$ ($\sim -10$~meV). This is in contrast with previous calculations \cite{jeon_emergent_2022} where the sign reversal occurs 330~meV above $E_\mathrm{F}$. This difference is due to the inclusion of the vdW corrections in our DFT calculations (see the Supplemental Material Fig.~S10). We consider three different mechanisms influencing the value and sign of the anomalous Hall conductivity: (a) thermal broadening around the Fermi level (of the order of $k_\mathrm{B}T$, i.e., 15 meV for $\Delta T = 180$~K), (b) charge transfer with the substrate (which we calculated was greatest on graphene inducing a Fermi level shift of $\approx +50$~meV), and (c) out-of-plane strain (see the Supplemental Material Fig.~S11). All these effects change the system energy in a range compatible with the calculations in Fig.~\ref{fig:DFT_AHE}.\\
We believe that the strain dependence of $\sigma^\mathrm{int.}_\mathrm{AH}$ is at the origin of the change of sign of the AHE reported in Fig.~\ref{electric-analysis}(a). Anisotropic lattice expansion with temperature was reported for Cr$_{1+\delta}$Te$_2$ \cite{li_diverse_2022}, which directly affects the AHE conductivity. To illustrate this qualitative argument, we chose in Fig.~\ref{fig:DFT_AHE} two reasonable strain values in agreement with the structural data we obtained, keeping in mind that films in this work have $c/a$ ranging between 1.797 and 1.868 (see Fig.~\ref{fig:DFT_anisotropy}). If the Fermi level of the sample lies in the red shaded area (between -8 and 0 meV), the evolution of $c/a$ from 1.79 to 1.82 with temperature would lead to a sign change of $\sigma^\mathrm{int.}_\mathrm{AH}$.\\
Another effect that could influence this picture is the thermal broadening of the Fermi-Dirac distribution upon heating. However, we obtained a mostly linear dependence of $\sigma^\mathrm{int.}_\mathrm{AH}$ on energy close to the Fermi level. When considering contributions above and below $E_F$, both would cancel out as the thermal broadening is symmetric.
Finally, the role of the substrate has to be also accounted for. As shown experimentally, charge transfer with the 2D materials was observed and leads to a shift of the Fermi level. This explains why the effect is present for samples standing on sapphire and not for the one on WSe$_2$. Indeed, for sapphire, the Fermi level is the lowest (carrier density of 4.5$\times$10$^{15}$ holes/cm$^2$ and 7.0$\times$10$^{15}$ holes/cm$^2$) so that we observe a sign change whereas the Fermi level is shifted up for WSe$_2$ (1.6$\times$10$^{15}$ holes/cm$^2$) and the sign change is absent. This observation is in agreement with the fact that the sign change in \textit{ab initio} calculations occurs for lower energies as shown in Fig.~\ref{fig:DFT_AHE}.

\begin{figure}[h]
    \center
    \includegraphics[width=8.6cm]{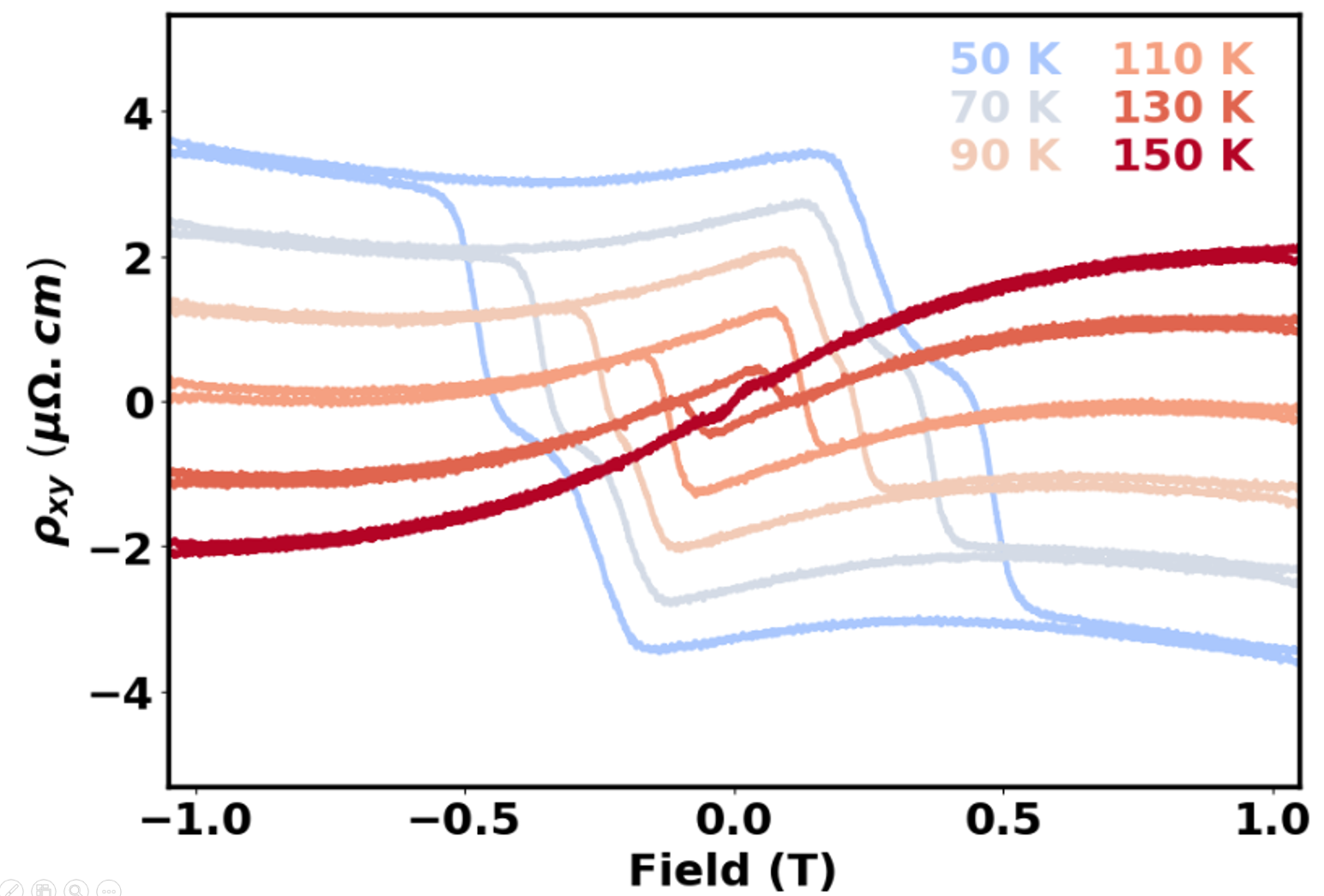}
    \caption{Temperature-dependent Hall resistivity of \CT/graphene/SiC. An arbitrary slope of 4 $\mu \Omega$.cm/T has been subtracted for comparison with Fig.~\ref{electric_wse2}.}
\label{electric_gr}
\end{figure}

Finally, in Fig.~\ref{electric_gr}, we present magnetotransport measurements on 5 layers of Cr$_2$Te$_3$ grown on graphene/SiC (sample 3). Both layers are metallic and contribute to conduction. The Hall resistivity is plotted as a function of the applied perpendicular magnetic field at different temperatures. No anomalous Hall resistivity sign change is measurable below the Curie temperature. This observation is in good agreement with \textit{ab initio} calculations since the extracted carrier density at 50~K, which is the lowest: 1.4$\times$10$^{14}$ holes/cm$^2$, corresponds to a Fermi level position shifted towards higher values. On top of the anomalous Hall contribution following the magnetization reversal at 0.5~T (at 50~K), another step close to 0.4~T is also present. This two-step signal behavior (absent in SQUID and XMCD measurements) vanishes progressively when increasing the temperature and disappears around 100~K, well below the Curie temperature. The origin of this effect needs further investigation and is out of the scope of the present work.

\section{Conclusion}

In conclusion, we reported the vdW epitaxy of \CT\ on three different 2D materials. We revealed the pristine interface and the preservation of the intrinsic properties of the underlying layers after the growth of the vdW ferromagnet. We demonstrated the free-standing character of \CT\ layers grown on these 2D materials after an annealing step at 400°C. Besides, the energy given to the system during the growth was identified as a way to control the crystal structure and tune the magnetic properties. We observed a correlation between the PMA energy of the system and the lattice parameters which was elucidated by \textit{ab initio} calculations. Finally, we theoretically predicted a strain-sensitive sign change of the Berry phase very close to the Fermi level explaining the measured sign change of AHE with temperature. Charge transfer between the 2D layers and \CT\ was shown to directly affect the temperature at which the AHE changes sign by shifting the Fermi level. To summarize, this system outputs highly tunable structural, magnetic and electrical properties, which presents an important asset for future spintronic applications.

\section*{Acknowledgments}
This project has received funding from the \textit{European Union's Horizon 2020 research and innovation programme} under grant agreement No \textit{800945} — NUMERICS — H2020-MSCA-COFUND-2017 and grant agreement 881603 (Graphene Flagship). We also acknowledge the French National Research Agency through the MAGICVALLEY project (ANR-18-CE24-0007) and the ESR-Equipex+ project 2D-MAG on two-dimensional magnetic materials.
We acknowledge the financial support from the ANR project ELMAX (ANR-20-CE24-0015) and from the LANEF framework (ANR-10-LABX-51-01) for its support with mutualized infrastructure. This work was partly supported by the French RENATECH network.
XMCD experiments were performed on the DEIMOS beamline at SOLEIL Synchrotron, France (proposal number 20220542). We are grateful to the SOLEIL staff for smoothly running the facility.

\bibliography{References}

\begin{thebibliography}{50}%
\makeatletter
\providecommand \@ifxundefined [1]{%
 \@ifx{#1\undefined}
}%
\providecommand \@ifnum [1]{%
 \ifnum #1\expandafter \@firstoftwo
 \else \expandafter \@secondoftwo
 \fi
}%
\providecommand \@ifx [1]{%
 \ifx #1\expandafter \@firstoftwo
 \else \expandafter \@secondoftwo
 \fi
}%
\providecommand \natexlab [1]{#1}%
\providecommand \enquote  [1]{``#1''}%
\providecommand \bibnamefont  [1]{#1}%
\providecommand \bibfnamefont [1]{#1}%
\providecommand \citenamefont [1]{#1}%
\providecommand \href@noop [0]{\@secondoftwo}%
\providecommand \href [0]{\begingroup \@sanitize@url \@href}%
\providecommand \@href[1]{\@@startlink{#1}\@@href}%
\providecommand \@@href[1]{\endgroup#1\@@endlink}%
\providecommand \@sanitize@url [0]{\catcode `\\12\catcode `\$12\catcode
  `\&12\catcode `\#12\catcode `\^12\catcode `\_12\catcode `\%12\relax}%
\providecommand \@@startlink[1]{}%
\providecommand \@@endlink[0]{}%
\providecommand \url  [0]{\begingroup\@sanitize@url \@url }%
\providecommand \@url [1]{\endgroup\@href {#1}{\urlprefix }}%
\providecommand \urlprefix  [0]{URL }%
\providecommand \Eprint [0]{\href }%
\providecommand \doibase [0]{https://doi.org/}%
\providecommand \selectlanguage [0]{\@gobble}%
\providecommand \bibinfo  [0]{\@secondoftwo}%
\providecommand \bibfield  [0]{\@secondoftwo}%
\providecommand \translation [1]{[#1]}%
\providecommand \BibitemOpen [0]{}%
\providecommand \bibitemStop [0]{}%
\providecommand \bibitemNoStop [0]{.\EOS\space}%
\providecommand \EOS [0]{\spacefactor3000\relax}%
\providecommand \BibitemShut  [1]{\csname bibitem#1\endcsname}%
\let\auto@bib@innerbib\@empty
\bibitem [{\citenamefont {Gong}\ \emph {et~al.}(2017)\citenamefont {Gong},
  \citenamefont {Li}, \citenamefont {Li}, \citenamefont {Ji}, \citenamefont
  {Stern}, \citenamefont {Xia}, \citenamefont {Cao}, \citenamefont {Bao},
  \citenamefont {Wang}, \citenamefont {Wang}, \citenamefont {Qiu},
  \citenamefont {Cava}, \citenamefont {Louie}, \citenamefont {Xia},\ and\
  \citenamefont {Zhang}}]{gong_discovery_2017}%
  \BibitemOpen
  \bibfield  {author} {\bibinfo {author} {\bibfnamefont {C.}~\bibnamefont
  {Gong}}, \bibinfo {author} {\bibfnamefont {L.}~\bibnamefont {Li}}, \bibinfo
  {author} {\bibfnamefont {Z.}~\bibnamefont {Li}}, \bibinfo {author}
  {\bibfnamefont {H.}~\bibnamefont {Ji}}, \bibinfo {author} {\bibfnamefont
  {A.}~\bibnamefont {Stern}}, \bibinfo {author} {\bibfnamefont
  {Y.}~\bibnamefont {Xia}}, \bibinfo {author} {\bibfnamefont {T.}~\bibnamefont
  {Cao}}, \bibinfo {author} {\bibfnamefont {W.}~\bibnamefont {Bao}}, \bibinfo
  {author} {\bibfnamefont {C.}~\bibnamefont {Wang}}, \bibinfo {author}
  {\bibfnamefont {Y.}~\bibnamefont {Wang}}, \bibinfo {author} {\bibfnamefont
  {Z.~Q.}\ \bibnamefont {Qiu}}, \bibinfo {author} {\bibfnamefont {R.~J.}\
  \bibnamefont {Cava}}, \bibinfo {author} {\bibfnamefont {S.~G.}\ \bibnamefont
  {Louie}}, \bibinfo {author} {\bibfnamefont {J.}~\bibnamefont {Xia}},\ and\
  \bibinfo {author} {\bibfnamefont {X.}~\bibnamefont {Zhang}},\ }\bibfield
  {title} {\bibinfo {title} {Discovery of intrinsic ferromagnetism in
  two-dimensional van der {Waals} crystals},\ }\href@noop {} {\bibfield
  {journal} {\bibinfo  {journal} {Nature}\ }\textbf {\bibinfo {volume} {546}},\
  \bibinfo {pages} {265} (\bibinfo {year} {2017})}\BibitemShut {NoStop}%
\bibitem [{\citenamefont {Huang}\ \emph {et~al.}(2017)\citenamefont {Huang},
  \citenamefont {Clark}, \citenamefont {Navarro-Moratalla}, \citenamefont
  {Klein}, \citenamefont {Cheng}, \citenamefont {Seyler}, \citenamefont
  {Zhong}, \citenamefont {Schmidgall}, \citenamefont {McGuire}, \citenamefont
  {Cobden}, \citenamefont {Yao}, \citenamefont {Xiao}, \citenamefont
  {Jarillo-Herrero},\ and\ \citenamefont {Xu}}]{huang_layer-dependent_2017}%
  \BibitemOpen
  \bibfield  {author} {\bibinfo {author} {\bibfnamefont {B.}~\bibnamefont
  {Huang}}, \bibinfo {author} {\bibfnamefont {G.}~\bibnamefont {Clark}},
  \bibinfo {author} {\bibfnamefont {E.}~\bibnamefont {Navarro-Moratalla}},
  \bibinfo {author} {\bibfnamefont {D.~R.}\ \bibnamefont {Klein}}, \bibinfo
  {author} {\bibfnamefont {R.}~\bibnamefont {Cheng}}, \bibinfo {author}
  {\bibfnamefont {K.~L.}\ \bibnamefont {Seyler}}, \bibinfo {author}
  {\bibfnamefont {D.}~\bibnamefont {Zhong}}, \bibinfo {author} {\bibfnamefont
  {E.}~\bibnamefont {Schmidgall}}, \bibinfo {author} {\bibfnamefont {M.~A.}\
  \bibnamefont {McGuire}}, \bibinfo {author} {\bibfnamefont {D.~H.}\
  \bibnamefont {Cobden}}, \bibinfo {author} {\bibfnamefont {W.}~\bibnamefont
  {Yao}}, \bibinfo {author} {\bibfnamefont {D.}~\bibnamefont {Xiao}}, \bibinfo
  {author} {\bibfnamefont {P.}~\bibnamefont {Jarillo-Herrero}},\ and\ \bibinfo
  {author} {\bibfnamefont {X.}~\bibnamefont {Xu}},\ }\bibfield  {title}
  {\bibinfo {title} {Layer-dependent ferromagnetism in a van der {Waals}
  crystal down to the monolayer limit},\ }\href
  {https://doi.org/10.1038/nature22391} {\bibfield  {journal} {\bibinfo
  {journal} {Nature}\ }\textbf {\bibinfo {volume} {546}},\ \bibinfo {pages}
  {270} (\bibinfo {year} {2017})}\BibitemShut {NoStop}%
\bibitem [{\citenamefont {Wang}\ \emph {et~al.}(2022)\citenamefont {Wang},
  \citenamefont {Bedoya-Pinto}, \citenamefont {Blei}, \citenamefont {Dismukes},
  \citenamefont {Hamo}, \citenamefont {Jenkins}, \citenamefont {Koperski},
  \citenamefont {Liu}, \citenamefont {Sun}, \citenamefont {Telford},
  \citenamefont {Kim}, \citenamefont {Augustin}, \citenamefont {Vool},
  \citenamefont {Yin}, \citenamefont {Li}, \citenamefont {Falin}, \citenamefont
  {Dean}, \citenamefont {Casanova}, \citenamefont {Evans}, \citenamefont
  {Chshiev}, \citenamefont {Mishchenko}, \citenamefont {Petrovic},
  \citenamefont {He}, \citenamefont {Zhao}, \citenamefont {Tsen}, \citenamefont
  {Gerardot}, \citenamefont {Brotons-Gisbert}, \citenamefont {Guguchia},
  \citenamefont {Roy}, \citenamefont {Tongay}, \citenamefont {Wang},
  \citenamefont {Hasan}, \citenamefont {Wrachtrup}, \citenamefont {Yacoby},
  \citenamefont {Fert}, \citenamefont {Parkin}, \citenamefont {Novoselov},
  \citenamefont {Dai}, \citenamefont {Balicas},\ and\ \citenamefont
  {Santos}}]{wang_magnetic_2022}%
  \BibitemOpen
  \bibfield  {author} {\bibinfo {author} {\bibfnamefont {Q.~H.}\ \bibnamefont
  {Wang}}, \bibinfo {author} {\bibfnamefont {A.}~\bibnamefont {Bedoya-Pinto}},
  \bibinfo {author} {\bibfnamefont {M.}~\bibnamefont {Blei}}, \bibinfo {author}
  {\bibfnamefont {A.~H.}\ \bibnamefont {Dismukes}}, \bibinfo {author}
  {\bibfnamefont {A.}~\bibnamefont {Hamo}}, \bibinfo {author} {\bibfnamefont
  {S.}~\bibnamefont {Jenkins}}, \bibinfo {author} {\bibfnamefont
  {M.}~\bibnamefont {Koperski}}, \bibinfo {author} {\bibfnamefont
  {Y.}~\bibnamefont {Liu}}, \bibinfo {author} {\bibfnamefont {Q.-C.}\
  \bibnamefont {Sun}}, \bibinfo {author} {\bibfnamefont {E.~J.}\ \bibnamefont
  {Telford}}, \bibinfo {author} {\bibfnamefont {H.~H.}\ \bibnamefont {Kim}},
  \bibinfo {author} {\bibfnamefont {M.}~\bibnamefont {Augustin}}, \bibinfo
  {author} {\bibfnamefont {U.}~\bibnamefont {Vool}}, \bibinfo {author}
  {\bibfnamefont {J.-X.}\ \bibnamefont {Yin}}, \bibinfo {author} {\bibfnamefont
  {L.~H.}\ \bibnamefont {Li}}, \bibinfo {author} {\bibfnamefont
  {A.}~\bibnamefont {Falin}}, \bibinfo {author} {\bibfnamefont {C.~R.}\
  \bibnamefont {Dean}}, \bibinfo {author} {\bibfnamefont {F.}~\bibnamefont
  {Casanova}}, \bibinfo {author} {\bibfnamefont {R.~F.~L.}\ \bibnamefont
  {Evans}}, \bibinfo {author} {\bibfnamefont {M.}~\bibnamefont {Chshiev}},
  \bibinfo {author} {\bibfnamefont {A.}~\bibnamefont {Mishchenko}}, \bibinfo
  {author} {\bibfnamefont {C.}~\bibnamefont {Petrovic}}, \bibinfo {author}
  {\bibfnamefont {R.}~\bibnamefont {He}}, \bibinfo {author} {\bibfnamefont
  {L.}~\bibnamefont {Zhao}}, \bibinfo {author} {\bibfnamefont {A.~W.}\
  \bibnamefont {Tsen}}, \bibinfo {author} {\bibfnamefont {B.~D.}\ \bibnamefont
  {Gerardot}}, \bibinfo {author} {\bibfnamefont {M.}~\bibnamefont
  {Brotons-Gisbert}}, \bibinfo {author} {\bibfnamefont {Z.}~\bibnamefont
  {Guguchia}}, \bibinfo {author} {\bibfnamefont {X.}~\bibnamefont {Roy}},
  \bibinfo {author} {\bibfnamefont {S.}~\bibnamefont {Tongay}}, \bibinfo
  {author} {\bibfnamefont {Z.}~\bibnamefont {Wang}}, \bibinfo {author}
  {\bibfnamefont {M.~Z.}\ \bibnamefont {Hasan}}, \bibinfo {author}
  {\bibfnamefont {J.}~\bibnamefont {Wrachtrup}}, \bibinfo {author}
  {\bibfnamefont {A.}~\bibnamefont {Yacoby}}, \bibinfo {author} {\bibfnamefont
  {A.}~\bibnamefont {Fert}}, \bibinfo {author} {\bibfnamefont {S.}~\bibnamefont
  {Parkin}}, \bibinfo {author} {\bibfnamefont {K.~S.}\ \bibnamefont
  {Novoselov}}, \bibinfo {author} {\bibfnamefont {P.}~\bibnamefont {Dai}},
  \bibinfo {author} {\bibfnamefont {L.}~\bibnamefont {Balicas}},\ and\ \bibinfo
  {author} {\bibfnamefont {E.~J.~G.}\ \bibnamefont {Santos}},\ }\bibfield
  {title} {\bibinfo {title} {The magnetic genome of two-dimensional van der
  waals materials},\ }\href {https://doi.org/10.1021/acsnano.1c09150}
  {\bibfield  {journal} {\bibinfo  {journal} {ACS Nano}\ }\textbf {\bibinfo
  {volume} {16}},\ \bibinfo {pages} {6960} (\bibinfo {year}
  {2022})}\BibitemShut {NoStop}%
\bibitem [{\citenamefont {C. Ferrari}\ \emph {et~al.}(2015)\citenamefont
  {C. Ferrari}, \citenamefont {Bonaccorso}, \citenamefont {Fal'ko},
  \citenamefont {S. Novoselov}, \citenamefont {Roche}, \citenamefont
  {Bøggild}, \citenamefont {Borini}, \citenamefont {L. Koppens},
  \citenamefont {Palermo}, \citenamefont {Pugno}, \citenamefont {A. Garrido},
  \citenamefont {Sordan}, \citenamefont {Bianco}, \citenamefont {Ballerini},
  \citenamefont {Prato}, \citenamefont {Lidorikis}, \citenamefont {Kivioja},
  \citenamefont {Marinelli}, \citenamefont {Ryhänen}, \citenamefont
  {Morpurgo}, \citenamefont {N. Coleman}, \citenamefont {Nicolosi},
  \citenamefont {Colombo}, \citenamefont {Fert}, \citenamefont
  {Garcia-Hernandez}, \citenamefont {Bachtold}, \citenamefont {F. Schneider},
  \citenamefont {Guinea}, \citenamefont {Dekker}, \citenamefont {Barbone},
  \citenamefont {Sun}, \citenamefont {Galiotis}, \citenamefont
  {N. Grigorenko}, \citenamefont {Konstantatos}, \citenamefont {Kis},
  \citenamefont {Katsnelson}, \citenamefont {Vandersypen}, \citenamefont
  {Loiseau}, \citenamefont {Morandi}, \citenamefont {Neumaier}, \citenamefont
  {Treossi}, \citenamefont {Pellegrini}, \citenamefont {Polini}, \citenamefont
  {Tredicucci}, \citenamefont {M. Williams}, \citenamefont {Hong},
  \citenamefont {Ahn}, \citenamefont {Kim}, \citenamefont {Zirath},
  \citenamefont {Wees}, \citenamefont {Zant}, \citenamefont {Occhipinti},
  \citenamefont {Matteo}, \citenamefont {A. Kinloch}, \citenamefont {Seyller},
  \citenamefont {Quesnel}, \citenamefont {Feng}, \citenamefont {Teo},
  \citenamefont {Rupesinghe}, \citenamefont {Hakonen}, \citenamefont
  {T. Neil}, \citenamefont {Tannock}, \citenamefont {Löfwander},\ and\
  \citenamefont {Kinaret}}]{cferrari_science_2015}%
  \BibitemOpen
  \bibfield  {author} {\bibinfo {author} {\bibfnamefont {A.}~\bibnamefont
  {C. Ferrari}}, \bibinfo {author} {\bibfnamefont {F.}~\bibnamefont
  {Bonaccorso}}, \bibinfo {author} {\bibfnamefont {V.}~\bibnamefont {Fal'ko}},
  \bibinfo {author} {\bibfnamefont {K.}~\bibnamefont {S. Novoselov}}, \bibinfo
  {author} {\bibfnamefont {S.}~\bibnamefont {Roche}}, \bibinfo {author}
  {\bibfnamefont {P.}~\bibnamefont {Bøggild}}, \bibinfo {author}
  {\bibfnamefont {S.}~\bibnamefont {Borini}}, \bibinfo {author} {\bibfnamefont
  {F.~H.}\ \bibnamefont {L. Koppens}}, \bibinfo {author} {\bibfnamefont
  {V.}~\bibnamefont {Palermo}}, \bibinfo {author} {\bibfnamefont
  {N.}~\bibnamefont {Pugno}}, \bibinfo {author} {\bibfnamefont
  {J.}~\bibnamefont {A. Garrido}}, \bibinfo {author} {\bibfnamefont
  {R.}~\bibnamefont {Sordan}}, \bibinfo {author} {\bibfnamefont
  {A.}~\bibnamefont {Bianco}}, \bibinfo {author} {\bibfnamefont
  {L.}~\bibnamefont {Ballerini}}, \bibinfo {author} {\bibfnamefont
  {M.}~\bibnamefont {Prato}}, \bibinfo {author} {\bibfnamefont
  {E.}~\bibnamefont {Lidorikis}}, \bibinfo {author} {\bibfnamefont
  {J.}~\bibnamefont {Kivioja}}, \bibinfo {author} {\bibfnamefont
  {C.}~\bibnamefont {Marinelli}}, \bibinfo {author} {\bibfnamefont
  {T.}~\bibnamefont {Ryhänen}}, \bibinfo {author} {\bibfnamefont
  {A.}~\bibnamefont {Morpurgo}}, \bibinfo {author} {\bibfnamefont
  {J.}~\bibnamefont {N. Coleman}}, \bibinfo {author} {\bibfnamefont
  {V.}~\bibnamefont {Nicolosi}}, \bibinfo {author} {\bibfnamefont
  {L.}~\bibnamefont {Colombo}}, \bibinfo {author} {\bibfnamefont
  {A.}~\bibnamefont {Fert}}, \bibinfo {author} {\bibfnamefont {M.}~\bibnamefont
  {Garcia-Hernandez}}, \bibinfo {author} {\bibfnamefont {A.}~\bibnamefont
  {Bachtold}}, \bibinfo {author} {\bibfnamefont {G.}~\bibnamefont
  {F. Schneider}}, \bibinfo {author} {\bibfnamefont {F.}~\bibnamefont
  {Guinea}}, \bibinfo {author} {\bibfnamefont {C.}~\bibnamefont {Dekker}},
  \bibinfo {author} {\bibfnamefont {M.}~\bibnamefont {Barbone}}, \bibinfo
  {author} {\bibfnamefont {Z.}~\bibnamefont {Sun}}, \bibinfo {author}
  {\bibfnamefont {C.}~\bibnamefont {Galiotis}}, \bibinfo {author}
  {\bibfnamefont {A.}~\bibnamefont {N. Grigorenko}}, \bibinfo {author}
  {\bibfnamefont {G.}~\bibnamefont {Konstantatos}}, \bibinfo {author}
  {\bibfnamefont {A.}~\bibnamefont {Kis}}, \bibinfo {author} {\bibfnamefont
  {M.}~\bibnamefont {Katsnelson}}, \bibinfo {author} {\bibfnamefont
  {L.}~\bibnamefont {Vandersypen}}, \bibinfo {author} {\bibfnamefont
  {A.}~\bibnamefont {Loiseau}}, \bibinfo {author} {\bibfnamefont
  {V.}~\bibnamefont {Morandi}}, \bibinfo {author} {\bibfnamefont
  {D.}~\bibnamefont {Neumaier}}, \bibinfo {author} {\bibfnamefont
  {E.}~\bibnamefont {Treossi}}, \bibinfo {author} {\bibfnamefont
  {V.}~\bibnamefont {Pellegrini}}, \bibinfo {author} {\bibfnamefont
  {M.}~\bibnamefont {Polini}}, \bibinfo {author} {\bibfnamefont
  {A.}~\bibnamefont {Tredicucci}}, \bibinfo {author} {\bibfnamefont
  {G.}~\bibnamefont {M. Williams}}, \bibinfo {author} {\bibfnamefont {B.~H.}\
  \bibnamefont {Hong}}, \bibinfo {author} {\bibfnamefont {J.-H.}\ \bibnamefont
  {Ahn}}, \bibinfo {author} {\bibfnamefont {J.~M.}\ \bibnamefont {Kim}},
  \bibinfo {author} {\bibfnamefont {H.}~\bibnamefont {Zirath}}, \bibinfo
  {author} {\bibfnamefont {B.~J.~v.}\ \bibnamefont {Wees}}, \bibinfo {author}
  {\bibfnamefont {H.~v.~d.}\ \bibnamefont {Zant}}, \bibinfo {author}
  {\bibfnamefont {L.}~\bibnamefont {Occhipinti}}, \bibinfo {author}
  {\bibfnamefont {A.~D.}\ \bibnamefont {Matteo}}, \bibinfo {author}
  {\bibfnamefont {I.}~\bibnamefont {A. Kinloch}}, \bibinfo {author}
  {\bibfnamefont {T.}~\bibnamefont {Seyller}}, \bibinfo {author} {\bibfnamefont
  {E.}~\bibnamefont {Quesnel}}, \bibinfo {author} {\bibfnamefont
  {X.}~\bibnamefont {Feng}}, \bibinfo {author} {\bibfnamefont {K.}~\bibnamefont
  {Teo}}, \bibinfo {author} {\bibfnamefont {N.}~\bibnamefont {Rupesinghe}},
  \bibinfo {author} {\bibfnamefont {P.}~\bibnamefont {Hakonen}}, \bibinfo
  {author} {\bibfnamefont {S.~R.}\ \bibnamefont {T. Neil}}, \bibinfo {author}
  {\bibfnamefont {Q.}~\bibnamefont {Tannock}}, \bibinfo {author} {\bibfnamefont
  {T.}~\bibnamefont {Löfwander}},\ and\ \bibinfo {author} {\bibfnamefont
  {J.}~\bibnamefont {Kinaret}},\ }\bibfield  {title} {\bibinfo {title} {Science
  and technology roadmap for graphene, related two-dimensional crystals, and
  hybrid systems},\ }\href@noop {} {\bibfield  {journal} {\bibinfo  {journal}
  {Nanoscale}\ }\textbf {\bibinfo {volume} {7}},\ \bibinfo {pages} {4598}
  (\bibinfo {year} {2015})}\BibitemShut {NoStop}%
\bibitem [{\citenamefont {Yang}\ \emph {et~al.}(2022)\citenamefont {Yang},
  \citenamefont {Valenzuela}, \citenamefont {Chshiev}, \citenamefont {Couet},
  \citenamefont {Dieny}, \citenamefont {Dlubak}, \citenamefont {Fert},
  \citenamefont {Garello}, \citenamefont {Jamet}, \citenamefont {Jeong},
  \citenamefont {Lee}, \citenamefont {Lee}, \citenamefont {Martin},
  \citenamefont {Kar}, \citenamefont {Sénéor}, \citenamefont {Shin},\ and\
  \citenamefont {Roche}}]{yang_two-dimensional_2022}%
  \BibitemOpen
  \bibfield  {author} {\bibinfo {author} {\bibfnamefont {H.}~\bibnamefont
  {Yang}}, \bibinfo {author} {\bibfnamefont {S.~O.}\ \bibnamefont
  {Valenzuela}}, \bibinfo {author} {\bibfnamefont {M.}~\bibnamefont {Chshiev}},
  \bibinfo {author} {\bibfnamefont {S.}~\bibnamefont {Couet}}, \bibinfo
  {author} {\bibfnamefont {B.}~\bibnamefont {Dieny}}, \bibinfo {author}
  {\bibfnamefont {B.}~\bibnamefont {Dlubak}}, \bibinfo {author} {\bibfnamefont
  {A.}~\bibnamefont {Fert}}, \bibinfo {author} {\bibfnamefont {K.}~\bibnamefont
  {Garello}}, \bibinfo {author} {\bibfnamefont {M.}~\bibnamefont {Jamet}},
  \bibinfo {author} {\bibfnamefont {D.-E.}\ \bibnamefont {Jeong}}, \bibinfo
  {author} {\bibfnamefont {K.}~\bibnamefont {Lee}}, \bibinfo {author}
  {\bibfnamefont {T.}~\bibnamefont {Lee}}, \bibinfo {author} {\bibfnamefont
  {M.-B.}\ \bibnamefont {Martin}}, \bibinfo {author} {\bibfnamefont {G.~S.}\
  \bibnamefont {Kar}}, \bibinfo {author} {\bibfnamefont {P.}~\bibnamefont
  {Sénéor}}, \bibinfo {author} {\bibfnamefont {H.-J.}\ \bibnamefont {Shin}},\
  and\ \bibinfo {author} {\bibfnamefont {S.}~\bibnamefont {Roche}},\ }\bibfield
   {title} {\bibinfo {title} {Two-dimensional materials prospects for
  non-volatile spintronic memories},\ }\href@noop {} {\bibfield  {journal}
  {\bibinfo  {journal} {Nature}\ }\textbf {\bibinfo {volume} {606}},\ \bibinfo
  {pages} {663} (\bibinfo {year} {2022})}\BibitemShut {NoStop}%
\bibitem [{\citenamefont {Liu}\ \emph {et~al.}(2020)\citenamefont {Liu},
  \citenamefont {Chen}, \citenamefont {Wang}, \citenamefont {Liu},
  \citenamefont {Jiang}, \citenamefont {Zhang}, \citenamefont {Liu},\ and\
  \citenamefont {Zhou}}]{liu_two-dimensional_2020}%
  \BibitemOpen
  \bibfield  {author} {\bibinfo {author} {\bibfnamefont {C.}~\bibnamefont
  {Liu}}, \bibinfo {author} {\bibfnamefont {H.}~\bibnamefont {Chen}}, \bibinfo
  {author} {\bibfnamefont {S.}~\bibnamefont {Wang}}, \bibinfo {author}
  {\bibfnamefont {Q.}~\bibnamefont {Liu}}, \bibinfo {author} {\bibfnamefont
  {Y.-G.}\ \bibnamefont {Jiang}}, \bibinfo {author} {\bibfnamefont {D.~W.}\
  \bibnamefont {Zhang}}, \bibinfo {author} {\bibfnamefont {M.}~\bibnamefont
  {Liu}},\ and\ \bibinfo {author} {\bibfnamefont {P.}~\bibnamefont {Zhou}},\
  }\bibfield  {title} {\bibinfo {title} {Two-dimensional materials for
  next-generation computing technologies},\ }\href
  {https://doi.org/10.1038/s41565-020-0724-3} {\bibfield  {journal} {\bibinfo
  {journal} {Nature Nanotechnology}\ }\textbf {\bibinfo {volume} {15}},\
  \bibinfo {pages} {545} (\bibinfo {year} {2020})}\BibitemShut {NoStop}%
\bibitem [{\citenamefont {Dieny}\ and\ \citenamefont
  {Chshiev}(2017)}]{dieny_perpendicular_2017}%
  \BibitemOpen
  \bibfield  {author} {\bibinfo {author} {\bibfnamefont {B.}~\bibnamefont
  {Dieny}}\ and\ \bibinfo {author} {\bibfnamefont {M.}~\bibnamefont
  {Chshiev}},\ }\bibfield  {title} {\bibinfo {title} {Perpendicular magnetic
  anisotropy at transition metal/oxide interfaces and applications},\ }\href
  {https://doi.org/10.1103/RevModPhys.89.025008} {\bibfield  {journal}
  {\bibinfo  {journal} {Rev. Mod. Phys.}\ }\textbf {\bibinfo {volume} {89}},\
  \bibinfo {pages} {025008} (\bibinfo {year} {2017})}\BibitemShut {NoStop}%
\bibitem [{\citenamefont {Ribeiro}\ \emph {et~al.}(2022)\citenamefont
  {Ribeiro}, \citenamefont {Gentile}, \citenamefont {Marty}, \citenamefont
  {Dosenovic}, \citenamefont {Okuno}, \citenamefont {Vergnaud}, \citenamefont
  {Jacquot}, \citenamefont {Jalabert}, \citenamefont {Longo}, \citenamefont
  {Ohresser}, \citenamefont {Hallal}, \citenamefont {Chshiev}, \citenamefont
  {Boulle}, \citenamefont {Bonell},\ and\ \citenamefont
  {Jamet}}]{ribeiro_large-scale_2022}%
  \BibitemOpen
  \bibfield  {author} {\bibinfo {author} {\bibfnamefont {M.}~\bibnamefont
  {Ribeiro}}, \bibinfo {author} {\bibfnamefont {G.}~\bibnamefont {Gentile}},
  \bibinfo {author} {\bibfnamefont {A.}~\bibnamefont {Marty}}, \bibinfo
  {author} {\bibfnamefont {D.}~\bibnamefont {Dosenovic}}, \bibinfo {author}
  {\bibfnamefont {H.}~\bibnamefont {Okuno}}, \bibinfo {author} {\bibfnamefont
  {C.}~\bibnamefont {Vergnaud}}, \bibinfo {author} {\bibfnamefont {J.-F.}\
  \bibnamefont {Jacquot}}, \bibinfo {author} {\bibfnamefont {D.}~\bibnamefont
  {Jalabert}}, \bibinfo {author} {\bibfnamefont {D.}~\bibnamefont {Longo}},
  \bibinfo {author} {\bibfnamefont {P.}~\bibnamefont {Ohresser}}, \bibinfo
  {author} {\bibfnamefont {A.}~\bibnamefont {Hallal}}, \bibinfo {author}
  {\bibfnamefont {M.}~\bibnamefont {Chshiev}}, \bibinfo {author} {\bibfnamefont
  {O.}~\bibnamefont {Boulle}}, \bibinfo {author} {\bibfnamefont
  {F.}~\bibnamefont {Bonell}},\ and\ \bibinfo {author} {\bibfnamefont
  {M.}~\bibnamefont {Jamet}},\ }\bibfield  {title} {\bibinfo {title}
  {Large-scale epitaxy of two-dimensional van der {Waals} room-temperature
  ferromagnet {Fe5GeTe2}},\ }\href@noop {} {\bibfield  {journal} {\bibinfo
  {journal} {npj 2D Materials and Applications}\ }\textbf {\bibinfo {volume}
  {6}},\ \bibinfo {pages} {1} (\bibinfo {year} {2022})}\BibitemShut {NoStop}%
\bibitem [{\citenamefont {Freitas}\ \emph {et~al.}(2015)\citenamefont
  {Freitas}, \citenamefont {Weht}, \citenamefont {Sulpice}, \citenamefont
  {Remenyi}, \citenamefont {Strobel}, \citenamefont {Gay}, \citenamefont
  {Marcus},\ and\ \citenamefont
  {Núñez-Regueiro}}]{freitas_ferromagnetism_2015}%
  \BibitemOpen
  \bibfield  {author} {\bibinfo {author} {\bibfnamefont {D.~C.}\ \bibnamefont
  {Freitas}}, \bibinfo {author} {\bibfnamefont {R.}~\bibnamefont {Weht}},
  \bibinfo {author} {\bibfnamefont {A.}~\bibnamefont {Sulpice}}, \bibinfo
  {author} {\bibfnamefont {G.}~\bibnamefont {Remenyi}}, \bibinfo {author}
  {\bibfnamefont {P.}~\bibnamefont {Strobel}}, \bibinfo {author} {\bibfnamefont
  {F.}~\bibnamefont {Gay}}, \bibinfo {author} {\bibfnamefont {J.}~\bibnamefont
  {Marcus}},\ and\ \bibinfo {author} {\bibfnamefont {M.}~\bibnamefont
  {Núñez-Regueiro}},\ }\bibfield  {title} {\bibinfo {title} {Ferromagnetism
  in layered metastable 1 \textit{{T}} -{CrTe} $_{\textrm{2}}$},\ }\href@noop
  {} {\bibfield  {journal} {\bibinfo  {journal} {Journal of Physics: Condensed
  Matter}\ }\textbf {\bibinfo {volume} {27}},\ \bibinfo {pages} {176002}
  (\bibinfo {year} {2015})}\BibitemShut {NoStop}%
\bibitem [{\citenamefont {Purbawati}\ \emph {et~al.}(2020)\citenamefont
  {Purbawati}, \citenamefont {Coraux}, \citenamefont {Vogel}, \citenamefont
  {Hadj-Azzem}, \citenamefont {Wu}, \citenamefont {Bendiab}, \citenamefont
  {Jegouso}, \citenamefont {Renard}, \citenamefont {Marty}, \citenamefont
  {Bouchiat}, \citenamefont {Sulpice}, \citenamefont {Aballe}, \citenamefont
  {Foerster}, \citenamefont {Genuzio}, \citenamefont {Locatelli}, \citenamefont
  {Menteş}, \citenamefont {Han}, \citenamefont {Sun}, \citenamefont
  {Núñez-Regueiro},\ and\ \citenamefont
  {Rougemaille}}]{purbawati_-plane_2020}%
  \BibitemOpen
  \bibfield  {author} {\bibinfo {author} {\bibfnamefont {A.}~\bibnamefont
  {Purbawati}}, \bibinfo {author} {\bibfnamefont {J.}~\bibnamefont {Coraux}},
  \bibinfo {author} {\bibfnamefont {J.}~\bibnamefont {Vogel}}, \bibinfo
  {author} {\bibfnamefont {A.}~\bibnamefont {Hadj-Azzem}}, \bibinfo {author}
  {\bibfnamefont {N.}~\bibnamefont {Wu}}, \bibinfo {author} {\bibfnamefont
  {N.}~\bibnamefont {Bendiab}}, \bibinfo {author} {\bibfnamefont
  {D.}~\bibnamefont {Jegouso}}, \bibinfo {author} {\bibfnamefont
  {J.}~\bibnamefont {Renard}}, \bibinfo {author} {\bibfnamefont
  {L.}~\bibnamefont {Marty}}, \bibinfo {author} {\bibfnamefont
  {V.}~\bibnamefont {Bouchiat}}, \bibinfo {author} {\bibfnamefont
  {A.}~\bibnamefont {Sulpice}}, \bibinfo {author} {\bibfnamefont
  {L.}~\bibnamefont {Aballe}}, \bibinfo {author} {\bibfnamefont
  {M.}~\bibnamefont {Foerster}}, \bibinfo {author} {\bibfnamefont
  {F.}~\bibnamefont {Genuzio}}, \bibinfo {author} {\bibfnamefont
  {A.}~\bibnamefont {Locatelli}}, \bibinfo {author} {\bibfnamefont {T.~O.}\
  \bibnamefont {Menteş}}, \bibinfo {author} {\bibfnamefont {Z.~V.}\
  \bibnamefont {Han}}, \bibinfo {author} {\bibfnamefont {X.}~\bibnamefont
  {Sun}}, \bibinfo {author} {\bibfnamefont {M.}~\bibnamefont
  {Núñez-Regueiro}},\ and\ \bibinfo {author} {\bibfnamefont {N.}~\bibnamefont
  {Rougemaille}},\ }\bibfield  {title} {\bibinfo {title} {In-{Plane} {Magnetic}
  {Domains} and {Néel}-like {Domain} {Walls} in {Thin} {Flakes} of the {Room}
  {Temperature} {CrTe2} {Van} der {Waals} {Ferromagnet}},\ }\href@noop {}
  {\bibfield  {journal} {\bibinfo  {journal} {ACS Applied Materials \&
  Interfaces}\ }\textbf {\bibinfo {volume} {12}},\ \bibinfo {pages} {30702}
  (\bibinfo {year} {2020})}\BibitemShut {NoStop}%
\bibitem [{\citenamefont {Zhang}\ \emph {et~al.}(2021)\citenamefont {Zhang},
  \citenamefont {Lu}, \citenamefont {Liu}, \citenamefont {Niu}, \citenamefont
  {Sun}, \citenamefont {Cook}, \citenamefont {Vaninger}, \citenamefont
  {Miceli}, \citenamefont {Singh}, \citenamefont {Lian}, \citenamefont {Chang},
  \citenamefont {He}, \citenamefont {Du}, \citenamefont {He}, \citenamefont
  {Zhang}, \citenamefont {Bian},\ and\ \citenamefont
  {Xu}}]{zhang_room-temperature_2021}%
  \BibitemOpen
  \bibfield  {author} {\bibinfo {author} {\bibfnamefont {X.}~\bibnamefont
  {Zhang}}, \bibinfo {author} {\bibfnamefont {Q.}~\bibnamefont {Lu}}, \bibinfo
  {author} {\bibfnamefont {W.}~\bibnamefont {Liu}}, \bibinfo {author}
  {\bibfnamefont {W.}~\bibnamefont {Niu}}, \bibinfo {author} {\bibfnamefont
  {J.}~\bibnamefont {Sun}}, \bibinfo {author} {\bibfnamefont {J.}~\bibnamefont
  {Cook}}, \bibinfo {author} {\bibfnamefont {M.}~\bibnamefont {Vaninger}},
  \bibinfo {author} {\bibfnamefont {P.~F.}\ \bibnamefont {Miceli}}, \bibinfo
  {author} {\bibfnamefont {D.~J.}\ \bibnamefont {Singh}}, \bibinfo {author}
  {\bibfnamefont {S.-W.}\ \bibnamefont {Lian}}, \bibinfo {author}
  {\bibfnamefont {T.-R.}\ \bibnamefont {Chang}}, \bibinfo {author}
  {\bibfnamefont {X.}~\bibnamefont {He}}, \bibinfo {author} {\bibfnamefont
  {J.}~\bibnamefont {Du}}, \bibinfo {author} {\bibfnamefont {L.}~\bibnamefont
  {He}}, \bibinfo {author} {\bibfnamefont {R.}~\bibnamefont {Zhang}}, \bibinfo
  {author} {\bibfnamefont {G.}~\bibnamefont {Bian}},\ and\ \bibinfo {author}
  {\bibfnamefont {Y.}~\bibnamefont {Xu}},\ }\bibfield  {title} {\bibinfo
  {title} {Room-temperature intrinsic ferromagnetism in epitaxial {CrTe2}
  ultrathin films},\ }\href@noop {} {\bibfield  {journal} {\bibinfo  {journal}
  {Nature Communications}\ }\textbf {\bibinfo {volume} {12}},\ \bibinfo {pages}
  {2492} (\bibinfo {year} {2021})}\BibitemShut {NoStop}%
\bibitem [{\citenamefont {Fujisawa}\ \emph {et~al.}(2020)\citenamefont
  {Fujisawa}, \citenamefont {Pardo-Almanza}, \citenamefont {Garland},
  \citenamefont {Yamagami}, \citenamefont {Zhu}, \citenamefont {Chen},
  \citenamefont {Araki}, \citenamefont {Takeda}, \citenamefont {Kobayashi},
  \citenamefont {Takeda}, \citenamefont {Hsu}, \citenamefont {Chuang},
  \citenamefont {Laskowski}, \citenamefont {Khoo}, \citenamefont
  {Soumyanarayanan},\ and\ \citenamefont {Okada}}]{fujisawa_tailoring_2020}%
  \BibitemOpen
  \bibfield  {author} {\bibinfo {author} {\bibfnamefont {Y.}~\bibnamefont
  {Fujisawa}}, \bibinfo {author} {\bibfnamefont {M.}~\bibnamefont
  {Pardo-Almanza}}, \bibinfo {author} {\bibfnamefont {J.}~\bibnamefont
  {Garland}}, \bibinfo {author} {\bibfnamefont {K.}~\bibnamefont {Yamagami}},
  \bibinfo {author} {\bibfnamefont {X.}~\bibnamefont {Zhu}}, \bibinfo {author}
  {\bibfnamefont {X.}~\bibnamefont {Chen}}, \bibinfo {author} {\bibfnamefont
  {K.}~\bibnamefont {Araki}}, \bibinfo {author} {\bibfnamefont
  {T.}~\bibnamefont {Takeda}}, \bibinfo {author} {\bibfnamefont
  {M.}~\bibnamefont {Kobayashi}}, \bibinfo {author} {\bibfnamefont
  {Y.}~\bibnamefont {Takeda}}, \bibinfo {author} {\bibfnamefont {C.~H.}\
  \bibnamefont {Hsu}}, \bibinfo {author} {\bibfnamefont {F.~C.}\ \bibnamefont
  {Chuang}}, \bibinfo {author} {\bibfnamefont {R.}~\bibnamefont {Laskowski}},
  \bibinfo {author} {\bibfnamefont {K.~H.}\ \bibnamefont {Khoo}}, \bibinfo
  {author} {\bibfnamefont {A.}~\bibnamefont {Soumyanarayanan}},\ and\ \bibinfo
  {author} {\bibfnamefont {Y.}~\bibnamefont {Okada}},\ }\bibfield  {title}
  {\bibinfo {title} {Tailoring magnetism in self-intercalated {Cr1}+{xTe} 2
  epitaxial films},\ }\href@noop {} {\bibfield  {journal} {\bibinfo  {journal}
  {Physical Review Materials}\ }\textbf {\bibinfo {volume} {4}},\ \bibinfo
  {pages} {114001} (\bibinfo {year} {2020})}\BibitemShut {NoStop}%
\bibitem [{\citenamefont {Dijkstrat}\ \emph {et~al.}(1989)\citenamefont
  {Dijkstrat}, \citenamefont {Weitering’i}, \citenamefont {van Bruggen},
  \citenamefont {Haast},\ and\ \citenamefont
  {de~Groot}}]{dijkstrat_band-structurecalculations_1989}%
  \BibitemOpen
  \bibfield  {author} {\bibinfo {author} {\bibfnamefont {J.}~\bibnamefont
  {Dijkstrat}}, \bibinfo {author} {\bibfnamefont {H.~H.}\ \bibnamefont
  {Weitering’i}}, \bibinfo {author} {\bibfnamefont {C.~F.}\ \bibnamefont {van
  Bruggen}}, \bibinfo {author} {\bibfnamefont {C.}~\bibnamefont {Haast}},\ and\
  \bibinfo {author} {\bibnamefont {de~Groot}},\ }\bibfield  {title} {\bibinfo
  {title} {Band-structurecalculations, and magnetic and transport properties of
  ferromagnetic chromium tellurides ({CrTe}, {Cr3Te4},{Cr},{Te},)},\
  }\href@noop {} {\bibfield  {journal} {\bibinfo  {journal} {Journal of
  Physics: Condensed Matter}\ }\textbf {\bibinfo {volume} {1}},\ \bibinfo
  {pages} {9141} (\bibinfo {year} {1989})}\BibitemShut {NoStop}%
\bibitem [{\citenamefont {Wen}\ \emph {et~al.}(2020)\citenamefont {Wen},
  \citenamefont {Liu}, \citenamefont {Zhang}, \citenamefont {Xia},
  \citenamefont {Zhai}, \citenamefont {Zhang}, \citenamefont {Zhai},
  \citenamefont {Shen}, \citenamefont {He}, \citenamefont {Cheng},
  \citenamefont {Yin}, \citenamefont {Yao}, \citenamefont {Getaye~Sendeku},
  \citenamefont {Wang}, \citenamefont {Ye}, \citenamefont {Liu}, \citenamefont
  {Jiang}, \citenamefont {Shan}, \citenamefont {Long},\ and\ \citenamefont
  {He}}]{wen_tunable_2020}%
  \BibitemOpen
  \bibfield  {author} {\bibinfo {author} {\bibfnamefont {Y.}~\bibnamefont
  {Wen}}, \bibinfo {author} {\bibfnamefont {Z.}~\bibnamefont {Liu}}, \bibinfo
  {author} {\bibfnamefont {Y.}~\bibnamefont {Zhang}}, \bibinfo {author}
  {\bibfnamefont {C.}~\bibnamefont {Xia}}, \bibinfo {author} {\bibfnamefont
  {B.}~\bibnamefont {Zhai}}, \bibinfo {author} {\bibfnamefont {X.}~\bibnamefont
  {Zhang}}, \bibinfo {author} {\bibfnamefont {G.}~\bibnamefont {Zhai}},
  \bibinfo {author} {\bibfnamefont {C.}~\bibnamefont {Shen}}, \bibinfo {author}
  {\bibfnamefont {P.}~\bibnamefont {He}}, \bibinfo {author} {\bibfnamefont
  {R.}~\bibnamefont {Cheng}}, \bibinfo {author} {\bibfnamefont
  {L.}~\bibnamefont {Yin}}, \bibinfo {author} {\bibfnamefont {Y.}~\bibnamefont
  {Yao}}, \bibinfo {author} {\bibfnamefont {M.}~\bibnamefont {Getaye~Sendeku}},
  \bibinfo {author} {\bibfnamefont {Z.}~\bibnamefont {Wang}}, \bibinfo {author}
  {\bibfnamefont {X.}~\bibnamefont {Ye}}, \bibinfo {author} {\bibfnamefont
  {C.}~\bibnamefont {Liu}}, \bibinfo {author} {\bibfnamefont {C.}~\bibnamefont
  {Jiang}}, \bibinfo {author} {\bibfnamefont {C.}~\bibnamefont {Shan}},
  \bibinfo {author} {\bibfnamefont {Y.}~\bibnamefont {Long}},\ and\ \bibinfo
  {author} {\bibfnamefont {J.}~\bibnamefont {He}},\ }\bibfield  {title}
  {\bibinfo {title} {Tunable {Room}-{Temperature} {Ferromagnetism} in
  {Two}-{Dimensional} {Cr} $_{\textrm{2}}$ {Te} $_{\textrm{3}}$},\ }\href@noop
  {} {\bibfield  {journal} {\bibinfo  {journal} {Nano Letters}\ }\textbf
  {\bibinfo {volume} {20}},\ \bibinfo {pages} {3130} (\bibinfo {year}
  {2020})}\BibitemShut {NoStop}%
\bibitem [{\citenamefont {Li}\ \emph {et~al.}(2019)\citenamefont {Li},
  \citenamefont {Wang}, \citenamefont {Chen}, \citenamefont {Yu}, \citenamefont
  {Zhou}, \citenamefont {Qiu}, \citenamefont {He}, \citenamefont {Ye},
  \citenamefont {Sou},\ and\ \citenamefont {Wang}}]{li_molecular_2019}%
  \BibitemOpen
  \bibfield  {author} {\bibinfo {author} {\bibfnamefont {H.}~\bibnamefont
  {Li}}, \bibinfo {author} {\bibfnamefont {L.}~\bibnamefont {Wang}}, \bibinfo
  {author} {\bibfnamefont {J.}~\bibnamefont {Chen}}, \bibinfo {author}
  {\bibfnamefont {T.}~\bibnamefont {Yu}}, \bibinfo {author} {\bibfnamefont
  {L.}~\bibnamefont {Zhou}}, \bibinfo {author} {\bibfnamefont {Y.}~\bibnamefont
  {Qiu}}, \bibinfo {author} {\bibfnamefont {H.}~\bibnamefont {He}}, \bibinfo
  {author} {\bibfnamefont {F.}~\bibnamefont {Ye}}, \bibinfo {author}
  {\bibfnamefont {I.~K.}\ \bibnamefont {Sou}},\ and\ \bibinfo {author}
  {\bibfnamefont {G.}~\bibnamefont {Wang}},\ }\bibfield  {title} {\bibinfo
  {title} {Molecular {Beam} {Epitaxy} {Grown} {Cr} $_{\textrm{2}}$ {Te}
  $_{\textrm{3}}$ {Thin} {Films} with {Tunable} {Curie} {Temperatures} for
  {Spintronic} {Devices}},\ }\href@noop {} {\bibfield  {journal} {\bibinfo
  {journal} {ACS Applied Nano Materials}\ }\textbf {\bibinfo {volume} {2}},\
  \bibinfo {pages} {6809} (\bibinfo {year} {2019})}\BibitemShut {NoStop}%
\bibitem [{\citenamefont {Zhou}\ \emph {et~al.}(2022)\citenamefont {Zhou},
  \citenamefont {Song}, \citenamefont {Chai}, \citenamefont {Wong},
  \citenamefont {Xu}, \citenamefont {Jiang}, \citenamefont {Feng},
  \citenamefont {Yang},\ and\ \citenamefont {Wang}}]{zhou_structure_2022}%
  \BibitemOpen
  \bibfield  {author} {\bibinfo {author} {\bibfnamefont {J.}~\bibnamefont
  {Zhou}}, \bibinfo {author} {\bibfnamefont {X.}~\bibnamefont {Song}}, \bibinfo
  {author} {\bibfnamefont {J.}~\bibnamefont {Chai}}, \bibinfo {author}
  {\bibfnamefont {N.~L.~M.}\ \bibnamefont {Wong}}, \bibinfo {author}
  {\bibfnamefont {X.}~\bibnamefont {Xu}}, \bibinfo {author} {\bibfnamefont
  {Y.}~\bibnamefont {Jiang}}, \bibinfo {author} {\bibfnamefont {Y.~P.}\
  \bibnamefont {Feng}}, \bibinfo {author} {\bibfnamefont {M.}~\bibnamefont
  {Yang}},\ and\ \bibinfo {author} {\bibfnamefont {S.}~\bibnamefont {Wang}},\
  }\bibfield  {title} {\bibinfo {title} {Structure dependent and strain tunable
  magnetic ordering in ultrathin chromium telluride},\ }\href@noop {}
  {\bibfield  {journal} {\bibinfo  {journal} {Journal of Alloys and Compounds}\
  }\textbf {\bibinfo {volume} {893}},\ \bibinfo {pages} {162223} (\bibinfo
  {year} {2022})}\BibitemShut {NoStop}%
\bibitem [{\citenamefont {Li}\ \emph {et~al.}(2021)\citenamefont {Li},
  \citenamefont {Li}, \citenamefont {Wu}, \citenamefont {Ding}, \citenamefont
  {Cao}, \citenamefont {Huang}, \citenamefont {Pan}, \citenamefont {Li},
  \citenamefont {Chen},\ and\ \citenamefont {Duan}}]{li_magnetic_2021}%
  \BibitemOpen
  \bibfield  {author} {\bibinfo {author} {\bibfnamefont {Q.-Q.}\ \bibnamefont
  {Li}}, \bibinfo {author} {\bibfnamefont {S.}~\bibnamefont {Li}}, \bibinfo
  {author} {\bibfnamefont {D.}~\bibnamefont {Wu}}, \bibinfo {author}
  {\bibfnamefont {Z.-K.}\ \bibnamefont {Ding}}, \bibinfo {author}
  {\bibfnamefont {X.-H.}\ \bibnamefont {Cao}}, \bibinfo {author} {\bibfnamefont
  {L.}~\bibnamefont {Huang}}, \bibinfo {author} {\bibfnamefont
  {H.}~\bibnamefont {Pan}}, \bibinfo {author} {\bibfnamefont {B.}~\bibnamefont
  {Li}}, \bibinfo {author} {\bibfnamefont {K.-Q.}\ \bibnamefont {Chen}},\ and\
  \bibinfo {author} {\bibfnamefont {X.-D.}\ \bibnamefont {Duan}},\ }\bibfield
  {title} {\bibinfo {title} {Magnetic properties manipulation of {CrTe2}
  bilayer through strain and self-intercalation},\ }\href@noop {} {\bibfield
  {journal} {\bibinfo  {journal} {Applied Physics Letters}\ }\textbf {\bibinfo
  {volume} {119}},\ \bibinfo {pages} {162402} (\bibinfo {year}
  {2021})}\BibitemShut {NoStop}%
\bibitem [{\citenamefont {Coughlin}\ \emph {et~al.}(2021)\citenamefont
  {Coughlin}, \citenamefont {Xie}, \citenamefont {Zhan}, \citenamefont {Yao},
  \citenamefont {Deng}, \citenamefont {Hewa-Walpitage}, \citenamefont {Bontke},
  \citenamefont {Chu}, \citenamefont {Li}, \citenamefont {Wang}, \citenamefont
  {Fertig},\ and\ \citenamefont {Zhang}}]{coughlin_van_2021}%
  \BibitemOpen
  \bibfield  {author} {\bibinfo {author} {\bibfnamefont {A.~L.}\ \bibnamefont
  {Coughlin}}, \bibinfo {author} {\bibfnamefont {D.}~\bibnamefont {Xie}},
  \bibinfo {author} {\bibfnamefont {X.}~\bibnamefont {Zhan}}, \bibinfo {author}
  {\bibfnamefont {Y.}~\bibnamefont {Yao}}, \bibinfo {author} {\bibfnamefont
  {L.}~\bibnamefont {Deng}}, \bibinfo {author} {\bibfnamefont {H.}~\bibnamefont
  {Hewa-Walpitage}}, \bibinfo {author} {\bibfnamefont {T.}~\bibnamefont
  {Bontke}}, \bibinfo {author} {\bibfnamefont {C.-W.}\ \bibnamefont {Chu}},
  \bibinfo {author} {\bibfnamefont {Y.}~\bibnamefont {Li}}, \bibinfo {author}
  {\bibfnamefont {J.}~\bibnamefont {Wang}}, \bibinfo {author} {\bibfnamefont
  {H.~A.}\ \bibnamefont {Fertig}},\ and\ \bibinfo {author} {\bibfnamefont
  {S.}~\bibnamefont {Zhang}},\ }\bibfield  {title} {\bibinfo {title} {Van der
  {Waals} {Superstructure} and {Twisting} in {Self}-{Intercalated} {Magnet}
  with {Near} {Room}-{Temperature} {Perpendicular} {Ferromagnetism}},\
  }\href@noop {} {\bibfield  {journal} {\bibinfo  {journal} {Nano Letters}\
  }\textbf {\bibinfo {volume} {21}},\ \bibinfo {pages} {9517} (\bibinfo {year}
  {2021})}\BibitemShut {NoStop}%
\bibitem [{\citenamefont {Chen}\ \emph {et~al.}(2019)\citenamefont {Chen},
  \citenamefont {Wang}, \citenamefont {Zhang}, \citenamefont {Zhou},
  \citenamefont {Zhang}, \citenamefont {Jin}, \citenamefont {Wang},
  \citenamefont {Qin}, \citenamefont {Qiu}, \citenamefont {Mei}, \citenamefont
  {Ye}, \citenamefont {Xi}, \citenamefont {He}, \citenamefont {Li},\ and\
  \citenamefont {Wang}}]{chen_evidence_2019}%
  \BibitemOpen
  \bibfield  {author} {\bibinfo {author} {\bibfnamefont {J.}~\bibnamefont
  {Chen}}, \bibinfo {author} {\bibfnamefont {L.}~\bibnamefont {Wang}}, \bibinfo
  {author} {\bibfnamefont {M.}~\bibnamefont {Zhang}}, \bibinfo {author}
  {\bibfnamefont {L.}~\bibnamefont {Zhou}}, \bibinfo {author} {\bibfnamefont
  {R.}~\bibnamefont {Zhang}}, \bibinfo {author} {\bibfnamefont
  {L.}~\bibnamefont {Jin}}, \bibinfo {author} {\bibfnamefont {X.}~\bibnamefont
  {Wang}}, \bibinfo {author} {\bibfnamefont {H.}~\bibnamefont {Qin}}, \bibinfo
  {author} {\bibfnamefont {Y.}~\bibnamefont {Qiu}}, \bibinfo {author}
  {\bibfnamefont {J.}~\bibnamefont {Mei}}, \bibinfo {author} {\bibfnamefont
  {F.}~\bibnamefont {Ye}}, \bibinfo {author} {\bibfnamefont {B.}~\bibnamefont
  {Xi}}, \bibinfo {author} {\bibfnamefont {H.}~\bibnamefont {He}}, \bibinfo
  {author} {\bibfnamefont {B.}~\bibnamefont {Li}},\ and\ \bibinfo {author}
  {\bibfnamefont {G.}~\bibnamefont {Wang}},\ }\bibfield  {title} {\bibinfo
  {title} {Evidence for {Magnetic} {Skyrmions} at the {Interface} of
  {Ferromagnet}/{Topological}-{Insulator} {Heterostructures}},\ }\href@noop {}
  {\bibfield  {journal} {\bibinfo  {journal} {Nano Letters}\ }\textbf {\bibinfo
  {volume} {19}},\ \bibinfo {pages} {6144} (\bibinfo {year}
  {2019})}\BibitemShut {NoStop}%
\bibitem [{\citenamefont {Chen}\ \emph {et~al.}(2022)\citenamefont {Chen},
  \citenamefont {Zhou}, \citenamefont {Wang}, \citenamefont {Yan},
  \citenamefont {Deng}, \citenamefont {Zhou}, \citenamefont {Mei},
  \citenamefont {Qiu}, \citenamefont {Xi}, \citenamefont {Wang}, \citenamefont
  {He},\ and\ \citenamefont {Wang}}]{chen_conformal_2022}%
  \BibitemOpen
  \bibfield  {author} {\bibinfo {author} {\bibfnamefont {J.}~\bibnamefont
  {Chen}}, \bibinfo {author} {\bibfnamefont {L.}~\bibnamefont {Zhou}}, \bibinfo
  {author} {\bibfnamefont {L.}~\bibnamefont {Wang}}, \bibinfo {author}
  {\bibfnamefont {Z.}~\bibnamefont {Yan}}, \bibinfo {author} {\bibfnamefont
  {X.}~\bibnamefont {Deng}}, \bibinfo {author} {\bibfnamefont {J.}~\bibnamefont
  {Zhou}}, \bibinfo {author} {\bibfnamefont {J.-w.}\ \bibnamefont {Mei}},
  \bibinfo {author} {\bibfnamefont {Y.}~\bibnamefont {Qiu}}, \bibinfo {author}
  {\bibfnamefont {B.}~\bibnamefont {Xi}}, \bibinfo {author} {\bibfnamefont
  {X.}~\bibnamefont {Wang}}, \bibinfo {author} {\bibfnamefont {H.}~\bibnamefont
  {He}},\ and\ \bibinfo {author} {\bibfnamefont {G.}~\bibnamefont {Wang}},\
  }\bibfield  {title} {\bibinfo {title} {Conformal {Growth} of {Cr}
  $_{\textrm{2}}$ {Te} $_{\textrm{3}}$ on {Bi} $_{\textrm{2}}$ {Te}
  $_{\textrm{3}}$ {Nanodots} with a {Topological} {Hall} {Effect}},\
  }\href@noop {} {\bibfield  {journal} {\bibinfo  {journal} {Crystal Growth \&
  Design}\ }\textbf {\bibinfo {volume} {22}},\ \bibinfo {pages} {140} (\bibinfo
  {year} {2022})}\BibitemShut {NoStop}%
\bibitem [{\citenamefont {Jeon}\ \emph {et~al.}(2022)\citenamefont {Jeon},
  \citenamefont {Na}, \citenamefont {Kim}, \citenamefont {Lee}, \citenamefont
  {Song}, \citenamefont {Kim}, \citenamefont {Park}, \citenamefont {Kim},
  \citenamefont {Noh}, \citenamefont {Kim}, \citenamefont {Jerng},\ and\
  \citenamefont {Chun}}]{jeon_emergent_2022}%
  \BibitemOpen
  \bibfield  {author} {\bibinfo {author} {\bibfnamefont {J.~H.}\ \bibnamefont
  {Jeon}}, \bibinfo {author} {\bibfnamefont {H.~R.}\ \bibnamefont {Na}},
  \bibinfo {author} {\bibfnamefont {H.}~\bibnamefont {Kim}}, \bibinfo {author}
  {\bibfnamefont {S.}~\bibnamefont {Lee}}, \bibinfo {author} {\bibfnamefont
  {S.}~\bibnamefont {Song}}, \bibinfo {author} {\bibfnamefont {J.}~\bibnamefont
  {Kim}}, \bibinfo {author} {\bibfnamefont {S.}~\bibnamefont {Park}}, \bibinfo
  {author} {\bibfnamefont {J.}~\bibnamefont {Kim}}, \bibinfo {author}
  {\bibfnamefont {H.}~\bibnamefont {Noh}}, \bibinfo {author} {\bibfnamefont
  {G.}~\bibnamefont {Kim}}, \bibinfo {author} {\bibfnamefont {S.-K.}\
  \bibnamefont {Jerng}},\ and\ \bibinfo {author} {\bibfnamefont {S.-H.}\
  \bibnamefont {Chun}},\ }\bibfield  {title} {\bibinfo {title} {Emergent
  {Topological} {Hall} {Effect} from {Exchange} {Coupling} in {Ferromagnetic}
  {Cr2Te3}/{Noncoplanar} {Antiferromagnetic} {Cr2Se3} {Bilayers}},\ }\href@noop
  {} {\bibfield  {journal} {\bibinfo  {journal} {ACS Nano}\ ,\ \bibinfo {pages}
  {9}} (\bibinfo {year} {2022})}\BibitemShut {NoStop}%
\bibitem [{\citenamefont {Pramanik}\ \emph {et~al.}(2017)\citenamefont
  {Pramanik}, \citenamefont {Roy}, \citenamefont {Dey}, \citenamefont {Rai},
  \citenamefont {Guchhait}, \citenamefont {Movva}, \citenamefont {Hsieh},\ and\
  \citenamefont {Banerjee}}]{pramanik_angular_2017}%
  \BibitemOpen
  \bibfield  {author} {\bibinfo {author} {\bibfnamefont {T.}~\bibnamefont
  {Pramanik}}, \bibinfo {author} {\bibfnamefont {A.}~\bibnamefont {Roy}},
  \bibinfo {author} {\bibfnamefont {R.}~\bibnamefont {Dey}}, \bibinfo {author}
  {\bibfnamefont {A.}~\bibnamefont {Rai}}, \bibinfo {author} {\bibfnamefont
  {S.}~\bibnamefont {Guchhait}}, \bibinfo {author} {\bibfnamefont {H.~C.}\
  \bibnamefont {Movva}}, \bibinfo {author} {\bibfnamefont {C.-C.}\ \bibnamefont
  {Hsieh}},\ and\ \bibinfo {author} {\bibfnamefont {S.~K.}\ \bibnamefont
  {Banerjee}},\ }\bibfield  {title} {\bibinfo {title} {Angular dependence of
  magnetization reversal in epitaxial chromium telluride thin films with
  perpendicular magnetic anisotropy},\ }\href@noop {} {\bibfield  {journal}
  {\bibinfo  {journal} {Journal of Magnetism and Magnetic Materials}\ }\textbf
  {\bibinfo {volume} {437}},\ \bibinfo {pages} {72} (\bibinfo {year}
  {2017})}\BibitemShut {NoStop}%
\bibitem [{\citenamefont {Ueno}\ \emph {et~al.}(1990)\citenamefont {Ueno},
  \citenamefont {Saiki}, \citenamefont {Shimada},\ and\ \citenamefont
  {Koma}}]{ueno_epitaxial_1990}%
  \BibitemOpen
  \bibfield  {author} {\bibinfo {author} {\bibfnamefont {K.}~\bibnamefont
  {Ueno}}, \bibinfo {author} {\bibfnamefont {K.}~\bibnamefont {Saiki}},
  \bibinfo {author} {\bibfnamefont {T.}~\bibnamefont {Shimada}},\ and\ \bibinfo
  {author} {\bibfnamefont {A.}~\bibnamefont {Koma}},\ }\bibfield  {title}
  {\bibinfo {title} {Epitaxial growth of transition metal dichalcogenides on
  cleaved faces of mica},\ }\href {https://doi.org/10.1116/1.576983} {\bibfield
   {journal} {\bibinfo  {journal} {Journal of Vacuum Science \& Technology A}\
  }\textbf {\bibinfo {volume} {8}},\ \bibinfo {pages} {68} (\bibinfo {year}
  {1990})}\BibitemShut {NoStop}%
\bibitem [{\citenamefont {Ohuchi}\ \emph {et~al.}(1990)\citenamefont {Ohuchi},
  \citenamefont {Parkinson}, \citenamefont {Ueno},\ and\ \citenamefont
  {Koma}}]{ohuchi_van_1990}%
  \BibitemOpen
  \bibfield  {author} {\bibinfo {author} {\bibfnamefont {F.~S.}\ \bibnamefont
  {Ohuchi}}, \bibinfo {author} {\bibfnamefont {B.~A.}\ \bibnamefont
  {Parkinson}}, \bibinfo {author} {\bibfnamefont {K.}~\bibnamefont {Ueno}},\
  and\ \bibinfo {author} {\bibfnamefont {A.}~\bibnamefont {Koma}},\ }\bibfield
  {title} {\bibinfo {title} {van der {Waals} epitaxial growth and
  characterization of {MoSe2} thin films on {SnS2}},\ }\href
  {https://doi.org/10.1063/1.346574} {\bibfield  {journal} {\bibinfo  {journal}
  {Journal of Applied Physics}\ }\textbf {\bibinfo {volume} {68}},\ \bibinfo
  {pages} {2168} (\bibinfo {year} {1990})}\BibitemShut {NoStop}%
\bibitem [{\citenamefont {Dau}\ \emph {et~al.}(2018)\citenamefont {Dau},
  \citenamefont {Gay}, \citenamefont {Di~Felice}, \citenamefont {Vergnaud},
  \citenamefont {Marty}, \citenamefont {Beigné}, \citenamefont {Renaud},
  \citenamefont {Renault}, \citenamefont {Mallet}, \citenamefont {Le~Quang},
  \citenamefont {Veuillen}, \citenamefont {Huder}, \citenamefont {Renard},
  \citenamefont {Chapelier}, \citenamefont {Zamborlini}, \citenamefont
  {Jugovac}, \citenamefont {Feyer}, \citenamefont {Dappe}, \citenamefont
  {Pochet},\ and\ \citenamefont {Jamet}}]{dau_beyond_2018}%
  \BibitemOpen
  \bibfield  {author} {\bibinfo {author} {\bibfnamefont {M.~T.}\ \bibnamefont
  {Dau}}, \bibinfo {author} {\bibfnamefont {M.}~\bibnamefont {Gay}}, \bibinfo
  {author} {\bibfnamefont {D.}~\bibnamefont {Di~Felice}}, \bibinfo {author}
  {\bibfnamefont {C.}~\bibnamefont {Vergnaud}}, \bibinfo {author}
  {\bibfnamefont {A.}~\bibnamefont {Marty}}, \bibinfo {author} {\bibfnamefont
  {C.}~\bibnamefont {Beigné}}, \bibinfo {author} {\bibfnamefont
  {G.}~\bibnamefont {Renaud}}, \bibinfo {author} {\bibfnamefont
  {O.}~\bibnamefont {Renault}}, \bibinfo {author} {\bibfnamefont
  {P.}~\bibnamefont {Mallet}}, \bibinfo {author} {\bibfnamefont
  {T.}~\bibnamefont {Le~Quang}}, \bibinfo {author} {\bibfnamefont {J.-Y.}\
  \bibnamefont {Veuillen}}, \bibinfo {author} {\bibfnamefont {L.}~\bibnamefont
  {Huder}}, \bibinfo {author} {\bibfnamefont {V.~T.}\ \bibnamefont {Renard}},
  \bibinfo {author} {\bibfnamefont {C.}~\bibnamefont {Chapelier}}, \bibinfo
  {author} {\bibfnamefont {G.}~\bibnamefont {Zamborlini}}, \bibinfo {author}
  {\bibfnamefont {M.}~\bibnamefont {Jugovac}}, \bibinfo {author} {\bibfnamefont
  {V.}~\bibnamefont {Feyer}}, \bibinfo {author} {\bibfnamefont {Y.~J.}\
  \bibnamefont {Dappe}}, \bibinfo {author} {\bibfnamefont {P.}~\bibnamefont
  {Pochet}},\ and\ \bibinfo {author} {\bibfnamefont {M.}~\bibnamefont
  {Jamet}},\ }\bibfield  {title} {\bibinfo {title} {Beyond van der {Waals}
  {Interaction}: {The} {Case} of {MoSe2} {Epitaxially} {Grown} on {Few}-{Layer}
  {Graphene}},\ }\href {https://doi.org/10.1021/acsnano.7b07446} {\bibfield
  {journal} {\bibinfo  {journal} {ACS Nano}\ }\textbf {\bibinfo {volume}
  {12}},\ \bibinfo {pages} {2319} (\bibinfo {year} {2018})}\BibitemShut
  {NoStop}%
\bibitem [{\citenamefont {Dappe}\ \emph {et~al.}(2020)\citenamefont {Dappe},
  \citenamefont {Almadori}, \citenamefont {Dau}, \citenamefont {Vergnaud},
  \citenamefont {Jamet}, \citenamefont {Paillet}, \citenamefont {Journot},
  \citenamefont {Hyot}, \citenamefont {Pochet},\ and\ \citenamefont
  {Grévin}}]{Dappe_2020}%
  \BibitemOpen
  \bibfield  {author} {\bibinfo {author} {\bibfnamefont {Y.~J.}\ \bibnamefont
  {Dappe}}, \bibinfo {author} {\bibfnamefont {Y.}~\bibnamefont {Almadori}},
  \bibinfo {author} {\bibfnamefont {M.~T.}\ \bibnamefont {Dau}}, \bibinfo
  {author} {\bibfnamefont {C.}~\bibnamefont {Vergnaud}}, \bibinfo {author}
  {\bibfnamefont {M.}~\bibnamefont {Jamet}}, \bibinfo {author} {\bibfnamefont
  {C.}~\bibnamefont {Paillet}}, \bibinfo {author} {\bibfnamefont
  {T.}~\bibnamefont {Journot}}, \bibinfo {author} {\bibfnamefont
  {B.}~\bibnamefont {Hyot}}, \bibinfo {author} {\bibfnamefont {P.}~\bibnamefont
  {Pochet}},\ and\ \bibinfo {author} {\bibfnamefont {B.}~\bibnamefont
  {Grévin}},\ }\bibfield  {title} {\bibinfo {title} {Charge transfers and
  charged defects in {WSe2}/graphene-{SiC} interfaces},\ }\href
  {https://doi.org/10.1088/1361-6528/ab8083} {\bibfield  {journal} {\bibinfo
  {journal} {Nanotechnology}\ }\textbf {\bibinfo {volume} {31}},\ \bibinfo
  {pages} {255709} (\bibinfo {year} {2020})}\BibitemShut {NoStop}%
\bibitem [{\citenamefont {Ohresser}\ \emph {et~al.}(2014)\citenamefont
  {Ohresser}, \citenamefont {Otero}, \citenamefont {Choueikani}, \citenamefont
  {Chen}, \citenamefont {Stanescu}, \citenamefont {Deschamps}, \citenamefont
  {Moreno}, \citenamefont {Polack}, \citenamefont {Lagarde}, \citenamefont
  {Daguerre}, \citenamefont {Marteau1}, \citenamefont {Scheurer}, \citenamefont
  {Joly}, \citenamefont {Kappler}, \citenamefont {Muller}, \citenamefont
  {Bunau},\ and\ \citenamefont {Sainctavit}}]{ohresser2014deimos}%
  \BibitemOpen
  \bibfield  {author} {\bibinfo {author} {\bibfnamefont {P.}~\bibnamefont
  {Ohresser}}, \bibinfo {author} {\bibfnamefont {E.}~\bibnamefont {Otero}},
  \bibinfo {author} {\bibfnamefont {F.}~\bibnamefont {Choueikani}}, \bibinfo
  {author} {\bibfnamefont {K.}~\bibnamefont {Chen}}, \bibinfo {author}
  {\bibfnamefont {S.}~\bibnamefont {Stanescu}}, \bibinfo {author}
  {\bibfnamefont {F.}~\bibnamefont {Deschamps}}, \bibinfo {author}
  {\bibfnamefont {T.}~\bibnamefont {Moreno}}, \bibinfo {author} {\bibfnamefont
  {F.}~\bibnamefont {Polack}}, \bibinfo {author} {\bibfnamefont
  {B.}~\bibnamefont {Lagarde}}, \bibinfo {author} {\bibfnamefont {J.-P.}\
  \bibnamefont {Daguerre}}, \bibinfo {author} {\bibfnamefont {F.}~\bibnamefont
  {Marteau1}}, \bibinfo {author} {\bibfnamefont {F.}~\bibnamefont {Scheurer}},
  \bibinfo {author} {\bibfnamefont {L.}~\bibnamefont {Joly}}, \bibinfo {author}
  {\bibfnamefont {J.-P.}\ \bibnamefont {Kappler}}, \bibinfo {author}
  {\bibfnamefont {B.}~\bibnamefont {Muller}}, \bibinfo {author} {\bibfnamefont
  {O.}~\bibnamefont {Bunau}},\ and\ \bibinfo {author} {\bibfnamefont
  {P.}~\bibnamefont {Sainctavit}},\ }\bibfield  {title} {\bibinfo {title}
  {Deimos: A beamline dedicated to dichroism measurements in the 350--2500 ev
  energy range},\ }\href@noop {} {\bibfield  {journal} {\bibinfo  {journal}
  {Review of Scientific Instruments}\ }\textbf {\bibinfo {volume} {85}},\
  \bibinfo {pages} {013106} (\bibinfo {year} {2014})}\BibitemShut {NoStop}%
\bibitem [{\citenamefont {Kresse}\ and\ \citenamefont
  {Hafner}(1993)}]{kresse_ab_1993}%
  \BibitemOpen
  \bibfield  {author} {\bibinfo {author} {\bibfnamefont {G.}~\bibnamefont
  {Kresse}}\ and\ \bibinfo {author} {\bibfnamefont {J.}~\bibnamefont
  {Hafner}},\ }\bibfield  {title} {\bibinfo {title} {\textit{{Ab} initio}
  molecular dynamics for liquid metals},\ }\href
  {https://doi.org/10.1103/PhysRevB.47.558} {\bibfield  {journal} {\bibinfo
  {journal} {Physical Review B}\ }\textbf {\bibinfo {volume} {47}},\ \bibinfo
  {pages} {558} (\bibinfo {year} {1993})}\BibitemShut {NoStop}%
\bibitem [{\citenamefont {Kresse}\ and\ \citenamefont
  {Furthmüller}(1996)}]{kresse_efficiency_1996}%
  \BibitemOpen
  \bibfield  {author} {\bibinfo {author} {\bibfnamefont {G.}~\bibnamefont
  {Kresse}}\ and\ \bibinfo {author} {\bibfnamefont {J.}~\bibnamefont
  {Furthmüller}},\ }\bibfield  {title} {\bibinfo {title} {Efficiency of
  ab-initio total energy calculations for metals and semiconductors using a
  plane-wave basis set},\ }\href {https://doi.org/10.1016/0927-0256(96)00008-0}
  {\bibfield  {journal} {\bibinfo  {journal} {Computational Materials Science}\
  }\textbf {\bibinfo {volume} {6}},\ \bibinfo {pages} {15} (\bibinfo {year}
  {1996})}\BibitemShut {NoStop}%
\bibitem [{\citenamefont {Perdew}\ \emph {et~al.}(1996)\citenamefont {Perdew},
  \citenamefont {Burke},\ and\ \citenamefont
  {Ernzerhof}}]{perdew_generalized_1996-1}%
  \BibitemOpen
  \bibfield  {author} {\bibinfo {author} {\bibfnamefont {J.~P.}\ \bibnamefont
  {Perdew}}, \bibinfo {author} {\bibfnamefont {K.}~\bibnamefont {Burke}},\ and\
  \bibinfo {author} {\bibfnamefont {M.}~\bibnamefont {Ernzerhof}},\ }\bibfield
  {title} {\bibinfo {title} {Generalized {Gradient} {Approximation} {Made}
  {Simple}},\ }\href {https://doi.org/10.1103/PhysRevLett.77.3865} {\bibfield
  {journal} {\bibinfo  {journal} {Physical Review Letters}\ }\textbf {\bibinfo
  {volume} {77}},\ \bibinfo {pages} {3865} (\bibinfo {year}
  {1996})}\BibitemShut {NoStop}%
\bibitem [{\citenamefont {Dudarev}\ \emph {et~al.}(1998)\citenamefont
  {Dudarev}, \citenamefont {Botton}, \citenamefont {Savrasov}, \citenamefont
  {Humphreys},\ and\ \citenamefont
  {Sutton}}]{dudarev_electron-energy-loss_1998}%
  \BibitemOpen
  \bibfield  {author} {\bibinfo {author} {\bibfnamefont {S.~L.}\ \bibnamefont
  {Dudarev}}, \bibinfo {author} {\bibfnamefont {G.~A.}\ \bibnamefont {Botton}},
  \bibinfo {author} {\bibfnamefont {S.~Y.}\ \bibnamefont {Savrasov}}, \bibinfo
  {author} {\bibfnamefont {C.~J.}\ \bibnamefont {Humphreys}},\ and\ \bibinfo
  {author} {\bibfnamefont {A.~P.}\ \bibnamefont {Sutton}},\ }\bibfield  {title}
  {\bibinfo {title} {Electron-energy-loss spectra and the structural stability
  of nickel oxide: {An} {LSDA}+{U} study},\ }\href
  {https://doi.org/10.1103/PhysRevB.57.1505} {\bibfield  {journal} {\bibinfo
  {journal} {Physical Review B}\ }\textbf {\bibinfo {volume} {57}},\ \bibinfo
  {pages} {1505} (\bibinfo {year} {1998})}\BibitemShut {NoStop}%
\bibitem [{\citenamefont {Grimme}\ \emph {et~al.}(2010)\citenamefont {Grimme},
  \citenamefont {Antony}, \citenamefont {Ehrlich},\ and\ \citenamefont
  {Krieg}}]{grimme_consistent_2010}%
  \BibitemOpen
  \bibfield  {author} {\bibinfo {author} {\bibfnamefont {S.}~\bibnamefont
  {Grimme}}, \bibinfo {author} {\bibfnamefont {J.}~\bibnamefont {Antony}},
  \bibinfo {author} {\bibfnamefont {S.}~\bibnamefont {Ehrlich}},\ and\ \bibinfo
  {author} {\bibfnamefont {H.}~\bibnamefont {Krieg}},\ }\bibfield  {title}
  {\bibinfo {title} {A consistent and accurate \textit{ab initio}
  parametrization of density functional dispersion correction ({DFT}-{D}) for
  the 94 elements {H}-{Pu}},\ }\href {https://doi.org/10.1063/1.3382344}
  {\bibfield  {journal} {\bibinfo  {journal} {The Journal of Chemical Physics}\
  }\textbf {\bibinfo {volume} {132}},\ \bibinfo {pages} {154104} (\bibinfo
  {year} {2010})}\BibitemShut {NoStop}%
\bibitem [{\citenamefont {Grimme}\ \emph {et~al.}(2011)\citenamefont {Grimme},
  \citenamefont {Ehrlich},\ and\ \citenamefont {Goerigk}}]{grimme_effect_2011}%
  \BibitemOpen
  \bibfield  {author} {\bibinfo {author} {\bibfnamefont {S.}~\bibnamefont
  {Grimme}}, \bibinfo {author} {\bibfnamefont {S.}~\bibnamefont {Ehrlich}},\
  and\ \bibinfo {author} {\bibfnamefont {L.}~\bibnamefont {Goerigk}},\
  }\bibfield  {title} {\bibinfo {title} {Effect of the damping function in
  dispersion corrected density functional theory},\ }\href
  {https://doi.org/10.1002/jcc.21759} {\bibfield  {journal} {\bibinfo
  {journal} {Journal of Computational Chemistry}\ }\textbf {\bibinfo {volume}
  {32}},\ \bibinfo {pages} {1456} (\bibinfo {year} {2011})}\BibitemShut
  {NoStop}%
\bibitem [{\citenamefont {Choudhary}\ \emph {et~al.}(2020)\citenamefont
  {Choudhary}, \citenamefont {Garrity}, \citenamefont {Hartman}, \citenamefont
  {Pilania},\ and\ \citenamefont {Tavazza}}]{choudhary_efficient_2020}%
  \BibitemOpen
  \bibfield  {author} {\bibinfo {author} {\bibfnamefont {K.}~\bibnamefont
  {Choudhary}}, \bibinfo {author} {\bibfnamefont {K.~F.}\ \bibnamefont
  {Garrity}}, \bibinfo {author} {\bibfnamefont {S.~T.}\ \bibnamefont
  {Hartman}}, \bibinfo {author} {\bibfnamefont {G.}~\bibnamefont {Pilania}},\
  and\ \bibinfo {author} {\bibfnamefont {F.}~\bibnamefont {Tavazza}},\
  }\bibfield  {title} {\bibinfo {title} {Efficient computational design of 2d
  van der waals heterostructures: Band-alignment, lattice-mismatch, web-app
  generation and machine-learning},\ }\href@noop {} {\bibfield  {journal}
  {\bibinfo  {journal} {arXiv: Materials Science}\ } (\bibinfo {year}
  {2020})}\BibitemShut {NoStop}%
\bibitem [{\citenamefont {Hallal}\ \emph {et~al.}(2014)\citenamefont {Hallal},
  \citenamefont {Dieny},\ and\ \citenamefont
  {Chshiev}}]{hallal_impurity-induced_2014}%
  \BibitemOpen
  \bibfield  {author} {\bibinfo {author} {\bibfnamefont {A.}~\bibnamefont
  {Hallal}}, \bibinfo {author} {\bibfnamefont {B.}~\bibnamefont {Dieny}},\ and\
  \bibinfo {author} {\bibfnamefont {M.}~\bibnamefont {Chshiev}},\ }\bibfield
  {title} {\bibinfo {title} {Impurity-induced enhancement of perpendicular
  magnetic anisotropy in {Fe}/{MgO} tunnel junctions},\ }\href
  {https://doi.org/10.1103/PhysRevB.90.064422} {\bibfield  {journal} {\bibinfo
  {journal} {Physical Review B}\ }\textbf {\bibinfo {volume} {90}},\ \bibinfo
  {pages} {064422} (\bibinfo {year} {2014})}\BibitemShut {NoStop}%
\bibitem [{\citenamefont {Wang}\ \emph {et~al.}(2006)\citenamefont {Wang},
  \citenamefont {Yates}, \citenamefont {Souza},\ and\ \citenamefont
  {Vanderbilt}}]{wang_ab_2006}%
  \BibitemOpen
  \bibfield  {author} {\bibinfo {author} {\bibfnamefont {X.}~\bibnamefont
  {Wang}}, \bibinfo {author} {\bibfnamefont {J.~R.}\ \bibnamefont {Yates}},
  \bibinfo {author} {\bibfnamefont {I.}~\bibnamefont {Souza}},\ and\ \bibinfo
  {author} {\bibfnamefont {D.}~\bibnamefont {Vanderbilt}},\ }\bibfield  {title}
  {\bibinfo {title} {Ab initio calculation of the anomalous {Hall} conductivity
  by {Wannier} interpolation},\ }\href
  {https://link.aps.org/doi/10.1103/PhysRevB.74.195118} {\bibfield  {journal}
  {\bibinfo  {journal} {Physical Review B}\ }\textbf {\bibinfo {volume} {74}},\
  \bibinfo {pages} {195118} (\bibinfo {year} {2006})}\BibitemShut {NoStop}%
\bibitem [{\citenamefont {Mostofi}\ \emph {et~al.}(2014)\citenamefont
  {Mostofi}, \citenamefont {Yates}, \citenamefont {Pizzi}, \citenamefont {Lee},
  \citenamefont {Souza}, \citenamefont {Vanderbilt},\ and\ \citenamefont
  {Marzari}}]{mostofi_updated_2014}%
  \BibitemOpen
  \bibfield  {author} {\bibinfo {author} {\bibfnamefont {A.~A.}\ \bibnamefont
  {Mostofi}}, \bibinfo {author} {\bibfnamefont {J.~R.}\ \bibnamefont {Yates}},
  \bibinfo {author} {\bibfnamefont {G.}~\bibnamefont {Pizzi}}, \bibinfo
  {author} {\bibfnamefont {Y.-S.}\ \bibnamefont {Lee}}, \bibinfo {author}
  {\bibfnamefont {I.}~\bibnamefont {Souza}}, \bibinfo {author} {\bibfnamefont
  {D.}~\bibnamefont {Vanderbilt}},\ and\ \bibinfo {author} {\bibfnamefont
  {N.}~\bibnamefont {Marzari}},\ }\bibfield  {title} {\bibinfo {title} {An
  updated version of wannier90: {A} tool for obtaining maximally-localised
  {Wannier} functions},\ }\href@noop {} {\bibfield  {journal} {\bibinfo
  {journal} {Computer Physics Communications}\ }\textbf {\bibinfo {volume}
  {185}},\ \bibinfo {pages} {2309} (\bibinfo {year} {2014})}\BibitemShut
  {NoStop}%
\bibitem [{\citenamefont {Tsirkin}(2021)}]{tsirkin_high_2021-1}%
  \BibitemOpen
  \bibfield  {author} {\bibinfo {author} {\bibfnamefont {S.~S.}\ \bibnamefont
  {Tsirkin}},\ }\bibfield  {title} {\bibinfo {title} {High performance
  {Wannier} interpolation of {Berry} curvature and related quantities with
  {WannierBerri} code},\ }\href {https://doi.org/10.1038/s41524-021-00498-5}
  {\bibfield  {journal} {\bibinfo  {journal} {npj Computational Materials}\
  }\textbf {\bibinfo {volume} {7}},\ \bibinfo {pages} {33} (\bibinfo {year}
  {2021})}\BibitemShut {NoStop}%
\bibitem [{\citenamefont {Destraz}\ \emph {et~al.}(2020)\citenamefont
  {Destraz}, \citenamefont {Das}, \citenamefont {Tsirkin}, \citenamefont {Xu},
  \citenamefont {Neupert}, \citenamefont {Chang}, \citenamefont {Schilling},
  \citenamefont {Grushin}, \citenamefont {Kohlbrecher}, \citenamefont {Keller},
  \citenamefont {Puphal}, \citenamefont {Pomjakushina},\ and\ \citenamefont
  {White}}]{destraz_magnetism_2020}%
  \BibitemOpen
  \bibfield  {author} {\bibinfo {author} {\bibfnamefont {D.}~\bibnamefont
  {Destraz}}, \bibinfo {author} {\bibfnamefont {L.}~\bibnamefont {Das}},
  \bibinfo {author} {\bibfnamefont {S.~S.}\ \bibnamefont {Tsirkin}}, \bibinfo
  {author} {\bibfnamefont {Y.}~\bibnamefont {Xu}}, \bibinfo {author}
  {\bibfnamefont {T.}~\bibnamefont {Neupert}}, \bibinfo {author} {\bibfnamefont
  {J.}~\bibnamefont {Chang}}, \bibinfo {author} {\bibfnamefont
  {A.}~\bibnamefont {Schilling}}, \bibinfo {author} {\bibfnamefont {A.~G.}\
  \bibnamefont {Grushin}}, \bibinfo {author} {\bibfnamefont {J.}~\bibnamefont
  {Kohlbrecher}}, \bibinfo {author} {\bibfnamefont {L.}~\bibnamefont {Keller}},
  \bibinfo {author} {\bibfnamefont {P.}~\bibnamefont {Puphal}}, \bibinfo
  {author} {\bibfnamefont {E.}~\bibnamefont {Pomjakushina}},\ and\ \bibinfo
  {author} {\bibfnamefont {J.~S.}\ \bibnamefont {White}},\ }\bibfield  {title}
  {\bibinfo {title} {Magnetism and anomalous transport in the {Weyl} semimetal
  {PrAlGe}: possible route to axial gauge fields},\ }\href
  {https://doi.org/10.1038/s41535-019-0207-7} {\bibfield  {journal} {\bibinfo
  {journal} {npj Quantum Materials}\ }\textbf {\bibinfo {volume} {5}},\
  \bibinfo {pages} {5} (\bibinfo {year} {2020})}\BibitemShut {NoStop}%
\bibitem [{\citenamefont {Kumar}\ \emph {et~al.}(2016)\citenamefont {Kumar},
  \citenamefont {Baraket}, \citenamefont {Paillet}, \citenamefont {Huntzinger},
  \citenamefont {Tiberj}, \citenamefont {Jansen}, \citenamefont {Vila},
  \citenamefont {Cubuku}, \citenamefont {Vergnaud}, \citenamefont {Jamet},
  \citenamefont {Lapertot}, \citenamefont {Rouchon}, \citenamefont {Zahab},
  \citenamefont {Sauvajol}, \citenamefont {Dubois}, \citenamefont {Lefloch},\
  and\ \citenamefont {Duclairoir}}]{kumar_growth_2016}%
  \BibitemOpen
  \bibfield  {author} {\bibinfo {author} {\bibfnamefont {B.}~\bibnamefont
  {Kumar}}, \bibinfo {author} {\bibfnamefont {M.}~\bibnamefont {Baraket}},
  \bibinfo {author} {\bibfnamefont {M.}~\bibnamefont {Paillet}}, \bibinfo
  {author} {\bibfnamefont {J.-R.}\ \bibnamefont {Huntzinger}}, \bibinfo
  {author} {\bibfnamefont {A.}~\bibnamefont {Tiberj}}, \bibinfo {author}
  {\bibfnamefont {A.}~\bibnamefont {Jansen}}, \bibinfo {author} {\bibfnamefont
  {L.}~\bibnamefont {Vila}}, \bibinfo {author} {\bibfnamefont {M.}~\bibnamefont
  {Cubuku}}, \bibinfo {author} {\bibfnamefont {C.}~\bibnamefont {Vergnaud}},
  \bibinfo {author} {\bibfnamefont {M.}~\bibnamefont {Jamet}}, \bibinfo
  {author} {\bibfnamefont {G.}~\bibnamefont {Lapertot}}, \bibinfo {author}
  {\bibfnamefont {D.}~\bibnamefont {Rouchon}}, \bibinfo {author} {\bibfnamefont
  {A.-A.}\ \bibnamefont {Zahab}}, \bibinfo {author} {\bibfnamefont {J.-L.}\
  \bibnamefont {Sauvajol}}, \bibinfo {author} {\bibfnamefont {L.}~\bibnamefont
  {Dubois}}, \bibinfo {author} {\bibfnamefont {F.}~\bibnamefont {Lefloch}},\
  and\ \bibinfo {author} {\bibfnamefont {F.}~\bibnamefont {Duclairoir}},\
  }\bibfield  {title} {\bibinfo {title} {Growth protocols and characterization
  of epitaxial graphene on {SiC} elaborated in a graphite enclosure},\ }\href
  {https://doi.org/https://doi.org/10.1016/j.physe.2015.07.022} {\bibfield
  {journal} {\bibinfo  {journal} {Physica E: Low-dimensional Systems and
  Nanostructures}\ }\textbf {\bibinfo {volume} {75}},\ \bibinfo {pages} {7}
  (\bibinfo {year} {2016})}\BibitemShut {NoStop}%
\bibitem [{\citenamefont {Pierucci}\ \emph {et~al.}(2022)\citenamefont
  {Pierucci}, \citenamefont {Mahmoudi}, \citenamefont {Silly}, \citenamefont
  {Bisti}, \citenamefont {Oehler}, \citenamefont {Patriarche}, \citenamefont
  {Bonell}, \citenamefont {Marty}, \citenamefont {Vergnaud}, \citenamefont
  {Jamet}, \citenamefont {Boukari}, \citenamefont {Lhuillier}, \citenamefont
  {Pala},\ and\ \citenamefont {Ouerghi}}]{pierucci_evidence_2022}%
  \BibitemOpen
  \bibfield  {author} {\bibinfo {author} {\bibfnamefont {D.}~\bibnamefont
  {Pierucci}}, \bibinfo {author} {\bibfnamefont {A.}~\bibnamefont {Mahmoudi}},
  \bibinfo {author} {\bibfnamefont {M.}~\bibnamefont {Silly}}, \bibinfo
  {author} {\bibfnamefont {F.}~\bibnamefont {Bisti}}, \bibinfo {author}
  {\bibfnamefont {F.}~\bibnamefont {Oehler}}, \bibinfo {author} {\bibfnamefont
  {G.}~\bibnamefont {Patriarche}}, \bibinfo {author} {\bibfnamefont
  {F.}~\bibnamefont {Bonell}}, \bibinfo {author} {\bibfnamefont
  {A.}~\bibnamefont {Marty}}, \bibinfo {author} {\bibfnamefont
  {C.}~\bibnamefont {Vergnaud}}, \bibinfo {author} {\bibfnamefont
  {M.}~\bibnamefont {Jamet}}, \bibinfo {author} {\bibfnamefont
  {H.}~\bibnamefont {Boukari}}, \bibinfo {author} {\bibfnamefont
  {E.}~\bibnamefont {Lhuillier}}, \bibinfo {author} {\bibfnamefont
  {M.}~\bibnamefont {Pala}},\ and\ \bibinfo {author} {\bibfnamefont
  {A.}~\bibnamefont {Ouerghi}},\ }\bibfield  {title} {\bibinfo {title}
  {Evidence for highly p-type doping and type {II} band alignment in large
  scale monolayer {WSe2}/{Se}-terminated {GaAs} heterojunction grown by
  molecular beam epitaxy},\ }\href {https://doi.org/10.1039/D2NR00458E}
  {\bibfield  {journal} {\bibinfo  {journal} {Nanoscale}\ }\textbf {\bibinfo
  {volume} {14}},\ \bibinfo {pages} {5859} (\bibinfo {year}
  {2022})}\BibitemShut {NoStop}%
\bibitem [{\citenamefont {Andresen}\ \emph {et~al.}(1970)\citenamefont
  {Andresen}, \citenamefont {Zeppezauer}, \citenamefont {Boive}, \citenamefont
  {Nordstr{\"o}m},\ and\ \citenamefont
  {Br{\"a}nd{\'e}n}}]{andresen1970magnetic}%
  \BibitemOpen
  \bibfield  {author} {\bibinfo {author} {\bibfnamefont {A.~F.}\ \bibnamefont
  {Andresen}}, \bibinfo {author} {\bibfnamefont {E.}~\bibnamefont
  {Zeppezauer}}, \bibinfo {author} {\bibfnamefont {T.}~\bibnamefont {Boive}},
  \bibinfo {author} {\bibfnamefont {B.}~\bibnamefont {Nordstr{\"o}m}},\ and\
  \bibinfo {author} {\bibfnamefont {C.}~\bibnamefont {Br{\"a}nd{\'e}n}},\
  }\bibfield  {title} {\bibinfo {title} {Magnetic structure of {Cr2Te3},
  {Cr3Te4}, and {Cr5Te6}},\ }\href@noop {} {\bibfield  {journal} {\bibinfo
  {journal} {Acta Chem. Scand.}\ }\textbf {\bibinfo {volume} {24}},\ \bibinfo
  {pages} {3495} (\bibinfo {year} {1970})}\BibitemShut {NoStop}%
\bibitem [{\citenamefont {Wang}\ \emph {et~al.}(2013)\citenamefont {Wang},
  \citenamefont {Zhu}, \citenamefont {Nilsson}, \citenamefont {Wen},
  \citenamefont {Wang}, \citenamefont {Shan}, \citenamefont {Zhang},
  \citenamefont {Zhang}, \citenamefont {Jia},\ and\ \citenamefont
  {Xue}}]{wang_situ_2013}%
  \BibitemOpen
  \bibfield  {author} {\bibinfo {author} {\bibfnamefont {C.}~\bibnamefont
  {Wang}}, \bibinfo {author} {\bibfnamefont {X.}~\bibnamefont {Zhu}}, \bibinfo
  {author} {\bibfnamefont {L.}~\bibnamefont {Nilsson}}, \bibinfo {author}
  {\bibfnamefont {J.}~\bibnamefont {Wen}}, \bibinfo {author} {\bibfnamefont
  {G.}~\bibnamefont {Wang}}, \bibinfo {author} {\bibfnamefont {X.}~\bibnamefont
  {Shan}}, \bibinfo {author} {\bibfnamefont {Q.}~\bibnamefont {Zhang}},
  \bibinfo {author} {\bibfnamefont {S.}~\bibnamefont {Zhang}}, \bibinfo
  {author} {\bibfnamefont {J.}~\bibnamefont {Jia}},\ and\ \bibinfo {author}
  {\bibfnamefont {Q.}~\bibnamefont {Xue}},\ }\bibfield  {title} {\bibinfo
  {title} {In situ {Raman} spectroscopy of topological insulator {Bi2Te3} films
  with varying thickness},\ }\href {https://doi.org/10.1007/s12274-013-0344-4}
  {\bibfield  {journal} {\bibinfo  {journal} {Nano Research}\ }\textbf
  {\bibinfo {volume} {6}},\ \bibinfo {pages} {688} (\bibinfo {year}
  {2013})}\BibitemShut {NoStop}%
\bibitem [{\citenamefont {Chi}\ \emph {et~al.}(2022)\citenamefont {Chi},
  \citenamefont {Ou}, \citenamefont {Eldred}, \citenamefont {Gao},
  \citenamefont {Kwon}, \citenamefont {Murray}, \citenamefont {Dreyer},
  \citenamefont {Butera}, \citenamefont {Foucher}, \citenamefont {Ambaye},
  \citenamefont {Keum}, \citenamefont {Greenberg}, \citenamefont {Liu},
  \citenamefont {Neupane}, \citenamefont {de~Coster}, \citenamefont {Vail},
  \citenamefont {Taylor}, \citenamefont {Folkes}, \citenamefont {Rong},
  \citenamefont {Yin}, \citenamefont {Lake}, \citenamefont {Ross},
  \citenamefont {Lauter}, \citenamefont {Heiman},\ and\ \citenamefont
  {Moodera}}]{chi2022strain}%
  \BibitemOpen
  \bibfield  {author} {\bibinfo {author} {\bibfnamefont {H.}~\bibnamefont
  {Chi}}, \bibinfo {author} {\bibfnamefont {Y.}~\bibnamefont {Ou}}, \bibinfo
  {author} {\bibfnamefont {T.~B.}\ \bibnamefont {Eldred}}, \bibinfo {author}
  {\bibfnamefont {W.}~\bibnamefont {Gao}}, \bibinfo {author} {\bibfnamefont
  {S.}~\bibnamefont {Kwon}}, \bibinfo {author} {\bibfnamefont {J.}~\bibnamefont
  {Murray}}, \bibinfo {author} {\bibfnamefont {M.}~\bibnamefont {Dreyer}},
  \bibinfo {author} {\bibfnamefont {R.~E.}\ \bibnamefont {Butera}}, \bibinfo
  {author} {\bibfnamefont {A.~C.}\ \bibnamefont {Foucher}}, \bibinfo {author}
  {\bibfnamefont {H.}~\bibnamefont {Ambaye}}, \bibinfo {author} {\bibfnamefont
  {J.}~\bibnamefont {Keum}}, \bibinfo {author} {\bibfnamefont {A.~T.}\
  \bibnamefont {Greenberg}}, \bibinfo {author} {\bibfnamefont {Y.}~\bibnamefont
  {Liu}}, \bibinfo {author} {\bibfnamefont {M.~R.}\ \bibnamefont {Neupane}},
  \bibinfo {author} {\bibfnamefont {G.~J.}\ \bibnamefont {de~Coster}}, \bibinfo
  {author} {\bibfnamefont {O.~A.}\ \bibnamefont {Vail}}, \bibinfo {author}
  {\bibfnamefont {P.~J.}\ \bibnamefont {Taylor}}, \bibinfo {author}
  {\bibfnamefont {P.~A.}\ \bibnamefont {Folkes}}, \bibinfo {author}
  {\bibfnamefont {C.}~\bibnamefont {Rong}}, \bibinfo {author} {\bibfnamefont
  {G.}~\bibnamefont {Yin}}, \bibinfo {author} {\bibfnamefont {R.~K.}\
  \bibnamefont {Lake}}, \bibinfo {author} {\bibfnamefont {F.~M.}\ \bibnamefont
  {Ross}}, \bibinfo {author} {\bibfnamefont {V.}~\bibnamefont {Lauter}},
  \bibinfo {author} {\bibfnamefont {D.}~\bibnamefont {Heiman}},\ and\ \bibinfo
  {author} {\bibfnamefont {J.~S.}\ \bibnamefont {Moodera}},\ }\bibfield
  {title} {\bibinfo {title} {Strain-tunable berry curvature in
  quasi-two-dimensional chromium telluride},\ }\href@noop {} {\bibfield
  {journal} {\bibinfo  {journal} {arXiv preprint arXiv:2207.02318}\ } (\bibinfo
  {year} {2022})}\BibitemShut {NoStop}%
\bibitem [{\citenamefont {Fujisawa}\ \emph {et~al.}(2022)\citenamefont
  {Fujisawa}, \citenamefont {Pardo-Almanza}, \citenamefont {Hsu}, \citenamefont
  {Mohamed}, \citenamefont {Yamagami}, \citenamefont {Krishnadas},
  \citenamefont {Chuang}, \citenamefont {Khoo}, \citenamefont {Zang},
  \citenamefont {Soumyanarayanan},\ and\ \citenamefont
  {Okada}}]{fujisawa2022widely}%
  \BibitemOpen
  \bibfield  {author} {\bibinfo {author} {\bibfnamefont {Y.}~\bibnamefont
  {Fujisawa}}, \bibinfo {author} {\bibfnamefont {M.}~\bibnamefont
  {Pardo-Almanza}}, \bibinfo {author} {\bibfnamefont {C.~H.}\ \bibnamefont
  {Hsu}}, \bibinfo {author} {\bibfnamefont {A.}~\bibnamefont {Mohamed}},
  \bibinfo {author} {\bibfnamefont {K.}~\bibnamefont {Yamagami}}, \bibinfo
  {author} {\bibfnamefont {A.}~\bibnamefont {Krishnadas}}, \bibinfo {author}
  {\bibfnamefont {F.~C.}\ \bibnamefont {Chuang}}, \bibinfo {author}
  {\bibfnamefont {K.~H.}\ \bibnamefont {Khoo}}, \bibinfo {author}
  {\bibfnamefont {J.}~\bibnamefont {Zang}}, \bibinfo {author} {\bibfnamefont
  {A.}~\bibnamefont {Soumyanarayanan}},\ and\ \bibinfo {author} {\bibfnamefont
  {Y.}~\bibnamefont {Okada}},\ }\bibfield  {title} {\bibinfo {title} {Widely
  tunable berry curvature in the magnetic semimetal cr1+ dte2},\ }\href@noop {}
  {\bibfield  {journal} {\bibinfo  {journal} {arXiv preprint arXiv:2204.02518}\
  } (\bibinfo {year} {2022})}\BibitemShut {NoStop}%
\bibitem [{\citenamefont {Zhang}\ \emph {et~al.}(2022)\citenamefont {Zhang},
  \citenamefont {Liu}, \citenamefont {Zhang}, \citenamefont {Yuan},
  \citenamefont {Wen}, \citenamefont {Li}, \citenamefont {Zheng}, \citenamefont
  {Zhang}, \citenamefont {Hou}, \citenamefont {Yin}, \citenamefont {Liu},
  \citenamefont {Peng},\ and\ \citenamefont {Zhang}}]{zhang_room_2022}%
  \BibitemOpen
  \bibfield  {author} {\bibinfo {author} {\bibfnamefont {C.}~\bibnamefont
  {Zhang}}, \bibinfo {author} {\bibfnamefont {C.}~\bibnamefont {Liu}}, \bibinfo
  {author} {\bibfnamefont {J.}~\bibnamefont {Zhang}}, \bibinfo {author}
  {\bibfnamefont {Y.}~\bibnamefont {Yuan}}, \bibinfo {author} {\bibfnamefont
  {Y.}~\bibnamefont {Wen}}, \bibinfo {author} {\bibfnamefont {Y.}~\bibnamefont
  {Li}}, \bibinfo {author} {\bibfnamefont {D.}~\bibnamefont {Zheng}}, \bibinfo
  {author} {\bibfnamefont {Q.}~\bibnamefont {Zhang}}, \bibinfo {author}
  {\bibfnamefont {Z.}~\bibnamefont {Hou}}, \bibinfo {author} {\bibfnamefont
  {G.}~\bibnamefont {Yin}}, \bibinfo {author} {\bibfnamefont {K.}~\bibnamefont
  {Liu}}, \bibinfo {author} {\bibfnamefont {Y.}~\bibnamefont {Peng}},\ and\
  \bibinfo {author} {\bibfnamefont {X.-X.}\ \bibnamefont {Zhang}},\ }\bibfield
  {title} {\bibinfo {title} {Room-temperature magnetic skyrmions and large
  topological hall effect in chromium telluride engineered by
  self-intercalation},\ }\href
  {https://doi.org/https://doi.org/10.1002/adma.202205967} {\bibfield
  {journal} {\bibinfo  {journal} {Advanced Materials}\ ,\ \bibinfo {pages}
  {2205967}} (\bibinfo {year} {2022})}\BibitemShut {NoStop}%
\bibitem [{\citenamefont {Tai}\ \emph {et~al.}(2022)\citenamefont {Tai},
  \citenamefont {Dai}, \citenamefont {Li}, \citenamefont {Huang}, \citenamefont
  {Chong}, \citenamefont {Wong}, \citenamefont {Zhang}, \citenamefont {Zhang},
  \citenamefont {Deng}, \citenamefont {Eckberg}, \citenamefont {Qiu},
  \citenamefont {He}, \citenamefont {Wu}, \citenamefont {Xu}, \citenamefont
  {Davydov}, \citenamefont {Wu},\ and\ \citenamefont
  {Wang}}]{tai_distinguishing_2022}%
  \BibitemOpen
  \bibfield  {author} {\bibinfo {author} {\bibfnamefont {L.}~\bibnamefont
  {Tai}}, \bibinfo {author} {\bibfnamefont {B.}~\bibnamefont {Dai}}, \bibinfo
  {author} {\bibfnamefont {J.}~\bibnamefont {Li}}, \bibinfo {author}
  {\bibfnamefont {H.}~\bibnamefont {Huang}}, \bibinfo {author} {\bibfnamefont
  {S.~K.}\ \bibnamefont {Chong}}, \bibinfo {author} {\bibfnamefont {K.~L.}\
  \bibnamefont {Wong}}, \bibinfo {author} {\bibfnamefont {H.}~\bibnamefont
  {Zhang}}, \bibinfo {author} {\bibfnamefont {P.}~\bibnamefont {Zhang}},
  \bibinfo {author} {\bibfnamefont {P.}~\bibnamefont {Deng}}, \bibinfo {author}
  {\bibfnamefont {C.}~\bibnamefont {Eckberg}}, \bibinfo {author} {\bibfnamefont
  {G.}~\bibnamefont {Qiu}}, \bibinfo {author} {\bibfnamefont {H.}~\bibnamefont
  {He}}, \bibinfo {author} {\bibfnamefont {D.}~\bibnamefont {Wu}}, \bibinfo
  {author} {\bibfnamefont {S.}~\bibnamefont {Xu}}, \bibinfo {author}
  {\bibfnamefont {A.}~\bibnamefont {Davydov}}, \bibinfo {author} {\bibfnamefont
  {R.}~\bibnamefont {Wu}},\ and\ \bibinfo {author} {\bibfnamefont {K.~L.}\
  \bibnamefont {Wang}},\ }\bibfield  {title} {\bibinfo {title} {Distinguishing
  the two-component anomalous hall effect from the topological hall effect},\
  }\href@noop {} {\bibfield  {journal} {\bibinfo  {journal} {ACS Nano}\
  }\textbf {\bibinfo {volume} {16}},\ \bibinfo {pages} {17336} (\bibinfo {year}
  {2022})}\BibitemShut {NoStop}%
\bibitem [{\citenamefont {Wang}\ \emph {et~al.}(2020)\citenamefont {Wang},
  \citenamefont {Feng}, \citenamefont {Lee}, \citenamefont {Ko}, \citenamefont
  {Lu},\ and\ \citenamefont {Noh}}]{wang_controllable_2020}%
  \BibitemOpen
  \bibfield  {author} {\bibinfo {author} {\bibfnamefont {L.}~\bibnamefont
  {Wang}}, \bibinfo {author} {\bibfnamefont {Q.}~\bibnamefont {Feng}}, \bibinfo
  {author} {\bibfnamefont {H.~G.}\ \bibnamefont {Lee}}, \bibinfo {author}
  {\bibfnamefont {E.~K.}\ \bibnamefont {Ko}}, \bibinfo {author} {\bibfnamefont
  {Q.}~\bibnamefont {Lu}},\ and\ \bibinfo {author} {\bibfnamefont {T.~W.}\
  \bibnamefont {Noh}},\ }\bibfield  {title} {\bibinfo {title} {Controllable
  thickness inhomogeneity and berry curvature engineering of anomalous hall
  effect in srruo3 ultrathin films},\ }\href@noop {} {\bibfield  {journal}
  {\bibinfo  {journal} {Nano Letters}\ }\textbf {\bibinfo {volume} {20}},\
  \bibinfo {pages} {2468} (\bibinfo {year} {2020})}\BibitemShut {NoStop}%
\bibitem [{\citenamefont {Kimbell}\ \emph {et~al.}(2022)\citenamefont
  {Kimbell}, \citenamefont {Kim}, \citenamefont {Wu}, \citenamefont {Cuoco},\
  and\ \citenamefont {Robinson}}]{kimbell_challenges_2022}%
  \BibitemOpen
  \bibfield  {author} {\bibinfo {author} {\bibfnamefont {G.}~\bibnamefont
  {Kimbell}}, \bibinfo {author} {\bibfnamefont {C.}~\bibnamefont {Kim}},
  \bibinfo {author} {\bibfnamefont {W.}~\bibnamefont {Wu}}, \bibinfo {author}
  {\bibfnamefont {M.}~\bibnamefont {Cuoco}},\ and\ \bibinfo {author}
  {\bibfnamefont {J.~W.~A.}\ \bibnamefont {Robinson}},\ }\bibfield  {title}
  {\bibinfo {title} {Challenges in identifying chiral spin textures via the
  topological {Hall} effect},\ }\href@noop {} {\bibfield  {journal} {\bibinfo
  {journal} {Communications Materials}\ }\textbf {\bibinfo {volume} {3}},\
  \bibinfo {pages} {1} (\bibinfo {year} {2022})}\BibitemShut {NoStop}%
\bibitem [{\citenamefont {Li}\ \emph {et~al.}(2022)\citenamefont {Li},
  \citenamefont {Liu}, \citenamefont {Jiang}, \citenamefont {Jin},
  \citenamefont {Pei}, \citenamefont {Wen}, \citenamefont {Yue},\ and\
  \citenamefont {Wang}}]{li_diverse_2022}%
  \BibitemOpen
  \bibfield  {author} {\bibinfo {author} {\bibfnamefont {C.}~\bibnamefont
  {Li}}, \bibinfo {author} {\bibfnamefont {K.}~\bibnamefont {Liu}}, \bibinfo
  {author} {\bibfnamefont {D.}~\bibnamefont {Jiang}}, \bibinfo {author}
  {\bibfnamefont {C.}~\bibnamefont {Jin}}, \bibinfo {author} {\bibfnamefont
  {T.}~\bibnamefont {Pei}}, \bibinfo {author} {\bibfnamefont {T.}~\bibnamefont
  {Wen}}, \bibinfo {author} {\bibfnamefont {B.}~\bibnamefont {Yue}},\ and\
  \bibinfo {author} {\bibfnamefont {Y.}~\bibnamefont {Wang}},\ }\bibfield
  {title} {\bibinfo {title} {Diverse {Thermal} {Expansion} {Behaviors} in
  {Ferromagnetic} {Cr1-dTe} with {NiAs}-{Type}, {Defective} {Structures}},\
  }\href {https://doi.org/10.1021/acs.inorgchem.2c01826} {\bibfield  {journal}
  {\bibinfo  {journal} {Inorganic Chemistry}\ }\textbf {\bibinfo {volume}
  {61}},\ \bibinfo {pages} {14641} (\bibinfo {year} {2022})}\BibitemShut
  {NoStop}%
\end{thebibliography}%


\begin{thebibliography}{6}%
\makeatletter
\providecommand \@ifxundefined [1]{%
 \@ifx{#1\undefined}
}%
\providecommand \@ifnum [1]{%
 \ifnum #1\expandafter \@firstoftwo
 \else \expandafter \@secondoftwo
 \fi
}%
\providecommand \@ifx [1]{%
 \ifx #1\expandafter \@firstoftwo
 \else \expandafter \@secondoftwo
 \fi
}%
\providecommand \natexlab [1]{#1}%
\providecommand \enquote  [1]{``#1''}%
\providecommand \bibnamefont  [1]{#1}%
\providecommand \bibfnamefont [1]{#1}%
\providecommand \citenamefont [1]{#1}%
\providecommand \href@noop [0]{\@secondoftwo}%
\providecommand \href [0]{\begingroup \@sanitize@url \@href}%
\providecommand \@href[1]{\@@startlink{#1}\@@href}%
\providecommand \@@href[1]{\endgroup#1\@@endlink}%
\providecommand \@sanitize@url [0]{\catcode `\\12\catcode `\$12\catcode
  `\&12\catcode `\#12\catcode `\^12\catcode `\_12\catcode `\%12\relax}%
\providecommand \@@startlink[1]{}%
\providecommand \@@endlink[0]{}%
\providecommand \url  [0]{\begingroup\@sanitize@url \@url }%
\providecommand \@url [1]{\endgroup\@href {#1}{\urlprefix }}%
\providecommand \urlprefix  [0]{URL }%
\providecommand \Eprint [0]{\href }%
\providecommand \doibase [0]{https://doi.org/}%
\providecommand \selectlanguage [0]{\@gobble}%
\providecommand \bibinfo  [0]{\@secondoftwo}%
\providecommand \bibfield  [0]{\@secondoftwo}%
\providecommand \translation [1]{[#1]}%
\providecommand \BibitemOpen [0]{}%
\providecommand \bibitemStop [0]{}%
\providecommand \bibitemNoStop [0]{.\EOS\space}%
\providecommand \EOS [0]{\spacefactor3000\relax}%
\providecommand \BibitemShut  [1]{\csname bibitem#1\endcsname}%
\let\auto@bib@innerbib\@empty
\bibitem [{\citenamefont {Wen}\ \emph {et~al.}(2020)\citenamefont {Wen},
  \citenamefont {Liu}, \citenamefont {Zhang}, \citenamefont {Xia},
  \citenamefont {Zhai}, \citenamefont {Zhang}, \citenamefont {Zhai},
  \citenamefont {Shen}, \citenamefont {He}, \citenamefont {Cheng},
  \citenamefont {Yin}, \citenamefont {Yao}, \citenamefont {Getaye~Sendeku},
  \citenamefont {Wang}, \citenamefont {Ye}, \citenamefont {Liu}, \citenamefont
  {Jiang}, \citenamefont {Shan}, \citenamefont {Long},\ and\ \citenamefont
  {He}}]{wen_tunable_2020}%
  \BibitemOpen
  \bibfield  {author} {\bibinfo {author} {\bibfnamefont {Y.}~\bibnamefont
  {Wen}}, \bibinfo {author} {\bibfnamefont {Z.}~\bibnamefont {Liu}}, \bibinfo
  {author} {\bibfnamefont {Y.}~\bibnamefont {Zhang}}, \bibinfo {author}
  {\bibfnamefont {C.}~\bibnamefont {Xia}}, \bibinfo {author} {\bibfnamefont
  {B.}~\bibnamefont {Zhai}}, \bibinfo {author} {\bibfnamefont {X.}~\bibnamefont
  {Zhang}}, \bibinfo {author} {\bibfnamefont {G.}~\bibnamefont {Zhai}},
  \bibinfo {author} {\bibfnamefont {C.}~\bibnamefont {Shen}}, \bibinfo {author}
  {\bibfnamefont {P.}~\bibnamefont {He}}, \bibinfo {author} {\bibfnamefont
  {R.}~\bibnamefont {Cheng}}, \bibinfo {author} {\bibfnamefont
  {L.}~\bibnamefont {Yin}}, \bibinfo {author} {\bibfnamefont {Y.}~\bibnamefont
  {Yao}}, \bibinfo {author} {\bibfnamefont {M.}~\bibnamefont {Getaye~Sendeku}},
  \bibinfo {author} {\bibfnamefont {Z.}~\bibnamefont {Wang}}, \bibinfo {author}
  {\bibfnamefont {X.}~\bibnamefont {Ye}}, \bibinfo {author} {\bibfnamefont
  {C.}~\bibnamefont {Liu}}, \bibinfo {author} {\bibfnamefont {C.}~\bibnamefont
  {Jiang}}, \bibinfo {author} {\bibfnamefont {C.}~\bibnamefont {Shan}},
  \bibinfo {author} {\bibfnamefont {Y.}~\bibnamefont {Long}},\ and\ \bibinfo
  {author} {\bibfnamefont {J.}~\bibnamefont {He}},\ }\bibfield  {title}
  {\bibinfo {title} {Tunable {Room}-{Temperature} {Ferromagnetism} in
  {Two}-{Dimensional} {Cr} $_{\textrm{2}}$ {Te} $_{\textrm{3}}$},\ }\href@noop
  {} {\bibfield  {journal} {\bibinfo  {journal} {Nano Letters}\ }\textbf
  {\bibinfo {volume} {20}},\ \bibinfo {pages} {3130} (\bibinfo {year}
  {2020})}\BibitemShut {NoStop}%
\bibitem [{\citenamefont {Lasek}\ \emph {et~al.}(2020)\citenamefont {Lasek},
  \citenamefont {Coelho}, \citenamefont {Zberecki}, \citenamefont {Xin},
  \citenamefont {Kolekar}, \citenamefont {Li},\ and\ \citenamefont
  {Batzill}}]{lasek_molecular_2020}%
  \BibitemOpen
  \bibfield  {author} {\bibinfo {author} {\bibfnamefont {K.}~\bibnamefont
  {Lasek}}, \bibinfo {author} {\bibfnamefont {P.~M.}\ \bibnamefont {Coelho}},
  \bibinfo {author} {\bibfnamefont {K.}~\bibnamefont {Zberecki}}, \bibinfo
  {author} {\bibfnamefont {Y.}~\bibnamefont {Xin}}, \bibinfo {author}
  {\bibfnamefont {S.~K.}\ \bibnamefont {Kolekar}}, \bibinfo {author}
  {\bibfnamefont {J.}~\bibnamefont {Li}},\ and\ \bibinfo {author}
  {\bibfnamefont {M.}~\bibnamefont {Batzill}},\ }\bibfield  {title} {\bibinfo
  {title} {Molecular {Beam} {Epitaxy} of {Transition} {Metal} ({Ti}-, {V}-, and
  {Cr}-) {Tellurides}: {From} {Monolayer} {Ditellurides} to {Multilayer}
  {Self}-{Intercalation} {Compounds}},\ }\href
  {https://doi.org/10.1021/acsnano.0c02712} {\bibfield  {journal} {\bibinfo
  {journal} {ACS Nano}\ }\textbf {\bibinfo {volume} {14}},\ \bibinfo {pages}
  {8473} (\bibinfo {year} {2020})}\BibitemShut {NoStop}%
\bibitem [{\citenamefont {Bian}\ \emph {et~al.}(2021)\citenamefont {Bian},
  \citenamefont {Kamenskii}, \citenamefont {Han}, \citenamefont {Li},
  \citenamefont {Wei}, \citenamefont {Tian}, \citenamefont {Eason},
  \citenamefont {Sun}, \citenamefont {He}, \citenamefont {Hui}, \citenamefont
  {Yao}, \citenamefont {Sabirianov}, \citenamefont {Bird}, \citenamefont
  {Yang}, \citenamefont {Miao}, \citenamefont {Lin}, \citenamefont {Crooker},
  \citenamefont {Hou},\ and\ \citenamefont {Zeng}}]{bian_covalent_2021}%
  \BibitemOpen
  \bibfield  {author} {\bibinfo {author} {\bibfnamefont {M.}~\bibnamefont
  {Bian}}, \bibinfo {author} {\bibfnamefont {A.~N.}\ \bibnamefont {Kamenskii}},
  \bibinfo {author} {\bibfnamefont {M.}~\bibnamefont {Han}}, \bibinfo {author}
  {\bibfnamefont {W.}~\bibnamefont {Li}}, \bibinfo {author} {\bibfnamefont
  {S.}~\bibnamefont {Wei}}, \bibinfo {author} {\bibfnamefont {X.}~\bibnamefont
  {Tian}}, \bibinfo {author} {\bibfnamefont {D.~B.}\ \bibnamefont {Eason}},
  \bibinfo {author} {\bibfnamefont {F.}~\bibnamefont {Sun}}, \bibinfo {author}
  {\bibfnamefont {K.}~\bibnamefont {He}}, \bibinfo {author} {\bibfnamefont
  {H.}~\bibnamefont {Hui}}, \bibinfo {author} {\bibfnamefont {F.}~\bibnamefont
  {Yao}}, \bibinfo {author} {\bibfnamefont {R.}~\bibnamefont {Sabirianov}},
  \bibinfo {author} {\bibfnamefont {J.~P.}\ \bibnamefont {Bird}}, \bibinfo
  {author} {\bibfnamefont {C.}~\bibnamefont {Yang}}, \bibinfo {author}
  {\bibfnamefont {J.}~\bibnamefont {Miao}}, \bibinfo {author} {\bibfnamefont
  {J.}~\bibnamefont {Lin}}, \bibinfo {author} {\bibfnamefont {S.~A.}\
  \bibnamefont {Crooker}}, \bibinfo {author} {\bibfnamefont {Y.}~\bibnamefont
  {Hou}},\ and\ \bibinfo {author} {\bibfnamefont {H.}~\bibnamefont {Zeng}},\
  }\bibfield  {title} {\bibinfo {title} {Covalent {2D} {Cr2Te3} ferromagnet},\
  }\href {https://doi.org/10.1080/21663831.2020.1865469} {\bibfield  {journal}
  {\bibinfo  {journal} {Materials Research Letters}\ }\textbf {\bibinfo
  {volume} {9}},\ \bibinfo {pages} {205} (\bibinfo {year} {2021})}\BibitemShut
  {NoStop}%
\bibitem [{\citenamefont {Wang}\ \emph {et~al.}(2008)\citenamefont {Wang},
  \citenamefont {Ni}, \citenamefont {Yu}, \citenamefont {Shen}, \citenamefont
  {Wang}, \citenamefont {Wu}, \citenamefont {Chen},\ and\ \citenamefont
  {Shen~Wee}}]{wang_raman_2008}%
  \BibitemOpen
  \bibfield  {author} {\bibinfo {author} {\bibfnamefont {Y.~y.}\ \bibnamefont
  {Wang}}, \bibinfo {author} {\bibfnamefont {Z.~h.}\ \bibnamefont {Ni}},
  \bibinfo {author} {\bibfnamefont {T.}~\bibnamefont {Yu}}, \bibinfo {author}
  {\bibfnamefont {Z.~X.}\ \bibnamefont {Shen}}, \bibinfo {author}
  {\bibfnamefont {H.~m.}\ \bibnamefont {Wang}}, \bibinfo {author}
  {\bibfnamefont {Y.~h.}\ \bibnamefont {Wu}}, \bibinfo {author} {\bibfnamefont
  {W.}~\bibnamefont {Chen}},\ and\ \bibinfo {author} {\bibfnamefont {A.~T.}\
  \bibnamefont {Shen~Wee}},\ }\bibfield  {title} {\bibinfo {title} {Raman
  {Studies} of {Monolayer} {Graphene}: {The} {Substrate} {Effect}},\ }\href
  {https://doi.org/10.1021/jp8008404} {\bibfield  {journal} {\bibinfo
  {journal} {The Journal of Physical Chemistry C}\ }\textbf {\bibinfo {volume}
  {112}},\ \bibinfo {pages} {10637} (\bibinfo {year} {2008})}\BibitemShut
  {NoStop}%
\bibitem [{\citenamefont {Kubota}\ \emph {et~al.}(2022)\citenamefont {Kubota},
  \citenamefont {Okamoto}, \citenamefont {Kanematsu}, \citenamefont {Yajima},
  \citenamefont {Hirai},\ and\ \citenamefont {Takenaka}}]{kubota2022large}%
  \BibitemOpen
  \bibfield  {author} {\bibinfo {author} {\bibfnamefont {Y.}~\bibnamefont
  {Kubota}}, \bibinfo {author} {\bibfnamefont {Y.}~\bibnamefont {Okamoto}},
  \bibinfo {author} {\bibfnamefont {T.}~\bibnamefont {Kanematsu}}, \bibinfo
  {author} {\bibfnamefont {T.}~\bibnamefont {Yajima}}, \bibinfo {author}
  {\bibfnamefont {D.}~\bibnamefont {Hirai}},\ and\ \bibinfo {author}
  {\bibfnamefont {K.}~\bibnamefont {Takenaka}},\ }\bibfield  {title} {\bibinfo
  {title} {Large magnetic-field-induced strains in sintered chromium
  tellurides},\ }\href@noop {} {\bibfield  {journal} {\bibinfo  {journal}
  {arXiv preprint arXiv:2211.13388}\ } (\bibinfo {year} {2022})}\BibitemShut
  {NoStop}%
\bibitem [{\citenamefont {Jeon}\ \emph {et~al.}(2022)\citenamefont {Jeon},
  \citenamefont {Na}, \citenamefont {Kim}, \citenamefont {Lee}, \citenamefont
  {Song}, \citenamefont {Kim}, \citenamefont {Park}, \citenamefont {Kim},
  \citenamefont {Noh}, \citenamefont {Kim}, \citenamefont {Jerng},\ and\
  \citenamefont {Chun}}]{jeon_emergent_2022}%
  \BibitemOpen
  \bibfield  {author} {\bibinfo {author} {\bibfnamefont {J.~H.}\ \bibnamefont
  {Jeon}}, \bibinfo {author} {\bibfnamefont {H.~R.}\ \bibnamefont {Na}},
  \bibinfo {author} {\bibfnamefont {H.}~\bibnamefont {Kim}}, \bibinfo {author}
  {\bibfnamefont {S.}~\bibnamefont {Lee}}, \bibinfo {author} {\bibfnamefont
  {S.}~\bibnamefont {Song}}, \bibinfo {author} {\bibfnamefont {J.}~\bibnamefont
  {Kim}}, \bibinfo {author} {\bibfnamefont {S.}~\bibnamefont {Park}}, \bibinfo
  {author} {\bibfnamefont {J.}~\bibnamefont {Kim}}, \bibinfo {author}
  {\bibfnamefont {H.}~\bibnamefont {Noh}}, \bibinfo {author} {\bibfnamefont
  {G.}~\bibnamefont {Kim}}, \bibinfo {author} {\bibfnamefont {S.-K.}\
  \bibnamefont {Jerng}},\ and\ \bibinfo {author} {\bibfnamefont {S.-H.}\
  \bibnamefont {Chun}},\ }\bibfield  {title} {\bibinfo {title} {Emergent
  {Topological} {Hall} {Effect} from {Exchange} {Coupling} in {Ferromagnetic}
  {Cr2Te3}/{Noncoplanar} {Antiferromagnetic} {Cr2Se3} {Bilayers}},\ }\href@noop
  {} {\bibfield  {journal} {\bibinfo  {journal} {ACS Nano}\ ,\ \bibinfo {pages}
  {9}} (\bibinfo {year} {2022})}\BibitemShut {NoStop}%
\end{thebibliography}%
\end{document}


\title{Supplemental Material: Epitaxial van~der~Waals heterostructures of Cr$_2$Te$_3$ on 2D materials}

\author{Quentin Guillet}
 \altaffiliation{QG and LV contributed equally to this work}
\author{Libor Vojáček}
 \altaffiliation{QG and LV contributed equally to this work}
\affiliation{Université Grenoble Alpes, CEA, CNRS, IRIG-SPINTEC, 38000 Grenoble, France
}
\author{Djordje Dosenovic}
\affiliation{Université Grenoble Alpes, CEA, IRIG-MEM, 38000 Grenoble, France
}
\author{Fatima Ibrahim}
\affiliation{Université Grenoble Alpes, CEA, CNRS, IRIG-SPINTEC, 38000 Grenoble, France
}
\author{Hervé Boukari}
\affiliation{Université Grenoble Alpes, CNRS, Institut Neel, 38000 Grenoble, France
}
\author{Jing Li}
\affiliation{Université Grenoble Alpes, CEA, Leti, F-38000 Grenoble, France
}
\author{Fadi Choueikani}
\affiliation{Synchrotron SOLEIL, L'Orme des Merisiers, 91190 Saint-Aubin, France
}
\author{Philippe Ohresser}
\affiliation{Synchrotron SOLEIL, L'Orme des Merisiers, 91190 Saint-Aubin, France
}
\author{Abdelkarim Ouerghi}
\affiliation{Université Paris-Saclay, CNRS, Centre de Nanosciences et de Nanotechnologies, Palaiseau, France
}
\author{Florie Mesple}
\affiliation{Université Grenoble Alpes, CEA, CNRS, IRIG-PHELIQS, 38000 Grenoble, France
}
\author{Vincent Renard}
\affiliation{Université Grenoble Alpes, CEA, CNRS, IRIG-PHELIQS, 38000 Grenoble, France
}
\author{Jean-François Jacquot}
\affiliation{Université Grenoble Alpes, CEA, CNRS, IRIG-SYMMES, 38000 Grenoble, France
}
\author{Denis Jalabert}
\affiliation{Université Grenoble Alpes, CEA, IRIG-MEM, 38000 Grenoble, France
}
\author{Hanako Okuno}
\affiliation{Université Grenoble Alpes, CEA, IRIG-MEM, 38000 Grenoble, France
}
\author{Mairbek Chshiev}
\affiliation{Université Grenoble Alpes, CEA, CNRS, IRIG-SPINTEC, 38000 Grenoble, France
}
\affiliation{Institut Universitaire de France, 75231 Paris, France}
\author{Céline Vergnaud}
\affiliation{Université Grenoble Alpes, CEA, CNRS, IRIG-SPINTEC, 38000 Grenoble, France
}
\author{Frédéric Bonell}
\affiliation{Université Grenoble Alpes, CEA, CNRS, IRIG-SPINTEC, 38000 Grenoble, France
}
\author{Alain Marty}
\affiliation{Université Grenoble Alpes, CEA, CNRS, IRIG-SPINTEC, 38000 Grenoble, France
}
\author{Matthieu Jamet}
\affiliation{Université Grenoble Alpes, CEA, CNRS, IRIG-SPINTEC, 38000 Grenoble, France
}
\date{\today}

\maketitle

\tableofcontents
\clearpage

\section{RHEED images of \CT\ layers on Graphene/SiC and Bi$_2$Te$_3$/Al$_2$O$_3$}

\begin{figure*}[ht!]
\renewcommand*\thesubfigure{\arabic{subfigure}} 
\subfloat[\centering On Graphene/SiC]{{\includegraphics[width=11cm]{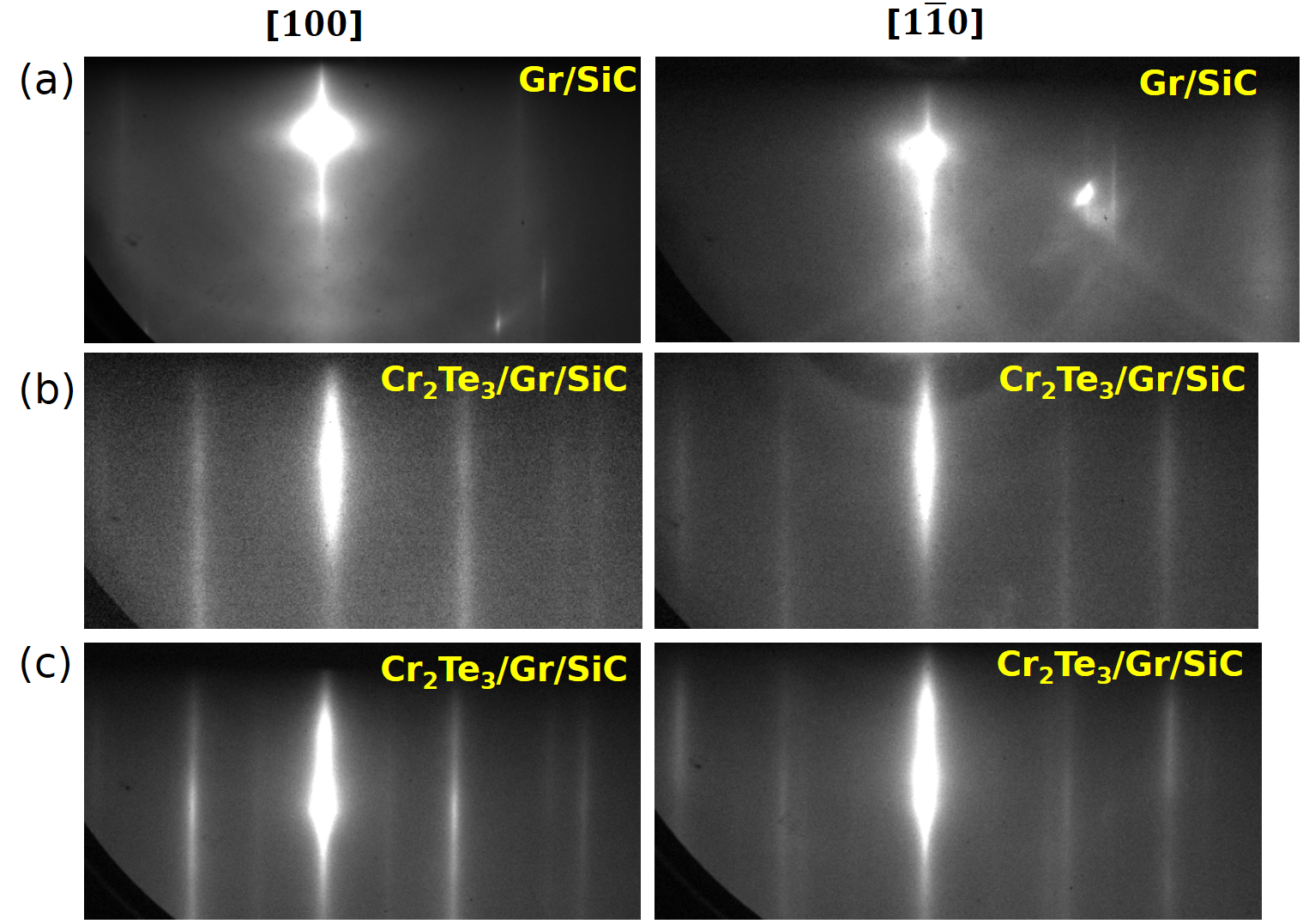} }}
\qquad
\subfloat[\centering On Bi$_2$Te$_3$/Al$_2$O$_3$]{{\includegraphics[width=11cm]{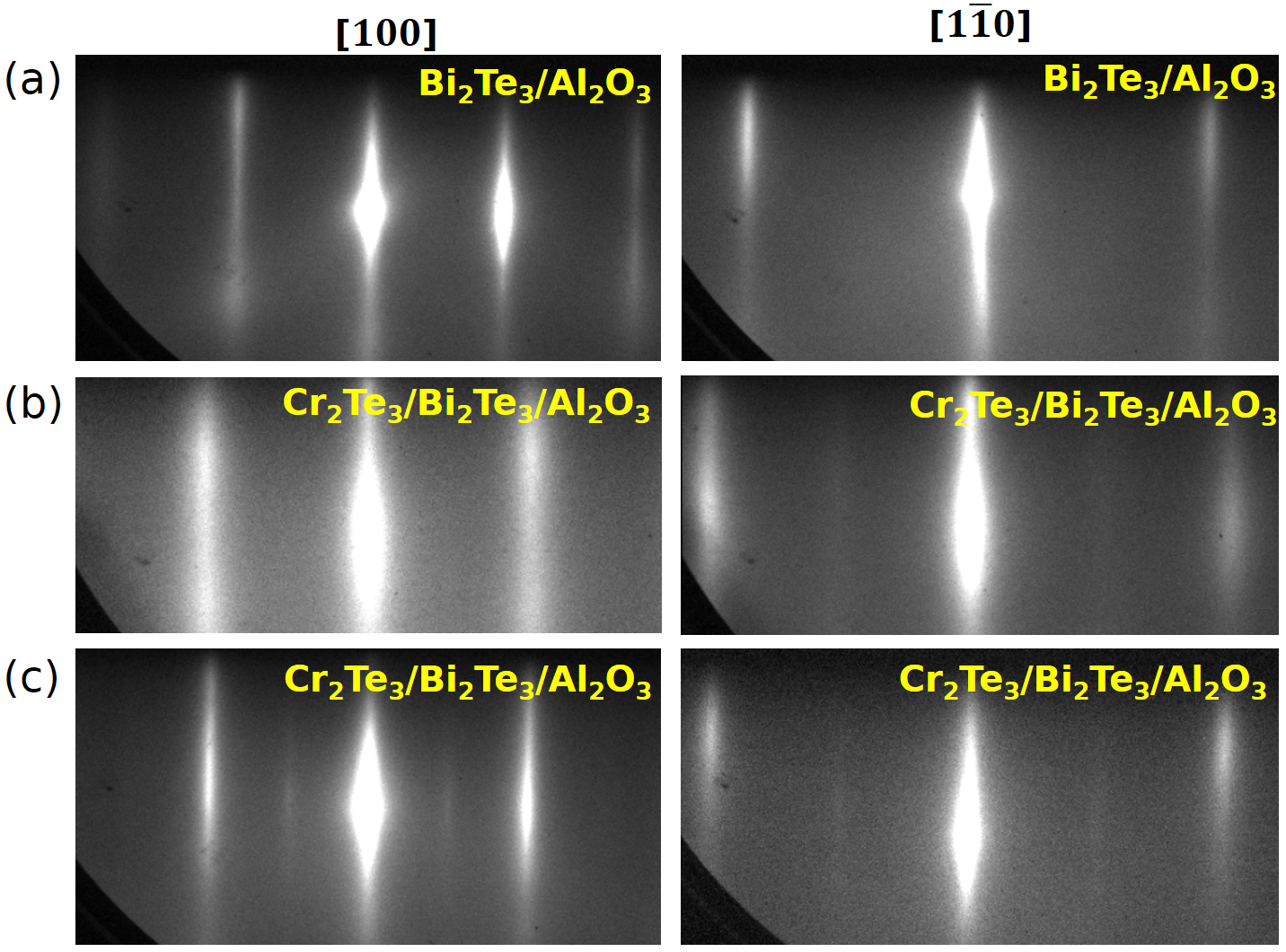} }}
\caption{(a) In-situ RHEED images of the 2D material along two main crystal directions. (b) RHEED pattern after the deposition of Cr$_2$Te$_3$ at 300°C. (c) RHEED pattern after annealing at 400°C.}
\label{RHEED-sup}
\end{figure*}
\pagebreak

\section{Lattice parameter of heterostructures with increasing \CT~thickness}

\begin{figure*}[ht!]
\includegraphics[width=17.2cm]{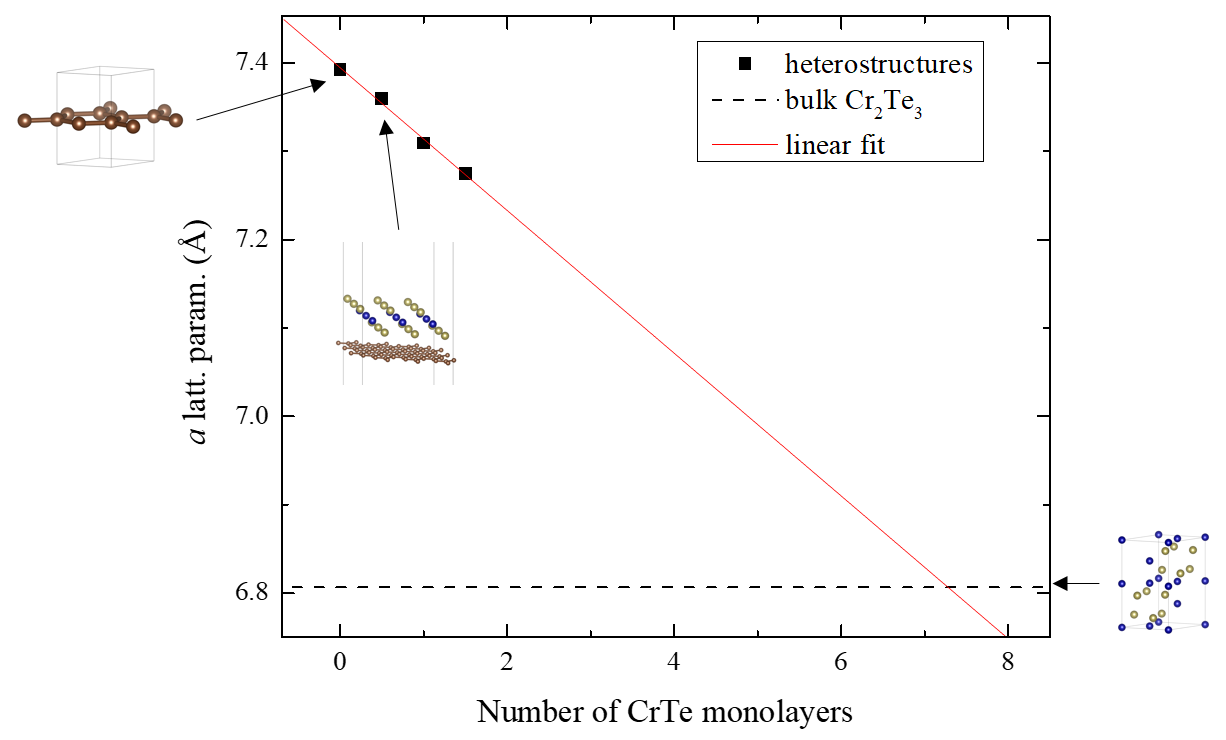}
\caption{The in-plane lattice parameter of Cr$_2$Te$_3$/graphene heterostructure with increasing \CT~thickness.}
\label{fig:CT_graphene_latt_param}
\end{figure*}

We argue that the choice of a particular 2D material has a negligible effect on the Cr$_2$Te$_3$/2D material lattice parameters. We support this by performing DFT relaxation of Cr$_2$Te$_3$/graphene heterostructures with different \CT~thicknesses (see Fig.~\ref{fig:CT_graphene_latt_param}). A linear trend is observed. The value interpolates to the one of \CT~bulk at $\approx 7$ MLs of Cr$_2$Te$_3$, while the experimentally grown films have a thickness of 5 MLs. Hence, the lattice parameter of \CT~is expected to retain its bulk value. Even more when considering that in the calculation the two materials are bound to occupy the same unit cell. This makes their lattice parameters coupled more strongly than in reality.

\section{Cr$_2$Te$_3$ termination stability}

\begin{figure}[h]
    \center
    \includegraphics[width=8.6cm]{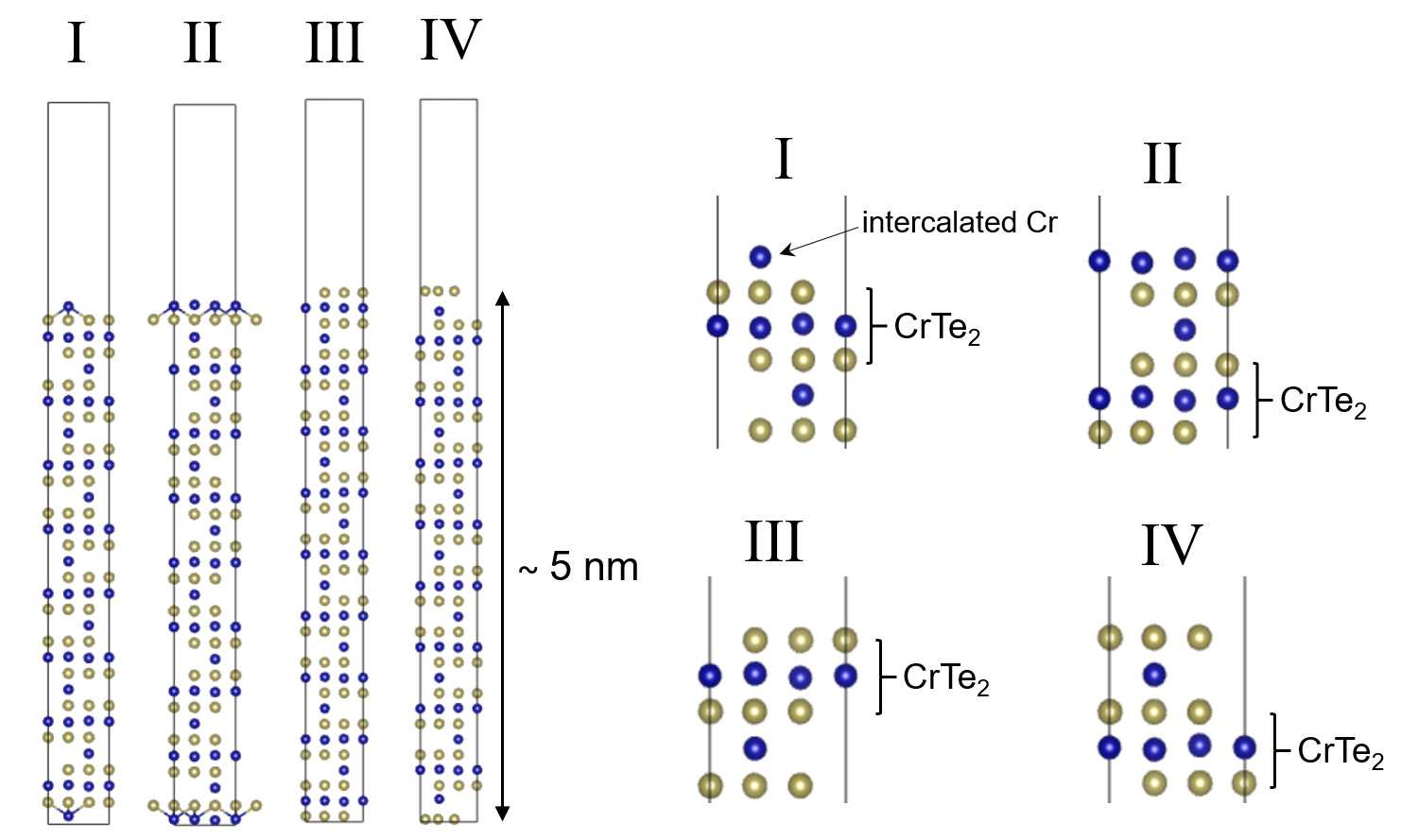}
    \caption{\CT\ thin films in vacuum used for the calculation of the surface-termination stability. Structure III (terminated by a complete CrTe$_2$ layer) is the most stable, followed by the intercalated Cr-terminated structure I. This corresponds to Ref.~\cite{wen_tunable_2020}.}
    \label{fig:DFT_termination}
\end{figure}

To build heterostructures made of \CT~and 2D materials, it is necessary to determine the stable termination of Cr$_2$Te$_3$, as there are 4 distinct possible terminations (Fig.~\ref{fig:DFT_termination}). It has been shown by Wen \textit{et al.} \cite{wen_tunable_2020} that \CT~should be terminated by Te, with a complete CrTe$_2$ layer at the boundary. The same conclusion was reached by Lasek \textit{et al.} \cite{lasek_molecular_2020} performing molecular dynamics of differently terminated \CT~on MoS$_2$. On the contrary, Bian \textit{et al.} \cite{bian_covalent_2021} concluded that the intercalated Cr is the one that should lie at the interface, arguing that this leads to strain relaxation. Due to this discordance in the literature we performed our own surface stability calculations.\\
Similar to \cite{wen_tunable_2020} we define the formation energy by:

\begin{equation}
    F_\mathrm{form} = ( F_\mathrm{tot} - n_\mathrm{Cr} \mu_\mathrm{Cr} - n_\mathrm{Te} \mu_\mathrm{Te}) / N_\mathrm{atoms}
\end{equation}

where $F_\mathrm{tot}$ is the total calculated free energy, $\mu_i$ is the chemical potential and $n_i$ is the number of atoms of element $i$ in the unit cell, and $N_\mathrm{atoms}$ is the total number of atoms per unit cell. The chemical potential $\mu$ is taken either as the free energy of an isolated atom (Cr or Te) or the free energy per atom in the bulk of the given material. Both cases give the same qualitative result regarding the surface termination (see Table~\ref{tab:DFT_termination}).
From our calculations, we conclude that the Te-terminated surface is indeed the most stable (Table~\ref{tab:DFT_termination}), although the formation energy of the intercalated Cr-terminated structure is very close - the difference is only 35 meV/atom (for $\mu$ from the Cr and Te bulk structures).

\begin{table}[htbp]
  \centering
  \caption{Chemical potentials of Cr and Te atoms and bulk. The formation energy of 4 differently-terminated \CT~thin films in vacuum. The most stable is structure III, see Fig.~\ref{fig:DFT_termination} for its definition.}
\setlength{\tabcolsep}{0pt}
  \begin{ruledtabular}
    \begin{tabular}{p{4.055em}crp{6.78em}rrrrr}
\multicolumn{1}{c}{} & \multicolumn{1}{p{4.055em}}{$\mu$ (eV)} &       & \multicolumn{1}{r}{} & \multicolumn{1}{p{5.61em}}{$F_\mathrm{tot}$ (eV)} & \multicolumn{1}{p{3.335em}}{$n_\mathrm{Cr}$} & \multicolumn{1}{p{2.72em}}{$n_\mathrm{Te}$} & \multicolumn{1}{p{5.72em}}{$F_\mathrm{form}^\mathrm{atoms}$ (eV)} & \multicolumn{1}{p{6.22em}}{$F_\mathrm{form}^\mathrm{bulk}$ (eV)} \\
\cmidrule{1-2}\cmidrule{4-9}    Cr atom & -4.89 &       & structure I & -462.477 & 33    & 48    & -3.318 & -0.365 \\
    Te atom & -0.68 &       & structure II & -474.406 & 35    & 48    & -3.264 & -0.305 \\
    Cr bulk & -8.10 &       & \textbf{structure III} & -504.478 & 35    & 54    & \textbf{-3.336} & \textbf{-0.390} \\
    Te bulk & -3.45 &       & structure IV & -483.289 & 33    & 54    & -3.282 & -0.342 \\
\end{tabular}%
\end{ruledtabular}
  \label{tab:DFT_termination}%
\end{table}%

\section{Stoichiometry of the layers}

\begin{figure*}[h]
    \center
    \includegraphics[width=17.2cm]{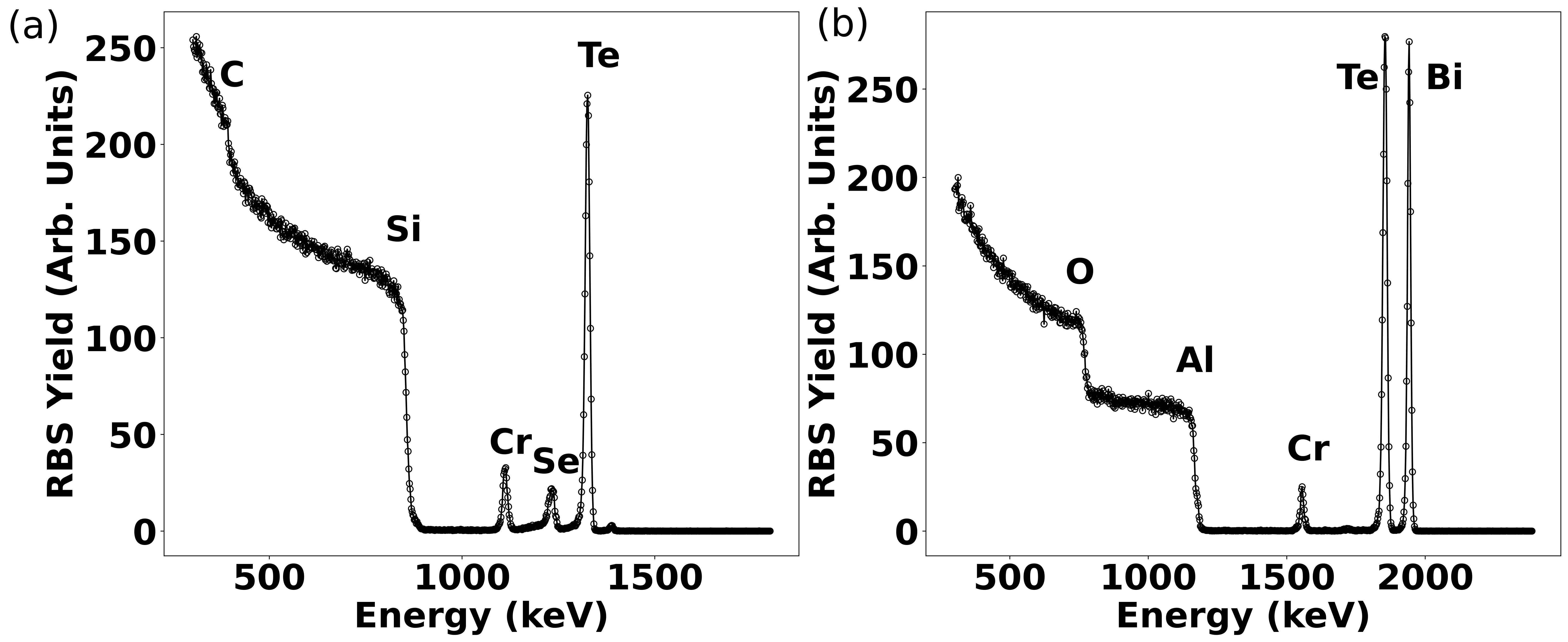}
    \caption{Stoichiometry of Cr$_x$Te$_3$
(a)	RBS of Cr$_{1.88}$Te$_3$ deposited on Gr/SiC (sample 3).
(b)	RBS of Cr$_{1.97}$Te$_3$ grown on Bi$_2$Te$_3$/ Al$_2$O$_3$ (sample 5). The exact stoichiometry of Bi$_2$Te$_3$ was considered to deduce the Cr/Te ratio in \CT.
}
\label{RBS}
\end{figure*}

To verify the stoichiometry of the grown samples, Rutherford Back Scattering (RBS) was performed (see Fig.~\ref{RBS}). The sample on graphene (n°3) was capped with amorphous Se and the one on Bi$_2$Te$_3$ (sample 5) was protected with Al. The ratio found is close to the expected 40\% Cr and 60\% Te.

\section{Raman scattering of monolayer graphene}

\begin{figure*}[h]
    \center
    \includegraphics[width=17.2cm]{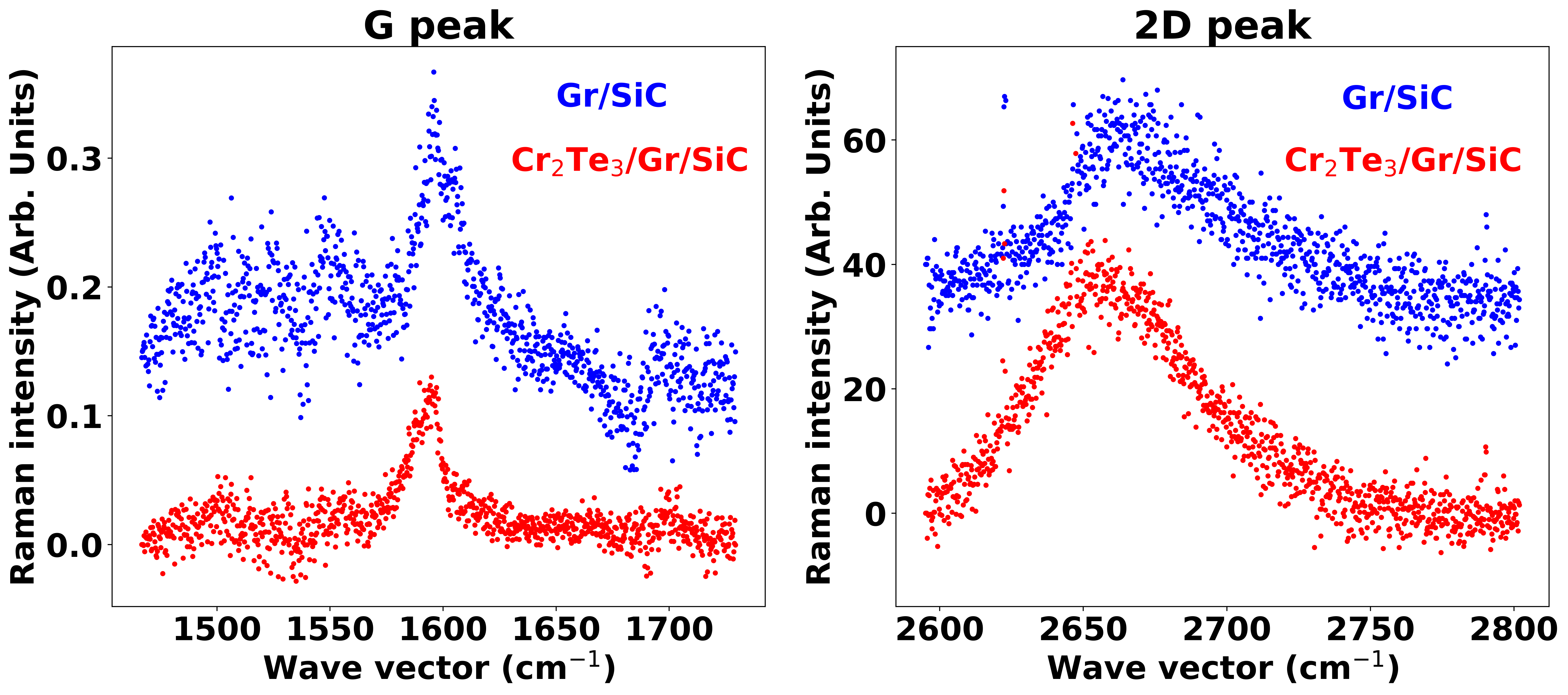}
    \caption{Two characteristic Raman scattering peaks of monolayer graphene before and after deposition of \CT.
}
\label{Raman}
\end{figure*}

Raman scattering is used to control the graphene quality \cite{wang_raman_2008}. We show in Fig.~\ref{Raman} that graphene is mostly unaffected by the deposition of \CT.

\newpage

\section{Hysteresis loop measured by XMCD}

To confirm the origin of the magnetic signal in \CT\ layers, we measured at the Soleil synchrotron source an hysteresis loop at the chromium L$_3$ edge and obtained a consistent signal with the experimental observations by SQUID magnetometry.

\begin{figure}[h]
    \center
    \includegraphics[width=8.6cm]{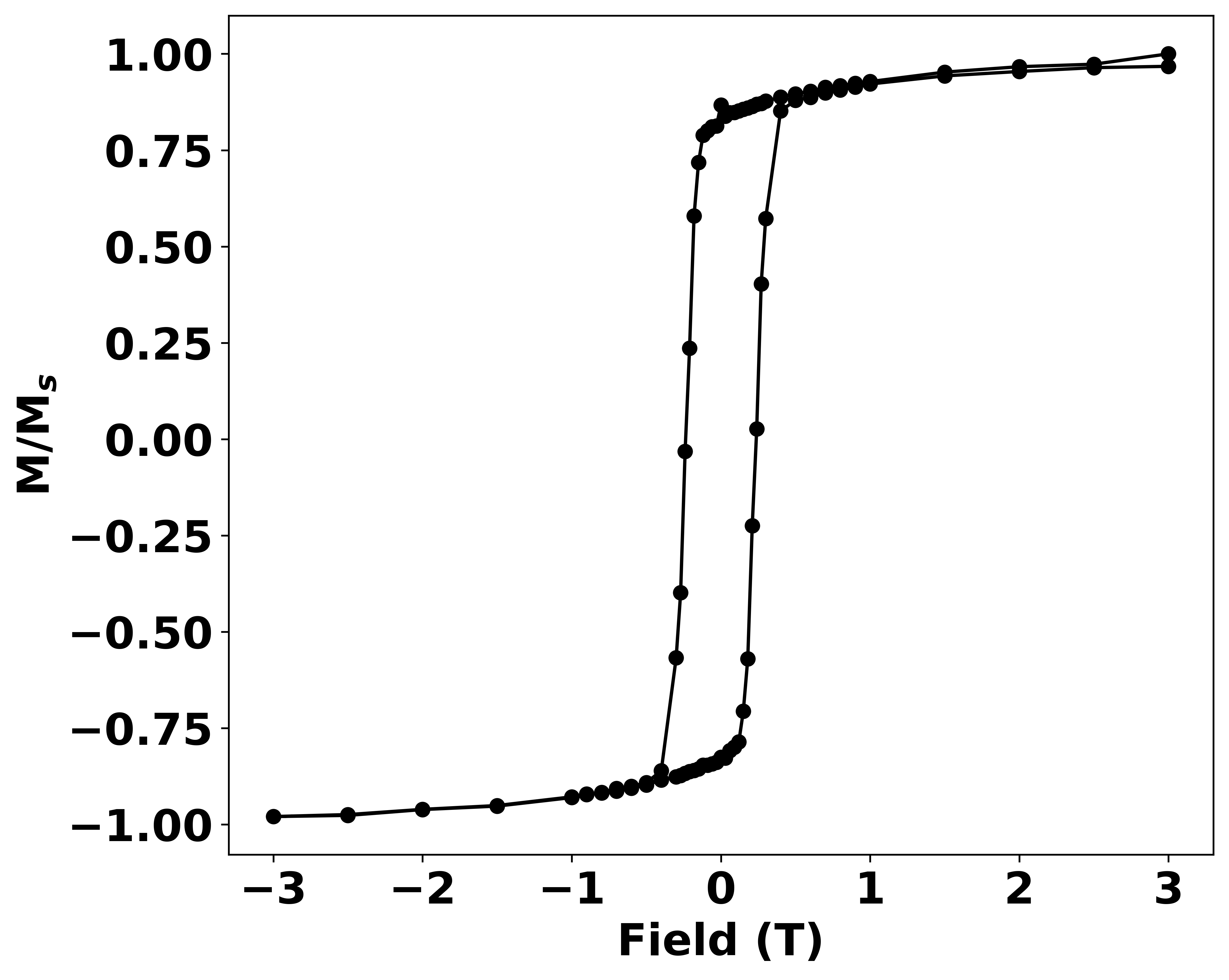}
    \caption{Hysteresis loop of \CT/Graphene/SiC measured at 5 K with a perpendicular applied magnetic field by recording the difference of XMCD signals at the chromium L$_3$ edge (corresponding to the maximum dichroic signal) and at 565 eV (pre-edge).
}
\label{xmcd_loop}
\end{figure}

\section{Layer-resolved magnetic moments in 2D material/\CT\\ heterostructures from \textit{ab initio} calculations}

\begin{figure*}[ht!]
\includegraphics[width=17.2cm]{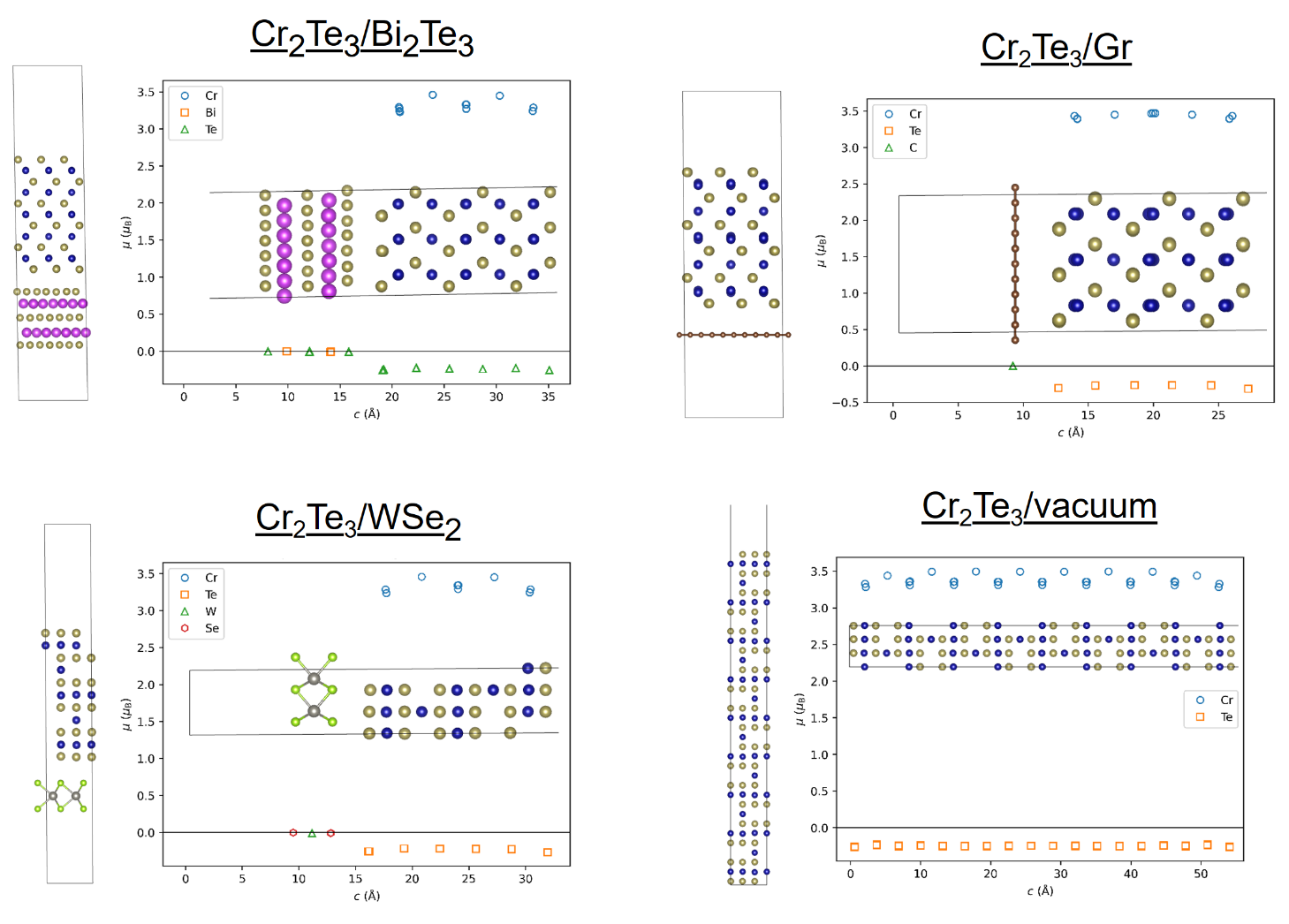}
\caption{Layer-resolved magnetic moments in \textit{ab initio}-calculated \CT/2D material~heterostructures. The magnetic moments of the 2D materials are negligible. Moreover, we do not observe any significant modification of the \CT~magnetic moments at the interface.}
\label{fig:magmom_heterostructures}
\end{figure*}

We simulate heterostructures of \CT~on Bi$_2$Te$_3$, graphene, and WSe$_2$, as well as \CT~slabs suspended in vacuum. None of these show a significant change of the interfacial magnetic moments in \CT~$-$ all the Cr and Te atoms keep their magnetic moments of $\mu_\mathrm{Cr} \approx 3.3$ $\mu_\mathrm{B}$ and $\mu_\mathrm{Te} \approx -0.3$ $\mu_\mathrm{B}$, see Fig.~\ref{fig:magmom_heterostructures}.

\newpage

\section{Metallic resistivity of \CT~layers}

\begin{figure}[h]
    \center
    \includegraphics[width=8.6cm]{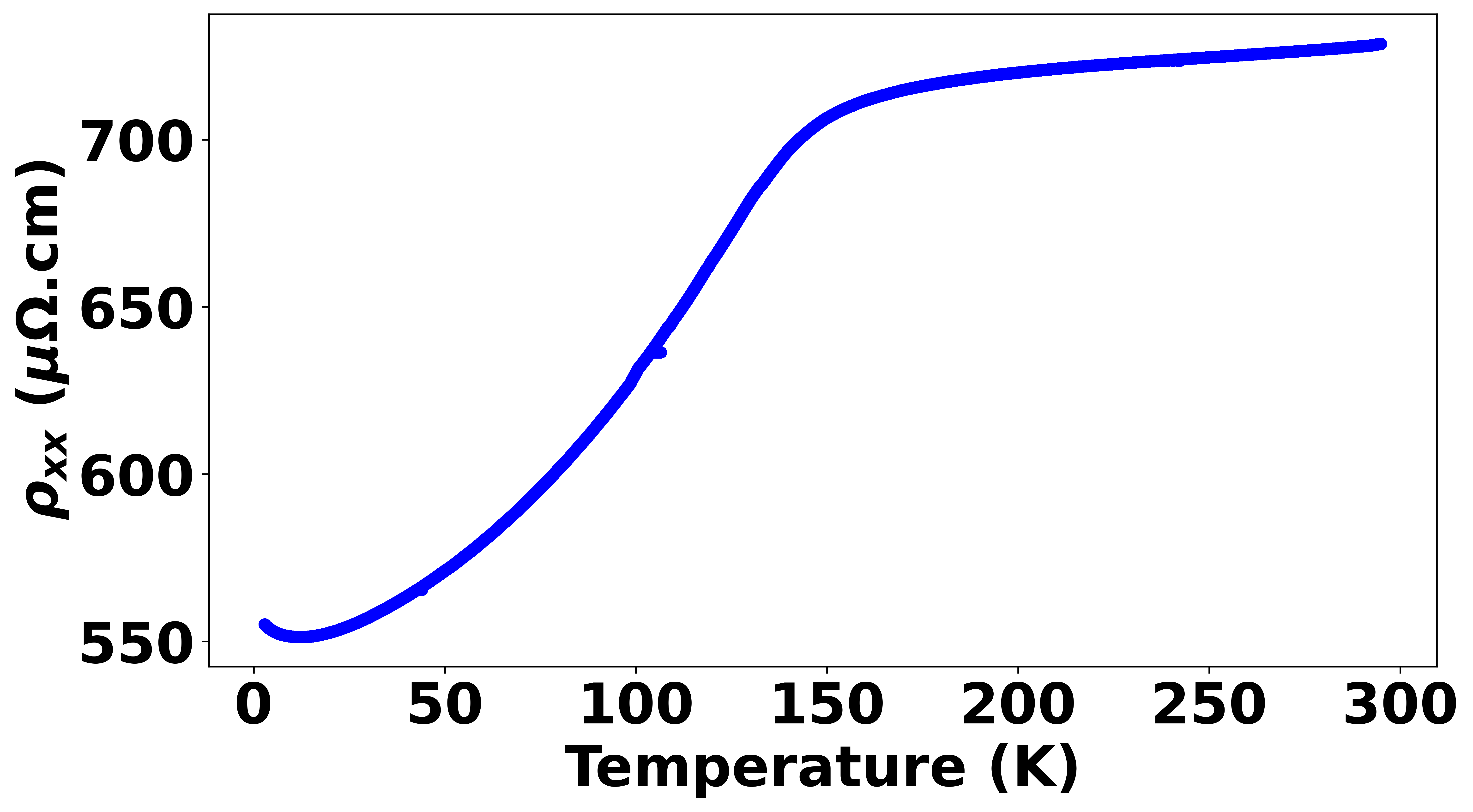}
    \caption{Longitudinal resistivity of \CT~deposited on sapphire versus temperature.
}
\label{Rxx}
\end{figure}

The magnetoresistance of 5 layers \CT~deposited on the insulating substrate Al$_2$O$_3$ is shown in Fig.~\ref{Rxx}. The resistivity is growing with temperature indicating the metallic nature of \CT~layers. A change in the slope is noticeable around the Curie temperature of the material.

\section{Hall resistivity of \CT~on sapphire}

\begin{figure*}[h]
    \center
    \includegraphics[width=17.2cm]{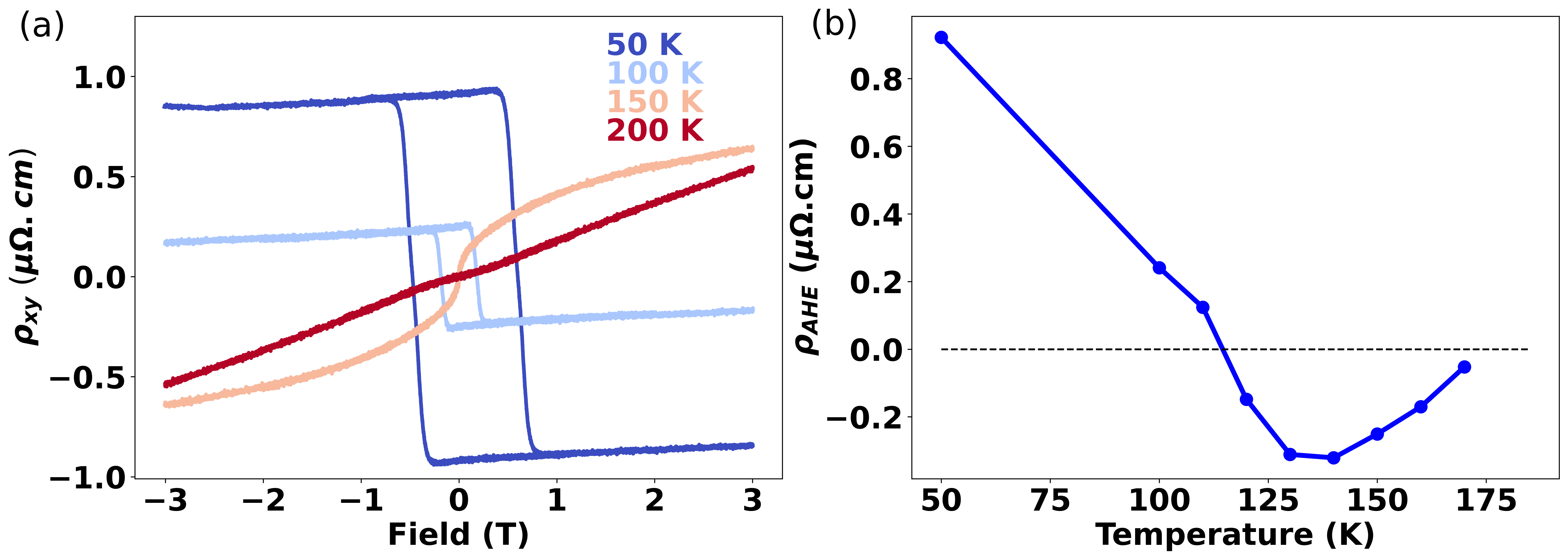}
    \caption{(a) Hall resistivity of \CT~deposited on sapphire at different temperatures. (b) Extracted anomalous Hall resistivity as a function of temperature.
}
\label{Rahe}
\end{figure*}

To understand the Hall signal measured on Cr$_2$Te$_3$/Bi$_2$Te$_3$/Al$_2$O$_3$ annealed at 400°C (when the \BT~layer evaporated), \CT~was directly deposited on sapphire and measured (see Fig.~\ref{Rahe}). The extracted carrier density was found higher (7.0*10$^{15}$ holes/cm$^2$ at 50 K) and a similar sign change of the anomalous Hall resistivity is noticeable below the Curie temperature. The temperature at which the resistivity switches sign occurs below the sample deposited on \BT. This difference is attributed to the presence of defects after removal of \BT~and therefore to different Fermi levels in these two samples.

\section{Anomalous Hall effect sign reversal}
\begin{figure*}
\includegraphics[width=17.2cm]{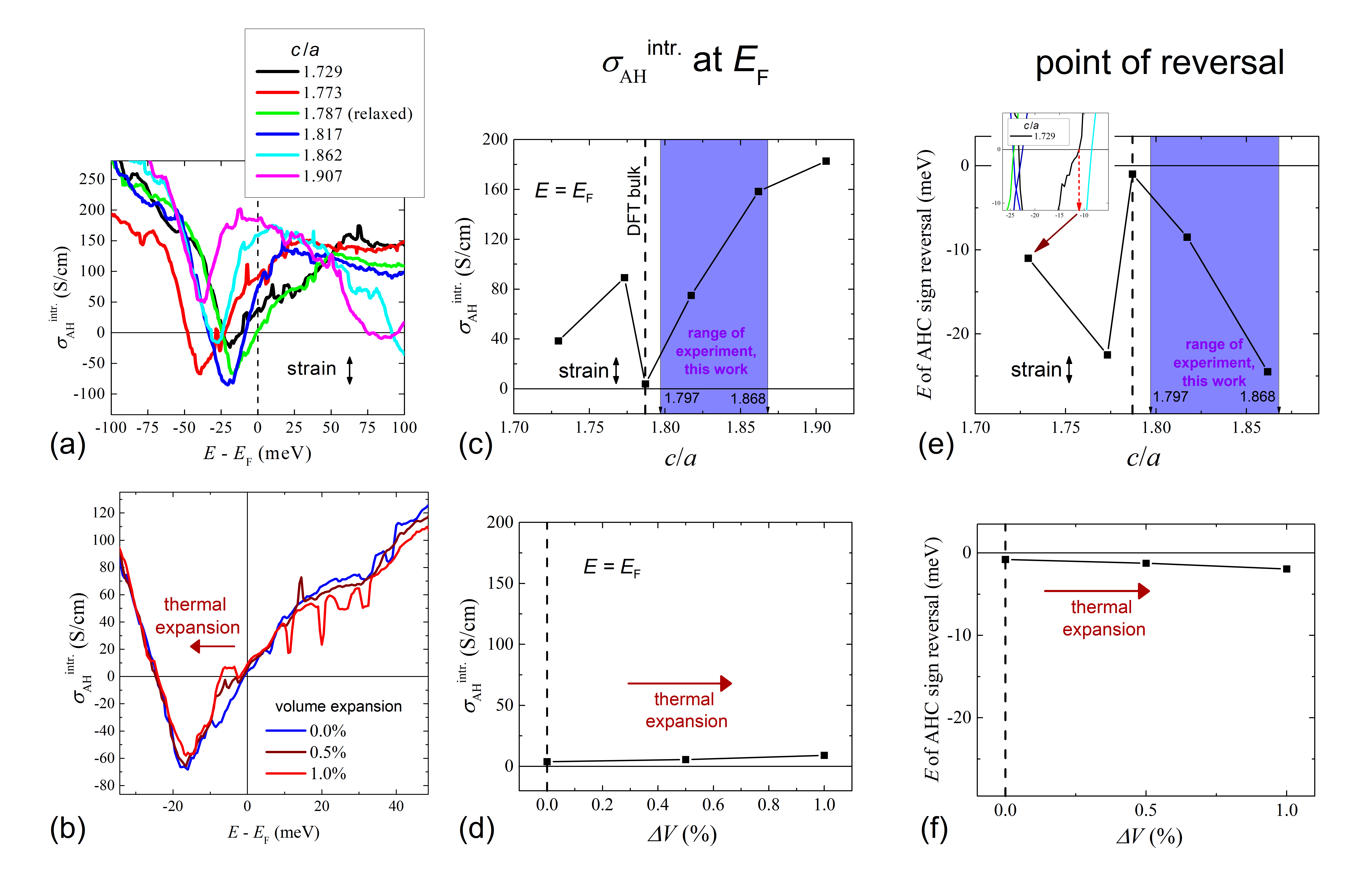}
\caption{(a) The intrinsic AHC $\sigma_\mathrm{AH}^\mathrm{intr.}$ as a function of the Fermi level position for bulk \CT\ under strain and (b) volume expansion. The strain can have a significant effect on AHC. (c) $\sigma_\mathrm{AH}^\mathrm{intr.}$ at the Fermi level as a function of strain and (d) volume expansion. (e) Shift of the (right-most) sign reversal energy as a function of strain and (f) volume expansion.}
\label{fig:AHE_complete}
\end{figure*}

\setlength{\parskip}{1pt plus2pt}
The anomalous Hall effect sign reversal around the Fermi level is sensitive to strain [Fig.~\ref{fig:AHE_complete}(a)], but not to volume expansion [Fig.~\ref{fig:AHE_complete}(d)]. The strain can significantly change both the value of $\sigma_\mathrm{AH}^\mathrm{intr.}$ at $E_\mathrm{F}$ [Fig.~\ref{fig:AHE_complete}(b)] and the Fermi level position at which the sign reversal occurs [Fig.~\ref{fig:AHE_complete}(c)]. In comparison, a realistic thermal volume expansion (0.5\% per 300K \cite{kubota2022large}) does not change the $\sigma_\mathrm{AH}^\mathrm{intr.}$ curve characteristics significantly [Fig.~\ref{fig:AHE_complete}(e)-(f)].

Note that an equivalent calculation in Ref.~\cite{jeon_emergent_2022} does not show the reversal around $E_\mathrm{F}$. We could reproduce this result by neglecting the vdW correction. The vdW correction is, however, important, due to the presence of the pseudo vdW gap in \CT. Otherwise, the relaxed unit cell volume is overestimated (by $\approx 8\%$) compared to both experiment and the calculation with vdW correction included.

A reversal is still present in Ref.~\cite{jeon_emergent_2022}, but at $E-E_\mathrm{F} = 300$~meV. We performed a charge transfer calculation and show that a charge transfer from the 2D materials to \CT\ should indeed increase its Fermi level, but only by 25~meV at most. Therefore a sign reversal region around 300 meV seems unrealistic.

\newpage

\section{Charge transfer between \CT\ and 2D materials}
\begin{figure*}
\includegraphics[width=17.2cm]{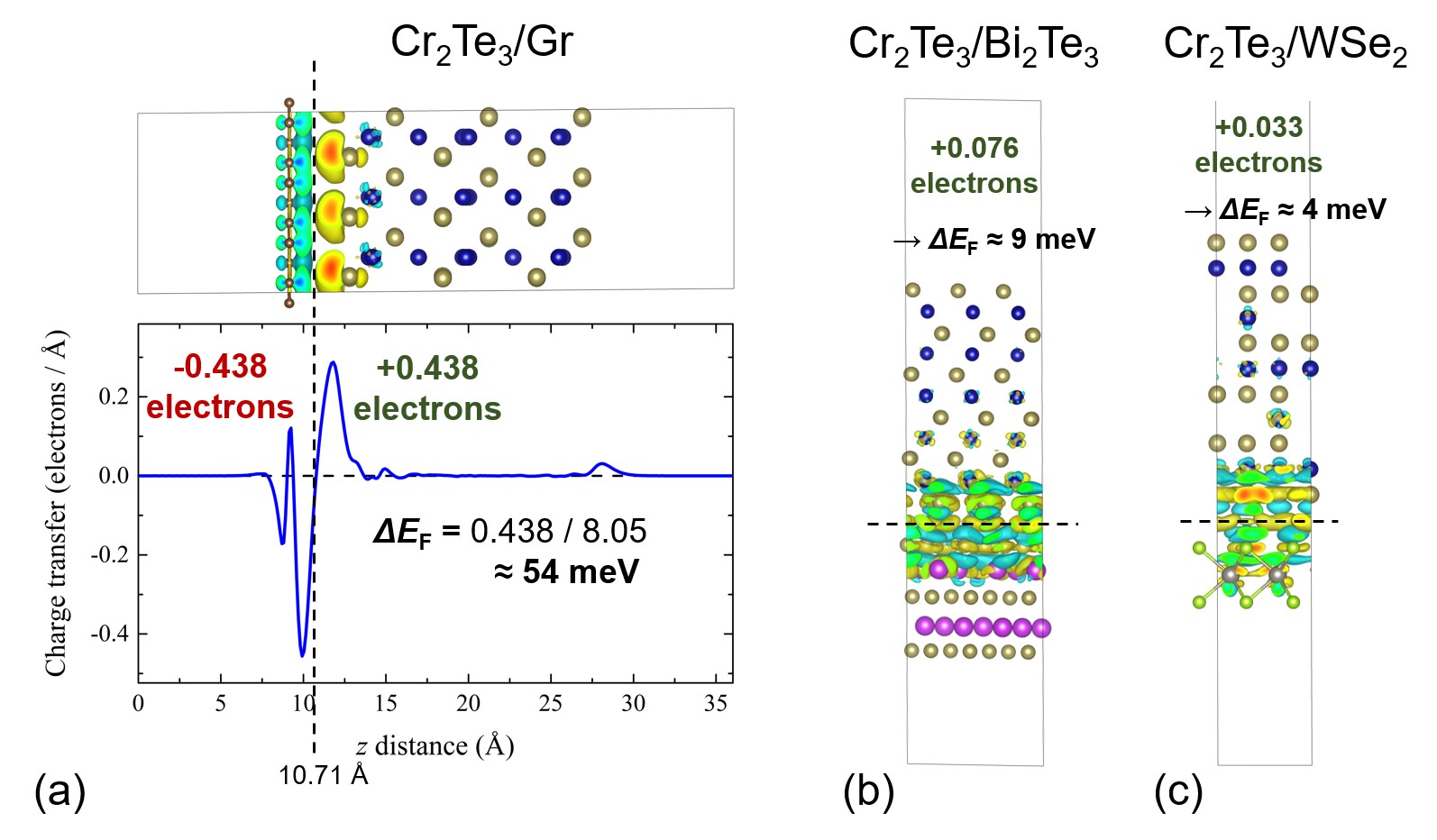}
\caption{Charge transfer calculation for \CT\ on top of (a) graphene, (b) Bi$_2$Te$_3$, and (c) WSe$_2$. Yellow cloud signifies positive and blue signifies negative electron transfer. In green is given the number of electrons (per unit cell) transferred from the 2D material to \CT. An estimation of the Fermi level shift $\Delta E_\mathrm{F}$ compared to bulk \CT\ resulting from this charge transfer is also given.}
\label{fig:charge_transfer}
\end{figure*}

We calculate heterostructures of \CT/2D materials to estimate the charge transfer between the studied 2D materials and \CT\ (see  Fig.~\ref{fig:charge_transfer}). We observe a positive electron transfer to \CT\ in all three cases. The resulting shift of the \CT\ Fermi level compared to its bulk value is estimated by dividing the number of transferred electrons/unit cell by the density of states of bulk \CT\ at $E_\mathrm{F}$, in this case, 8.05~electrons/eV for the considered unit cell.

In the case of graphene, the charge transfer is one order of magnitude larger than for WSe$_2$ or Bi$_2$Te$_3$, leading to an estimated $\Delta E_\mathrm{F} \approx 54$~meV.

\pagebreak

\bibliography{References}